\documentclass[11pt]{article}
\usepackage{amsmath,amssymb}

\begin{document}

\tolerance=5000
\def\be{\begin{equation}}
\def\ee{\end{equation}}
\def\bea{\begin{eqnarray}}
\def\eea{\end{eqnarray}}
\def\nn{\nonumber \\}
\def\bt{\beta}
\def\al{\alpha}
\def\pl{\partial}
\def\gm{\gamma}
\def\dl{\delta}
\def\th{\theta}
\def\ep{\epsilon}
\def\ad{\al\bt\gm\dl}
\def\ms{\mu\nu\rho\sigma}
\newcommand{\s}{\hspace{0.15cm}}
\newcommand{\bc}{\bigcirc}
\newcommand{\mc}{\mathcal}

\renewcommand{\thefootnote}{\fnsymbol{footnote}}

\begin{titlepage}
\begin{flushright}
    NSF-KITP-06-01 \\
    hep-th/0601092 \\
    January 2006
\end{flushright}
\begin{center}
  \vspace{3cm}
  {\bf \Large Higher Derivative Corrections to Eleven Dimensional \\[0.2cm]
              Supergravity via Local Supersymmetry}
  \\ \vspace{2cm}
  Yoshifumi Hyakutake\footnote{E-mail: hyaku@kitp.ucsb.edu}
  and Sachiko Ogushi\footnote{E-mail: ogushi@kitp.ucsb.edu}
   \\ \vspace{1cm}
   {\it Kavli Institute for Theoretical Physics, \\
   University of California, Santa Barbara, CA 93106, USA}
\end{center}

\vspace{2cm}
\begin{abstract}

In this paper we derive higher derivative corrections to the eleven 
dimensional supergravity by applying the Noether method with respect to
the $\mc{N}=1$ local supersymmetry.
An ansatz for the higher derivative effective action, which includes quartic terms of the
Riemann tensor, is parametrized by 132 parameters.
Then we show that by the requirement of the local supersymmetry,
the higher derivative effective action is essentially described by two parameters.
The bosonic parts of these two superinvariants completely match with 
the known results obtained by the perturbative calculations in the type IIA superstring theory.

Since the calculations are long and systematic, we build the computer programming to check
the cancellation of the variations under the local supersymmetry. 
This is an extended version of our previous paper\cite{HO}.

\end{abstract}
\end{titlepage}

\setlength{\baselineskip}{0.65cm}

\setcounter{page}{1}
\tableofcontents
\makeatletter
\renewcommand{\theequation}{\thesection.\arabic{equation}}
\@addtoreset{equation}{section}
\makeatother

\renewcommand{\thefootnote}{\arabic{footnote}}

\addtocounter{footnote}{-2}

\section{Introduction}

The requirement of the local supersymmetry is a powerful tool to restrict the form of an effective action.
Especially the action of the supergravity theory in eleven dimensions is uniquely determined
by imposing the $\mc{N}=1$ local supersymmetry\cite{CJS}.
The supergravity theory in eleven dimensions is well recognized as the low energy effective
theory of the M-theory which contains a membrane as a fundamental object.
Despite the importance of the M-theory, the quantization of the membrane
has not been accomplished yet. So it is impossible to calculate scattering amplitudes
of membranes, and we do not know so much about the M-theory beyond the supergravity approximation.

In this paper we pursue the effective action of the M-theory beyond the supergravity
by imposing the $\mc{N}=1$ local supersymmetry rather than quantizing the membrane.
Especially we want to examine whether the local supersymmetry is powerful
enough to determine the effective action of the M-theory.

Let us briefly review the expected form of the higher derivative corrections 
to the eleven dimensional supergravity, which is obtained by using the duality 
between the M-theory and the type IIA superstring theory.
The perturbative analyses of the scattering amplitudes in the type IIA superstring theory 
is one of the important methods to determine the structure of 
the higher derivative corrections to the supergravity. 
According to these analyses the bosonic part of the corrections to
the type IIA supergravity is given by
\begin{alignat}{3}
  &\mathcal{L}_{(\alpha')^3} \sim e^{-2\phi} I_{\text{tree}} + c \, I_{\text{1-loop}}, \label{eq:IIA}
  \\
  &I_{\text{tree}} = t_8 t_8 e R^4 + \tfrac{1}{4\cdot 2!} \epsilon_{10}\epsilon_{10} e R^4, \notag
  \\
  &I_{\text{1-loop}} = t_8 t_8 e R^4 - \tfrac{1}{4\cdot 2!} \epsilon_{10}\epsilon_{10} e R^4
  - \tfrac{1}{6} \epsilon_{10}t_8 BR^4, \notag
\end{alignat} 
where $c$ is some known constant. The definition of a tensor $t_8$ is given in the appendix \ref{CC}, 
and a tensor $\ep_{10}$ is a completely antisymmetric tensor with 10 indices.
The tree level effective action is obtained by
the four graviton amplitude and the sigma-model computation\cite{GW,GSl,GVZ}. The first two terms of
the one-loop effective action is found by the four graviton amplitude\cite{KP}.
The last term in the one-loop effective action is introduced to ensure the string-string
duality between type IIA on K3 and heterotic string on $T^4$\cite{VW,DLM}.
Under this duality, the last term is related to the Green-Schwarz anomaly cancellation term
in the heterotic string effective action\cite{GS}.

The bosonic part of the higher derivative corrections to the eleven dimensional supergravity 
is obtained by lifting the result~(\ref{eq:IIA}) to eleven dimensions.
Thus there are two candidates which will be invariant under the $\mc{N}=1$ local 
supersymmetry\cite{KP,Ts}.
\begin{alignat}{3}
  t_8 t_8 e R^4 - \tfrac{1}{12} \epsilon_{11}t_8 AR^4, \qquad
  t_8 t_8 e R^4 + \tfrac{1}{4!} \epsilon_{11}\epsilon_{11} e R^4, \label{eq:sinv}
\end{alignat}
where a tensor $\ep_{11}$ is a completely antisymmetric tensor with 11 indices.
For the first part, the supersymmetric counter terms, which are bilinear to the Majorana gravitino,
are derived by applying the 
Noether method with respect to the local supersymmetry\cite{BR}-\cite{PVW}.
For the second part, a complete expression for the supersymmetric counter terms is
not know yet. See ref.~\cite{PVW2} for some discussions.

Besides the approaches by the string perturbation analyses, the dualities and the Noether method 
discussed above, there are other methods to derive the higher derivative effective action of 
the M-theory\cite{GGV}-\cite{DS2}. 
The famous methods are the analyses by computing the scattering amplitudes of superparticles 
in eleven dimensions and the superfield method.
Among these approaches we employ the Noether method because this method maximally respect the local
supersymmetry.

The procedure of the Noether method is well known and quite simple.
First we prepare the ansatz for the higher derivative effective action
in which each term has some unknown coefficient.
Then we consider the variations of the ansatz under the $\mc{N}=1$ local supersymmetry.
The cancellation of these variations gives simultaneous equations among the unknown coefficients
in the ansatz. By solving these equations, we can determine the possible forms of the higher derivative 
effective action. At the same time we also obtain modifications of the supersymmetric
transformation rules. This is the merit of employing the Noether method.

As we are interested in the question whether the local supersymmetry is useful
enough to determine the effective action of the M-theory, basically we have to prepare the ansatz 
as general as possible. Although it is an ideal case, it is impractical to apply the Noether method
to that case because of the enormous calculations.
So it is essential to reduce the number of the 
terms in the ansatz. In fact we will classify the ansatz by examining the number of the covariant 
derivatives after the variation, and drop the terms from the ansatz which include more than one covariant 
derivative explicitly.

After the above prescription there still remain so many terms in the ansatz.
Actually we have 132 terms in the ansatz, and the variations are expanded by 264 bases.
So it is almost impossible to execute this program by hand.
Thus in order to complete this task, we often employ a computer programming.

With the aid of the computer programming, we complete the cancellation of the variations
under the local supersymmetry. As a result the number of the parameters in the ansatz are essentially 
reduced to only two, and the higher derivative effective action is given by a linear combination
of two superinvariants. Surprisingly these two superinvariants completely match with those
in the eq.~(\ref{eq:sinv}). Therefore at this stage it seems that the $\mc{N}=1$ local supersymmetry is powerful
enough to determine the structure of the higher derivative effective action of the M-theory.

The content of our paper is as follows.
In section \ref{sec:sugra}, we review the derivation of the eleven dimensional supergravity in detail.
In section \ref{sec:ov}, an overview of the procedure to determine the higher derivative corrections
is explained.
We give detailed explanations of the higher derivative corrections in section \ref{sec:dt}.
The 132 terms in the ansatz and the variations of them are explicitly written down.
These variations are expanded by the 264 bases, but 20 of them depend on the other terms by nontrivial identities.
Almost all the terms in the ansatz and the bases in the variations are derived both by hand and 
by the computer programming independently.
In section \ref{sec:res} we show that the local supersymmetry essentially reduce the number of 
the parameters to only two, and these two superinvariants are exactly match with the eq.(\ref{eq:sinv}).
We also mention about the modifications of the supersymmetric transformation rules. 
Section \ref{sec:con} is devoted to the conclusions and discussions.

\section{Review of the Eleven Dimensional Supergravity}\label{sec:sugra}

\subsection{The Ansatz for the Action and the Supersymmetric Transformations}

The eleven dimensional supergravity is the low energy effective theory of the M-theory
which is considered to be a strong coupling limit of the type IIA superstring theory.
The local supersymmetry of this theory possesses 32 super charges and 
determines the structure of the effective action uniquely\cite{CJS}.

In this section we briefly review the derivation of the eleven dimensional supergravity.
First we write down all possible terms for the effective action and the supersymmetric transformations
with arbitrary coefficients. Then we employ the Noether method to fix these coefficients
by the local supersymmetry. This section is placed to be preliminaries of later sections
where higher derivative corrections to the supergravity are discussed.
So readers who are only interested in the higher derivative corrections may skip this section
and refer this part like an appendix.

The field contents of the supergravity in eleven dimensions are quite simple. 
First of all we begin with a vielbein $e^\mu{}_a$ and a Majorana gravitino $\psi_\mu$
which have 44 and 128 physical degrees of freedom, respectively. In order to balance the numbers
of bosonic and fermionic states, we need a three-form potential $A_{\mu\nu\rho}$ which has 84 
physical degrees of freedom.
The Greek indices, $\mu, \nu, \rho, \cdots$, label the coordinates in the curved space-time and the Latin indices, $a,b,c,\cdots,$ refer to the local Lorentz coordinates, both of them run from 0 to 10.  
Spinor indices are neglected to make expressions simple. 

The action of the supergravity in eleven dimensions should be constrained by
the local symmetries and the parity invariance, which are listed as
\begin{alignat}{3}
  &\text{1. The general coordinate invariance and the local Lorentz invariance.} \notag \\ 
  &\text{2. The abelian gauge symmetry :}  \; A \to A + d\Lambda. \notag \\ 
  &\text{3. The $\mathcal{N}=1$ local supersymmetry.} \label{sym} \\
  &\text{4. The parity invariance :} \; x^{10} \to - x^{10},\; 
  A \to -A,\; \psi \to \gamma^{10} \psi, \notag
\end{alignat}
where $A$, $\Lambda$ and $\psi$ are 3-form, 2-form and 1-form, respectively.
The normalization of the forms is defined so as the sum of all the coefficient becomes one,
i.e., $A = \frac{1}{3!}A_{\mu\nu\rho}dx^\mu \wedge dx^\nu \wedge dx^\rho$ for example.
The matrices $\gamma^\mu$ are the gamma matrices in eleven dimensions and the $\gamma^{10}$
generates the parity transformation for the spinor indices.
The notation $\gamma^{\rho_1 \cdots \rho_n}$ is used to represent the product of 
$n$ gamma matrices whose indices are completely antisymmetrized.
The coefficient of each term is $\frac{1}{n!}$, so $\gamma^{\rho_1\rho_2}=\tfrac{1}{2!}
(\gamma^{\rho_1}\gamma^{\rho_2}-\gamma^{\rho_2}\gamma^{\rho_1})$ for instance.

Due to the local symmetries of 1 and 2 in the eq.~(\ref{sym}), 
the building blocks of the supergravity action should be
a Riemann tensor $R^{ab}{}_{\mu\nu}$, a 4-form field strength $F_{\mu\nu\rho\sigma}$, a covariant
derivative $D_\mu$ which acts only on the local Lorentz indices, and a bilinear term of the Majorana 
gravitino $\bar{\psi}_\mu \gamma^{\rho_1 \cdots \rho_n} \psi_\nu$.  
A Chern-Simons term $A \wedge F \wedge F$ is an exception, which 
include the 3-form potential explicitly but is still gauge invariant. 

The requirement of the local supersymmetry 3 in the eq.~(\ref{sym}) is useful to determine the action
of the eleven dimensional supergravity. In fact we derive the action uniquely 
by employing the Noether method. 
The parity invariance 4 in the eq.~(\ref{sym}) ensures the duality 
between the M-theory on $S^1/\mathbb{Z}_2$ and 
the heterotic superstring theory. As we will see later, the action obtained by imposing the local
supersymmetry satisfies the parity invariance automatically.
For the supergravity this symmetry is not so helpful to determine the action, but becomes
useful to restrict the ansatz for the higher derivative corrections.

Let us consider the ansatz for the Lagrangian.
Since the gravitational coupling constant has the length dimension of $[L]^{9}$,
the ansatz consists of terms with the length dimension of $[L]^{-2}$. 
The building blocks have dimensions like
$R^{ab}{}_{\mu\nu}=[L]^{-2}$, $F_{\mu\nu\rho\sigma}=[L]^{-1}$, $D_{\mu}=[L]^{-1}$ and
$\bar{\psi}_\mu \gamma^{\rho_1 \cdots \rho_n} \psi_\nu=[L]^{-1}$.
So the ansatz for the Lagrangian is written by
\begin{alignat}{3}
  \mathcal{L}_0 &= \mathcal{L}[eR] + \mathcal{L}[e \bar{\psi}\psi_{(2)}] 
  + \mathcal{L}[eF^2] + \mathcal{L}[eF\bar{\psi}\psi] + \mathcal{L}[e\epsilon_{11} AF^2]
  + \mathcal{O}(\psi^4), \label{eq:actsugra}
\end{alignat}
where we used a simple notation $[X]$. This represents a set of terms which become $X$ after 
ignoring their indices, coefficients and gamma matrices.
For example, ${\cal L}[eF\bar{\psi}\psi]$ includes a term
$e F^{\ms} \bar{\psi}_\mu \gm_{\rho\sigma} \psi_\nu$.
Since we examine the cancellation of variations under supersymmetry transformations
up to $\mathcal{O}(\psi^2)$, we only write the Lagrangian up to $\mathcal{O}(\psi^4)$.
The first three parts are the usual kinetic terms and last two parts
are the interaction terms, whose explicit expressions are given by
\begin{alignat}{3}
  &{\cal L}[eR] &&= e R, 
  \notag\\
  &{\cal L}[e\bar{\psi}\psi_{(2)}] &&= 
  - \tfrac{1}{2} e \bar{\psi}_\rho \gamma^{\rho\mu\nu} \psi_{\mu\nu} , 
  \notag\\
  &{\cal L}[eF^2] &&= - \tfrac{1}{48} e F_{\ms} F^{\ms}, \label{eq:actsugra2}
  \\
  &{\cal L}[eF\bar{\psi}\psi] &&= 
  c_1 e F^{\ms} \bar{\psi}_\mu \gm_{\rho\sigma} \psi_\nu
  + c_2 e F_{\ad} \bar{\psi}_\mu \gm^{\mu\nu\ad} \psi_\nu 
  \notag\\&&&\quad\,
  + c_3 e F_{\ad} \bar{\psi}_\mu \gm^{\ad} \psi^{\mu}
  + c_4 e F_{\al\bt\gm}{}^{\mu} \bar{\psi}_{(\mu} \gm^{\nu\al\bt\gm} \psi_{\nu)}, 
  \notag\\
  &{\cal L}[e\epsilon_{11}AF^2] &&= 
  c_5 e \epsilon_{11}^{\mu_1 \cdots \mu_{11}} 
  A_{\mu_1\mu_2\mu_3} F_{\mu_4\cdots\mu_7} F_{\mu_8\cdots\mu_{11}}.
  \notag
\end{alignat}
The coefficients of the kinetic terms are fixed by rescaling the fields.
On the other hand, the coefficients $c_n (n=1,\cdots,5)$
should be fixed by the local supersymmetry.
The indices in the brackets $(\;)$ or $[\;]$ are completely symmetrized or 
antisymmetrized respectively.
The $\epsilon_{11}^{\mu_1\cdots \mu_{11}}$ is an antisymmetric tensor defined as
$\epsilon_{11}^{\mu_1\cdots \mu_{11}} = e^{[\mu_1}{}_{a_1} \cdots e^{\mu_{11}]}{}_{a_{11}}
\epsilon_{11}^{a_1 \cdots a_{11}}$, and $\epsilon_{11}^{01\cdots 10}=1$ for local Lorentz indices.
Other notations we employed here are summarized in the appendix \ref{notations}.
It is easy to see that the Lagrangian is invariant under the parity transformation.

Now we introduce a space-time dependent parameter $\epsilon$ which transforms as a 
Majorana spinor. The dimension of $\epsilon$ is $[L]^{1/2}$, and
the supersymmetric transformations are assumed as
\begin{alignat}{3}
  \delta_0 e^{\mu}{}_a &= -\bar{\ep} \gamma^{\mu} \psi_a,
  \notag\\
  \delta_0 \psi_\mu &= d_1 D_\mu \ep + d_2 F_{\mu jkl} \gamma^{jkl} \epsilon
  + d_3 F_{ijkl} \gm^{ijkl}{}_\mu \epsilon + \mathcal{O}(\psi^2),  \label{eq:susytr}
  \\
  \delta_0 A_{\mu\nu\rho} &= d_4 \bar{\epsilon}\gamma_{[\mu\nu}\psi_{\rho]}
  + d_5 \bar{\epsilon} \gamma_{\mu\nu\rho\sigma} \psi^{\sigma}.
  \notag
\end{alignat}
Again terms which do not contribute to the cancellation up to $\mathcal{O}(\psi^2)$ are neglected.
The coefficients $d_n (n=1,\cdots,5)$ can be fixed by the local supersymmetry.
Note that the right sides of the above equations behave correctly under 
the parity transformation. This means that the supersymmetry transformations do not
mix parity even terms and parity odd terms. Therefore the local supersymmetry is 
compatible with the parity invariance.

\subsection{The Local Supersymmetry}

In this subsection we review the variations under the local supersymmetry.
The requirement of the local supersymmetry in eleven dimensions 
completely determines the coefficients $c_n$ and $d_n$ in the ansatz
in eqs.(\ref{eq:actsugra2}) and the transformation rules (\ref{eq:susytr}). 

In order to examine the local supersymmetry, 
it is convenient to employ the 1.5 order formalism. 
That is, the variations of the Lagrangian by the supersymmetric 
transformations are understood as
\begin{alignat}{3}
  \delta_0 \mathcal{L}_0 &= e \delta_0 e^\mu{}_a E(e)^a{}_\mu
  + e \delta_0 \bar{\psi}_\mu E(\psi)^\mu
  + e \delta_0 A_{\mu\nu\rho} E(A)^{\mu\nu\rho}
  + e \delta_0 \omega_\mu{}^{ab} E(\omega)^\mu{}_{ab} \notag
  \\
  &= e \delta_0 e^\mu{}_a E(e)^a{}_\mu |_{\omega(e,\psi)} 
  + e \delta_0 \bar{\psi}_\mu E(\psi)^\mu |_{\omega(e,\psi)} 
  + e \delta_0 A_{\mu\nu\rho} E(A)^{\mu\nu\rho} |_{\omega(e,\psi)}, \label{eq:varLag}
\end{alignat}
where $E(e)^a{}_\mu = e^{-1} \frac{\delta \mathcal{L}}{\delta e^\mu{}_a}$,
$E(\psi)^\mu = e^{-1} \frac{\delta \mathcal{L}}{\delta \bar{\psi}_\mu}$,
$E(A)^{\mu\nu\rho} = e^{-1} \frac{\delta \mathcal{L}}{\delta A_{\mu\nu\rho}}$ and
$E(\omega)^\mu{}_{ab} = e^{-1} \frac{\delta \mathcal{L}}{\delta \omega_\mu{}^{ab}}$
are field equations.
In the first line the spin connection $\omega_\mu{}^{ab}$ is treated as an independent field.
To go to the second line the vielbein postulate and the field equation for the spin connection,
\begin{alignat}{3}
  &D_\nu e^\mu{}_a + \Gamma^\mu{}_{\nu\rho} e^\rho{}_a = 0,
  \notag\\
  &E(\omega)^\mu{}_{ab}=- D_\nu (2 e e^{[\nu}{}_a e^{\mu]}{}_b)
  - \tfrac{1}{8} e \bar{\psi}_\rho \gamma^{\rho\mu\nu} \gamma^{ab} \psi_\nu=0, \label{spin}
\end{alignat}
are solved. From these equations the spin connection is expressed like
\begin{alignat}{3}
  (\omega_{\rho})^{ab} &= - e^{a\mu} e^{b\nu} e_{\rho c} \partial_{[\mu} e^c{}_{\nu]} +
  e^{b\nu} \partial_{[\nu} e^a{}_{\rho]} - e^{a\mu} \partial_{[\mu} e^b{}_{\rho]} 
  \notag\\&\quad\,
  + \tfrac{1}{4} \bar{\psi}^a \gamma_\rho \psi^b
  - \tfrac{1}{4} \bar{\psi}^b \gamma^a \psi_{\rho}
  + \tfrac{1}{4} \bar{\psi}^a \gamma^b \psi_{\rho}
  + \tfrac{1}{8} \bar{\psi}^\alpha \gamma^{ab}{}_{\rho\alpha\beta} \psi^\beta.
\end{alignat}
Thus the variations for the spin connection are trivially canceled in the case of the supergravity.
Note, however, that the variations fo the spin connection
in the case of higher derivative effective action do not cancel automatically.
We will mention this in section \ref{sec:dt}.
The field equations for the vielbein, 
the Majorana gravitino and the 3-form potential are calculated as
\begin{alignat}{3}
  E(e)^a{}_\mu &= 
  2 R^a{}_\mu - e^a{}_\mu R - \tfrac{1}{6} F^{aijk} F_{\mu ijk} 
  + \tfrac{1}{48} e^a{}_\mu F_{ijkl} F^{ijkl} + \mathcal{O}(\psi^2) ,
  \notag\\
  E(\psi)^\mu &=
  - \gamma^{\mu ab} \psi_{ab}
  + 2 c_1 F^{\mu ijk} \gm_{ij} \psi_k
  + 2 c_2 F_{ijkl} \gm^{\mu ijklm} \psi_m 
  \notag\\&\quad\,
  + 2 c_3 F_{ijkl} \gm^{ijkl} \psi^\mu
  + 2 c_4 F_{ijk}{}^{(\mu} \gm^{l)ijk} \psi_l + \mathcal{O}(\psi^3), \label{eq:EoM}
  \\
  E(A)^{\mu\nu\rho} &=
  \tfrac{1}{6} e^\mu{}_a e^\nu{}_b e^\rho{}_c D_d F^{dabc} 
  + c_5 \ep_{11}^{\mu\nu\rho ijklmnop} F_{ijkl} F_{mnop}
  + \mathcal{O}(\psi^2) .
  \notag
\end{alignat}
To derive these equations we neglect the torsion parts which are the order of $\mathcal{O}(\psi^2)$.
%\begin{alignat}{3}
%  D_\mu (e e^\mu{}_{[a} e^{\mu_1}{}_{a_1} \cdots e^{\mu_n}{}_{a_n]}) 
%  &= e T^\mu{}_{\mu [a} e^{\mu_1}{}_{a_1} \cdots e^{\mu_n}{}_{a_n]}
%  + e \Gamma^{\mu_1}{}_{[aa_1} e^{\mu_2}{}_{a_2} \cdots e^{\mu_n}{}_{a_n]}
%  \notag\\&\quad\,
%  + \cdots + e \Gamma^{\mu_n}{}_{[aa_n} e^{\mu_1}{}_{a_1} \cdots e^{\mu_{n-1}}{}_{a_{n-1}]}
%  \\
%  &= \mathcal{O}(\psi^2),
%  \notag
%\end{alignat} 

Now let us consider the supersymmetric variations of the ansatz.
From the eqs.~(\ref{eq:susytr}) and (\ref{eq:EoM}), the variations of
the Lagrangian (\ref{eq:varLag}) under the local supersymmetry transformations 
are sketched as
\begin{alignat}{3}
  \delta_0 \mathcal{L}_0 &= [eR\bar{\ep}\psi] \oplus [eF\bar{\ep}D\psi] \oplus [eDF\bar{\ep}\psi]
  \oplus [eF^2\bar{\ep}\psi],
\end{alignat}
where $[X]$ represents a set of terms which become $X$ after 
ignoring their indices, coefficients and gamma matrices.
The cancellation of the variation, $\delta_0 \mathcal{L}_0=0$, 
gives linear equations among the coefficients $c_n$ and $d_n$. 

The calculations in detail are executed as follows. 
First of all, the terms in $[eR\bar{\ep}\psi]$
come from the variations of ${\cal L} [eR]$ and ${\cal L }
[e\bar{\psi}\psi_{(2)}]$ in the eq.~(\ref{eq:actsugra2}).
These terms are free from the 4-form field strength and calculated as
\begin{alignat}{3}
  [eR\bar{\ep}\psi] = (- 2 + d_1) e (R_{ab} - \tfrac{1}{2} \eta_{ab} R) 
  \bar{\epsilon} \gamma^b \psi^a, \label{eq:cancel1}
\end{alignat}
where we used the relation
\begin{alignat}{3}
  D_{[e} \psi_{cd]} &= \tfrac{1}{4} R_{ab[cd} \gamma^{ab} \psi_{e]} + \mc{O}(\psi^3).
\end{alignat}
The vanishing of the eq.~(\ref{eq:cancel1}) leads to $d_1 = 2$. 

Next, the terms in $[eF\bar{\ep}D\psi]$ are derived from the variations
of ${\cal L} [e\bar{\psi}\psi_{(2)}]$ and ${\cal L }
[e F \bar{\psi}\psi]$ in the eq.~(\ref{eq:actsugra2}).
These terms are linear to the 4-form field strength as 
\begin{alignat}{3}
  [eF\bar{\ep}D\psi] &= 
  + (- 6 d_2 - 84 d_3 - 2 c_1) e F_{ijkl} \bar{\epsilon} \gm^{kl} \psi^{ij} 
  \notag\\&\quad\,
  + (6 d_2 + 48 d_3) e F_{ijkl} \bar{\epsilon} \gm^{mjkl} \psi^i{}_m 
  \notag\\&\quad\,
  + (d_2 + 5 d_3 - 2 c_2) e F_{ijkl} \bar{\epsilon} \gm^{ijklab} \psi_{ab} 
  \\&\quad\,
  - 4 c_3 F_{ijkl} \gm^{ijkl} D_a \psi^a
  \notag\\&\quad\,
  - 4 c_4 F_{ijk}{}^{l} \gm^{mijk} D_{(l} \psi_{m)}.
  \notag
\end{alignat}
These variations are canceled when $c_1=-18d_3$, $c_2=-\frac{3}{2}d_3$, $c_3=c_4=0$ and $d_2=-8d_3$.
The terms in $[eDF\bar{\ep}\psi]$ which are also linear to the 4-form field strength
come from the variations of ${\cal L} [eF^2]$ and ${\cal L }
[e F \bar{\psi}\psi]$ in the eq.~(\ref{eq:actsugra2}).
\begin{alignat}{3}
  [eDF\bar{\ep}\psi] &= + (72 d_3 + \tfrac{1}{6}d_4) e D^i F_{ijkl} \bar{\ep} \gm^{jk} \psi^l 
  \notag\\&\quad\,
  + \tfrac{1}{6} d_5 e D^i F_{ijkl} \bar{\ep} \gm^{jklm} \psi_m.
\end{alignat}
The cancellation of the right hand sides fixes the coefficients as
$d_4 = -432 d_3$ and $d_5=0$.

Finally, after tedious gamma calculations, 
the terms in $[eF^2\bar{\ep}\psi]$ which are quadratic to
the 4-form field strength come from the variations of ${\cal L} [eF^2]$, ${\cal L }
[e F \bar{\psi}\psi]$ and ${\cal L }[e \epsilon_{11} A F^2 ]$ in the eq.~(\ref{eq:actsugra2}).
\begin{alignat}{3}
  [eF^2\bar{\ep}\psi] &=
  + (\tfrac{1}{6} - 3456 d_3^2) e F_{aijk} F^{bijk} \bar{\epsilon} \gm^a \psi_b
  \notag\\&\quad\,
  +(-\tfrac{1}{48} + 432 d_3^2) e F_{ijkl} F^{ijkl} \bar{\epsilon} \gm^a \psi_a
  \\&\quad\,
  + (- 18 d_3^2 - 2592 d_3 c_5) e F_{abcd} F_{efgh} \bar{\epsilon} \gm^{abcdefghi} \psi_i.
  \notag
\end{alignat}
These terms vanish when $d_3=\frac{1}{144}$ and $c_5=-\frac{1}{(144)^2}$.

To summarize so far, we started from the ansatz (\ref{eq:actsugra}) and (\ref{eq:susytr})
with the coefficients $c_n$ and $d_n$. The requirement of the local supersymmetry
fixes these coefficients uniquely as
\begin{alignat}{5}
  &c_1 = - \tfrac{1}{8}, \qquad&& c_2 = - \tfrac{1}{96}, \qquad&& c_3 = 0, \qquad&&
  c_4 = 0, \qquad&& c_5 = - \tfrac{1}{(144)^2},
  \notag\\
  &d_1 = 2, \qquad&& d_2 = - \tfrac{1}{18}, \qquad&& d_3 = \tfrac{1}{144}, \qquad&& d_4 = -3 ,\qquad&&
  d_5 = 0.
\end{alignat}
In a similar way we will determine the structure of higher derivative effective action 
in subsequent sections.

Before ending this section, for later use, we summarize the relations obtained by
deforming the field equations.
\begin{alignat}{3}
  \gamma^{ab} \psi_{ab} &= - \tfrac{1}{9} \Phi + \mc{O}(\psi^3), \notag
  \\
  \gamma^b \psi_{ab} &= \tfrac{1}{2} \Phi_a - \tfrac{1}{18} \gamma_a \Phi + \mc{O}(\psi^3), \notag
  \\
  \gamma^c D_c \psi_{ab} &= \tfrac{1}{4} \gamma^e \gamma^{cd} \psi_e R_{ab cd}
  - \gamma^c \psi_{[a} R_{b]c} - D_{[a} \Phi_{b]} - \tfrac{1}{9} \gamma_{[a} D_{b]} \Phi 
  + \mc{O}(\psi^3), \label{eom}
  \\
  D^b \psi_{ab} &= \tfrac{1}{4} \gamma^{cd} \psi^b R_{ab cd}
  - \tfrac{1}{2} R_{ab} (\gamma^{bc} \psi_c - \psi^b) - \tfrac{1}{4} R \psi_a \notag
  \\
  &\quad\;+ \tfrac{1}{2} \gamma^b D_b \Phi_a + \tfrac{1}{18} \gamma_{ab} D^b \Phi 
  + \mc{O}(\psi^3), \notag
\end{alignat}
where we defined
\begin{alignat}{3}
\label{PE}
  \Phi^a &= E(\psi)^a + \tfrac{1}{4} F^{aijk} \gm_{ij} \psi_k
  + \tfrac{1}{48} F_{ijkl} \gm^{aijklm} \psi_m , \notag
  \\
  \Phi &= \gamma_a \Phi^a.
\end{alignat}
Note that all indices are local Lorentz ones and we neglect the torsion parts.
The Ricci tensor, the scalar curvature, the vector $\Phi^a$ and the scalar $\Phi$ are 
proportional to the field equations $E(e)^a{}_\mu$ or $E(\psi)^a$
when the four-form field strengths are neglected. Thus up to the field equations,
we obtain the following relations.
\begin{alignat}{3}
  \gamma^c D_c \psi_{ab} &\sim \tfrac{1}{4} \gamma^e \gamma^{cd} \psi_e R_{ab cd}, \notag
  \\
  D^b \psi_{ab} &\sim \tfrac{1}{4} \gamma^{cd} \psi^b R_{ab cd}. \label{feq}
\end{alignat}
The parts which depend only on the field equation of the Majorana gravitino are given by
\begin{alignat}{3}
  \gamma^c D_c \psi_{ab} &\sim \big( - \eta_{d[a} \eta_{b]c} 
  + \tfrac{1}{9} \eta_{d[a} \gm_{b]} \gm_c \big) D^d E(\psi)^c, \notag
  \\
  D^b \psi_{ab} &\sim \tfrac{1}{2} \big( \eta_{ac}\gm_b + \tfrac{1}{9}\gm_{ab}\gm_c \big)
  D^b E(\psi)^c. \label{feq2}
\end{alignat}
These relations will often be used to evaluate variations of the higher derivative
corrections.

\section{Higher Derivative Corrections : An Overview}\label{sec:ov}

Since the higher derivative corrections to the supergravity are too complicated, before discussing
details, let us have an overview of the construction of the higher derivative effective action.
The explicit forms of the ansatz for the action and the variations under the
local supersymmetry will be explained in the next section.

Let us consider the ansatz for the higher derivative corrections to the supergravity.
Since there are so many ways to contract indices,
in general we have enormous terms in the action.
Of course such situation is not suitable to apply the Noether method,
and we should make the ansatz for the action as simple as possible.

First of all we restrict ourselves to investigate leading order corrections to the
supergravity which start from the order of $\ell_p^6$.
\begin{alignat}{3}
  \mathcal{L} = \mathcal{L}_0 + \ell_p^6 \mathcal{L}_1. \label{eq:laghd}
\end{alignat}
Here $\ell_p$ is the unit length in eleven dimensions, and 
the field contents are the vielbein, the Majorana gravitino and the 3-form potential. 
The assumption that the leading corrections 
start from the order of $\ell_p^6$ is consistent with the duality between the M-theory 
and the type IIA superstring theory. For the IIA superstring theory 
the leading corrections are obtained from the scattering amplitude of closed strings, 
and contain quartic terms of the Riemann tensor, $\ell_s^6 R^4$.
Here $\ell_s$ is the string length and related to the unit length of the 11 dimensions
as $\ell_p = g_s^{1/3} \ell_s$.

Now let us consider the field redefinitions of 
\begin{alignat}{3}
  e^\mu{}_a &\to {e'}^\mu{}_a = e^\mu{}_a + \ell_p^6 \Delta e^\mu{}_a, \notag
  \\
  \psi_\mu &\to {\psi'}_\mu = \psi_\mu + \ell_p^6 \Delta \psi_\mu, 
  \\
  A_{\mu\nu\rho} &\to {A'}_{\mu\nu\rho} = A_{\mu\nu\rho} + \ell_p^6 \Delta A_{\mu\nu\rho}. \notag
\end{alignat}
Then the Lagrangian changes like
\begin{alignat}{3}
  \mc{L} &\;\rightarrow\; \mc{L}' = \mc{L}_0 + \ell_p^6 \big\{ \mc{L}_1 
  + e \Delta e^\mu{}_a E(e)^a{}_\mu + e \Delta \bar{\psi}_\mu E(\psi)^\mu 
  + e \Delta A_{\mu\nu\rho} E(A)^{\mu\nu\rho} \big\} 
  + \mc{O}(\ell_p^{12}), \label{eq:redef}
\end{alignat}
where $E(e)^a{}_\mu$, $E(\psi)^\mu$ and $E(A)^{\mu\nu\rho}$ are the field equations 
for $e^\mu{}_a$, $\psi_\mu$ and $A_{\mu\nu\rho}$, respectively.
This means that the higher derivative terms which depend on the field equations of the supergravity
can be removed by the appropriate field redefinitions\cite{Ts2}.
Thus in order to make the ansatz as simple as possible,
we only consider the higher derivative terms which are independent of the field equations.

The supersymmetric transformation rules for the fields should also be modified from 
the order of the $\ell_p^6$.
\begin{alignat}{3}
  \delta e^\mu{}_a &= \delta_0 e^\mu{}_a + \ell_p^6 \delta_1 e^\mu{}_a, \notag
  \\
  \delta \psi_\mu &= \delta_0 \psi_\mu + \ell_p^6 \delta_1 \psi_\mu, \label{eq:trhd}
  \\
  \delta A_{\mu\nu\rho} &= \delta_0 A_{\mu\nu\rho} + \ell_p^6 \delta_1 A_{\mu\nu\rho}. \notag
\end{alignat}
Then the variation of the Lagrangian (\ref{eq:laghd}) under the supersymmetric transformations
(\ref{eq:trhd}) is expressed as
\begin{alignat}{3}
\label{eomc}
  \delta \mc{L} &= \delta_0 \mc{L}_0 + \ell_p^6 \Big\{ \delta_0 \mc{L}_1
  + e \delta_1 e^\mu{}_a E(e)^a{}_\mu + e \delta_1 \bar{\psi}_\mu E(\psi)^\mu 
  + e \delta_1 A_{\mu\nu\rho} E(A)^{\mu\nu\rho} 
  \Big\} + \mc{O}(\ell_p^{12}) \notag
  \\
  &=  \delta_0 \mc{L}_0 + \ell_p^6 \Big\{ V 
  + e (X^a{}_\mu + \delta_1 e^\mu{}_a) E(e)^a{}_\mu 
  + e (\bar{X}_\mu + \delta_1 \bar{\psi}_\mu) E(\psi)^\mu 
  \\
  &\qquad\qquad\qquad\quad+ e (X_{\mu\nu\rho} + \delta_1 A_{\mu\nu\rho}) E(A)^{\mu\nu\rho} 
  \Big\} + \mc{O}(\ell_p^{12}), \notag
\end{alignat}
where $V$ is the variations of the higher derivative terms which are independent of
the field equations,
\begin{alignat}{3}
  V = \delta_0 \mc{L}_1 
  - e X^\mu{}_a E(e)^a{}_\mu - e \bar{X}_\mu E(\psi)^\mu 
  - e X_{\mu\nu\rho} E(A)^{\mu\nu\rho} . \label{V}
\end{alignat}
The cancellation of the leading term, i.e., the supergravity part 
is just checked in the previous section. 
Note that, though the higher derivative terms in the ansatz are independent of the field equations,
generally their variations under the local supersymmetry depend on them.
So the variations in $\delta_0 \mc{L}_1$ can be decomposed into the terms $V$ 
which are independent of the field equations and
the other terms which depend on them.

The order of the $\ell_p^6$ part in the second line should vanish under the requirement of the 
local supersymmetry. Thus we have the following conditions: 
\begin{alignat}{3}
  &V=0, \label{cond1}
  \\
  &\delta_1 e^\mu{}_a = - X^a{}_\mu, \quad
  \delta_1 \psi_\mu = - X_\mu, \quad
  \delta_1 A_{\mu\nu\rho} = - X_{\mu\nu\rho}. \label{cond}
\end{alignat}
The first line is used to determine the coefficients of the higher derivative terms in $\mc{L}_1$.
The second line gives the modifications of supersymmetric transformation rules.

\newpage
\section{Higher Derivative Corrections : Details}\label{sec:dt}

\subsection{The Ansatz for the Higher Derivative Terms}

In this paper we take up the cancellation of variations via the
local supersymmetry which are linearly dependent on the Majorana gravitino 
and independent of the 3-from potential.
This means that the ansatz for the higher derivative effective action 
is made out of the terms which are linearly dependent on the 3-form potential
or the bilinear of the Majorana gravitino at most.
(In the case of the supergravity, this corresponds to consider the 
variation of $[eR\bar{\ep}\psi]$ and the ansatz of $\mc{L}[eR]$ and 
$\mc{L}[e\bar{\psi}\psi_{(2)}]$.)

Since the building blocks for the ansatz are $R^{ab}{}_{\mu\nu}=[L]^{-2}$,
$F_{\mu\nu\rho\sigma}=[L]^{-1}$, $D_{\mu}=[L]^{-1}$ and 
$\bar{\psi}_\mu \gamma^{\rho_1\cdots\rho_n} \psi_{\nu}=[L]^{-1}$ and
the integrand of the higher derivative effective action should have the length dimension
of $[L]^{-8}$, the possible forms of the ansatz are given by
\begin{alignat}{3}
  &\mc{L}[eR^4], \quad \mc{L}[e\epsilon_{11} AR^4], \quad
  \mc{L}[eD^2R^3], \quad \mc{L}[eD^4R^2], \quad \mc{L}[eD^6R], \notag
  \\
  &\mc{L}[eDR^3\bar{\psi}\psi], \quad \mc{L}[eD^3 R^2 \bar{\psi}\psi], \quad 
  \mc{L}[eD^5R\bar{\psi}\psi], \quad \mc{L}[eD^7\bar{\psi}\psi], \label{ansatz}
\end{alignat}
where the covariant derivatives for the local Lorentz indices act on the Riemann tensor
or the Majorana gravitino. 
Note that the linear terms of the four-form field strength are
dropped by imposing the parity invariance. 
The Chern-Simons like terms are included, however.

The ansatz (\ref{ansatz}) is quite general but contains so many terms.
In order to apply the Noether method it is necessary to reduce the number of terms in practice.
One way is to classify the variations of the ansatz under the number of the covariant
derivatives. Let us restrict the ansatz to those whose variations contain one covariant 
derivative explicitly at most.
First the transformation rules for the vielbein, the three-form potential and the Riemann tensor are
given by $\delta_0 e \sim [\bar{\ep}\psi]$, $\delta_0 A \sim [\bar{\ep}\psi]$ 
and $\delta_0 R \sim [R\bar{\epsilon}\psi] \oplus [D(\bar{\ep}\psi_{(2)})]$, 
so in the first line of the ansatz (\ref{ansatz}),
\begin{alignat}{3}
  \mc{L}[eR^4], \quad \mc{L}[e\epsilon_{11} AR^4], \label{ansatzb}
\end{alignat}
are the parts whose variations have one covariant derivative at most.
Next the transformation rules for the Majorana gravitino and
its field strength are estimated as $\delta_0 \psi \sim [D\ep]$ and
$\delta_0 \psi_{(2)} \sim [R\ep]$, so in the second line of the ansatz (\ref{ansatz}),
the forms of
\begin{alignat}{3}
  \mc{L}[eR^3\bar{\psi}\psi_{(2)}], \quad \mc{L}[eR^2 \bar{\psi}_{(2)}D\psi_{(2)}], \quad
  (\mc{L}[eRDR \bar{\psi}_{(2)}\psi_{(2)}]), \label{ansatzf}
\end{alignat}
contain one covariant derivative at most after the supersymmetric variations.
Of course, there is no guarantee that the supersymmetric transformations cancel completely
among (\ref{ansatzb}) and (\ref{ansatzf}).
As we will see later, however, it is enough to take account of this ansatz
to close the supersymmetry cancellation whose variations are linearly dependent on the Majorana 
gravitino and independent of the 3-from potential.

In this paper we neglect the third part in the ansatz (\ref{ansatzf}).
There is no reason to drop this part at this stage, but the result is that this part does
not contribute to cancel the variations of the ansatz (\ref{ansatzb}).
Therefore we only consider the first two parts in the ansatz (\ref{ansatzf}).
Note that this ansatz is also employed in the ref.~\cite{PVW} where the ansatz is
obtained by estimating the scattering amplitude of four or five massless closed strings.

\subsubsection{$\mathcal{L}[e R^4]$ terms}

The $\mathcal{L}[e R^4]$ part represents the quartic terms of the Riemann tensor. 
Note that the terms which include Ricci tensor and the scalar curvature 
are removed by using the field redefinition (\ref{eq:redef}),
so the $\mathcal{L}[e R^4]$ part consists of only the Riemann tensors $R_{ab\mu\nu}(\omega)$.
It seems that there are many ways of contractions, but by using 
the antisymmetry $R_{abcd}=-R_{bacd}=-R_{abdc}$, the symmetry 
$R_{abcd}=R_{cdab}$ and the cyclicity $R_{a[bcd]}=0$,
the $\mathcal{L}[e R^4]$ part can eventually be expanded  by 7 terms. 
(See an eq.~(\ref{eq:pboson}) below.) 

The above argument is only for purely bosonic case.
Now we replace the spin connection $\omega$ to
the supercovariant spin connection $\hat\omega$.
That is, the Riemann tensor is defined by using this 
supercovariant spin connection $\hat{\omega}(e,\psi)$ as
\begin{alignat}{3}
  {\hat R}^{ab}{}_{\mu\nu}(\hat{\omega}) &= \partial_\mu \hat{\omega}_\nu{}^{ab}
  - \partial_\nu \hat{\omega}_\mu{}^{ab}
  + \hat{\omega}_\mu{}^a{}_c \hat{\omega}_\nu{}^{cb}
  - \hat{\omega}_\nu{}^a{}_c \hat{\omega}_\mu{}^{cb}.
\end{alignat}
Thus the $\mathcal{L}[e {R}^4]$ part includes the bilinear terms of the Majorana gravitino
through the spin connection $\hat \omega$.
This operation is often used to check the cancellation of the supersymmetric
variations of the higher derivative corrections\cite{BR}.
The variation of the supercovariant spin connection does not include the derivative
of the supersymmetric parameter, and the cancellation mechanism becomes
similar to that of the supergravity coupled to non-abelian gauge field\cite{BR,RSW1,PVW}.

The ${\hat R}_{abcd}$ only hold the antisymmetry property, and the terms 
in $\mathcal{L}[e {R}^4]$ are expanded by more than seven terms.
Since there are so many ways to contract the indices, we employ the computer
programming. As a result, the terms in $\mathcal{L}[e {R}^4]$ are expanded by 13 terms
whose variations are completely independent.
\begin{alignat}{3}
  &\mathcal{L}[e {R}^4] =
  + b^1_1 e {\hat R}_{abcd}{\hat R}_{abcd}{\hat R}_{efgh}{\hat R}_{efgh}
  + b^1_2 e {\hat R}_{abcd}{\hat R}_{agfh}{\hat R}_{becd}{\hat R}_{efgh} \notag
  \\&\quad\qquad\quad
  + b^1_3 e {\hat R}_{abcd}{\hat R}_{abdh}{\hat R}_{efcg}{\hat R}_{efgh}
  + b^1_4 e {\hat R}_{abcd}{\hat R}_{aedg}{\hat R}_{bcfh}{\hat R}_{efgh} \notag
  \\&\quad\qquad\quad
  + b^1_5 e {\hat R}_{abcd}{\hat R}_{agdh}{\hat R}_{bcef}{\hat R}_{efgh}
  + b^1_6 e {\hat R}_{abcd}{\hat R}_{ahdf}{\hat R}_{becg}{\hat R}_{efgh} \notag
  \\&\quad\qquad\quad
  + b^1_7 e {\hat R}_{abcf}{\hat R}_{adgh}{\hat R}_{bdce}{\hat R}_{efgh}
  + b^1_8 e {\hat R}_{abch}{\hat R}_{adef}{\hat R}_{bdcg}{\hat R}_{efgh} \label{b1}
  \\&\quad\qquad\quad
  + b^1_9 e {\hat R}_{abch}{\hat R}_{aedg}{\hat R}_{bdcf}{\hat R}_{efgh}
  + b^1_{10} e {\hat R}_{abch}{\hat R}_{aedf}{\hat R}_{bcdg}{\hat R}_{efgh} \notag
  \\&\quad\qquad\quad
  + b^1_{11} e {\hat R}_{abch}{\hat R}_{abdg}{\hat R}_{cedf}{\hat R}_{efgh}
  + b^1_{12} e {\hat R}_{adgh}{\hat R}_{afbc}{\hat R}_{debc}{\hat R}_{efgh} \notag
  \\&\quad\qquad\quad
  + b^1_{13} e {\hat R}_{abch}{\hat R}_{agde}{\hat R}_{bdcf}{\hat R}_{efgh}. \notag
\end{alignat}
When the torsion terms are neglected, because of the symmetry and the cyclicity of 
the Riemann tensor, 13 terms of $\mathcal{L}[eR^4]$ are redundant and classified 
by purely bosonic 7 terms as
\begin{alignat}{3}
  \mathcal{L}[eR^4]_{\text{pure}} = &
  + e R_{abcd} R_{abcd} R_{efgh} R_{efgh} \times (b^1_1) &\qquad& A_1 \notag
  \\&
  + e R_{abcd} R_{abce} R_{dfgh} R_{efgh} \times (- \tfrac{1}{2}b^1_2 + b^1_3) &\qquad& A_2 \notag
  \\&
  + e R_{abcd} R_{abef} R_{cdgh} R_{efgh} \times
  (- \tfrac{1}{8}b^1_4 - \tfrac{1}{4}b^1_5 + \tfrac{1}{8}b^1_6) &\qquad& A_3 \notag
  \\&
  + e R_{abcd} R_{aecg} R_{bfdh} R_{efgh} \times (- b^1_6) &\qquad& A_6 \label{eq:pboson}
  \\&
  + e R_{abce} R_{abdg} R_{cfdh} R_{efgh} \times (b^1_7 + b^1_8 + \tfrac{1}{2}b^1_9) &\qquad& A_5 \notag
  \\&
  + e R_{abce} R_{abdf} R_{cdgh} R_{efgh} \times
  (\tfrac{1}{4}b^1_{10} - \tfrac{1}{2}b^1_{11} - b^1_{12}) &\qquad& A_4 \notag
  \\&
  + e R_{abce} R_{adcg} R_{bfdh} R_{efgh} \times (- b^1_9 - b^1_{13}). &\qquad& A_7 \notag
\end{alignat}
The notations $A_1, \cdots, A_7$ on the right hand side are those used in ref.~\cite{RSW1}.

\subsubsection{$\mathcal{L}[e\epsilon_{11} A R^4]$ terms}

Let us consider the ansatz of the $\mathcal{L}[e\epsilon_{11} A R^4]$ part.
Since the 3-form potential is odd under the flip of the 11th coordinate,
the ansatz should be accompanied with the antisymmetric tensor as 
$\epsilon_{11}^{\mu_1\mu_2\mu_3\mu_4\cdots\mu_{11}}A_{\mu_1\mu_2\mu_3}$.
The remaining 8 indices $\mu_4, \cdots, \mu_{11}$ are completely antisymmetric and 
should be contracted with the 4 Riemann tensors. The way of the contraction is unique 
because of the cyclicity of the Riemann tensor, and is written as
$\epsilon_{11}^{\mu_1\cdots\mu_{11}} A_{\mu_1\mu_2\mu_3}
R_{\s\s\mu_4\mu_5}R_{\s\s\mu_6\mu_7}R_{\s\s\mu_8\mu_9}R_{\s\s\mu_{10}\mu_{11}}$.

The remaining work is to insert the indices $a$, $b$, $c$ and $d$ into the 4 Riemann tensors
which are contracted by the flat metric.
There are two possibilities.
\begin{alignat}{3}
  &\mathcal{L}[e\epsilon_{11} A R^4] =
  -\tfrac{1}{6} b^2_1 e \epsilon_{11}^{\mu_1\cdots\mu_{11}} A_{\mu_1\mu_2\mu_3}
  R_{ab\mu_4\mu_5}R_{ab\mu_6\mu_7}R_{cd\mu_8\mu_9}R_{cd\mu_{10}\mu_{11}}  \label{b2}
  \\&\qquad\qquad\quad\quad
  -\tfrac{1}{6} b^2_2 e \epsilon_{11}^{\mu_1\cdots\mu_{11}} A_{\mu_1\mu_2\mu_3}
  R_{ab\mu_4\mu_5}R_{bc\mu_6\mu_7}R_{cd\mu_8\mu_9}R_{da\mu_{10}\mu_{11}}. \notag
\end{alignat}
It is known that a linear combination of these terms are required to cancel the 
gravitational anomaly on a M5-brane.
So these two terms are topological and can be expressed by using the forms as
$A \wedge \text{tr}(R^2) \wedge \text{tr}(R^2)$ and $A \wedge \text{tr}(R^4)$.
Notice that the torsion of the Riemann tensor in $\mathcal{L}[e\epsilon_{11} A R^4]$ are 
neglected since we consider the cancellation of the order of $\mathcal{O}(A^0,\psi)$.

\subsubsection{$\mathcal{L}[eR^3\bar{\psi}\psi_{(2)}]$ terms}

Here we write down the bilinear terms of the Majorana gravitino
which are in the category of $\mathcal{L}[eR^3\bar{\psi}\psi_{(2)}]$.

First of all let us classify the types of $[\bar{\psi}\psi_{(2)}]$.
Since all indices are contracted, the total number of the indices is even
and the number of the indices for $[\bar{\psi}\psi_{(2)}]$ is also even.
Then the number of the indices for the gamma matrix should be odd and
the types of $[\bar{\psi}\psi_{(2)}]$ are classified as
\begin{alignat}{3}
  [\bar{\psi}\psi_{(2)}] \sim \;
  &\bar{\psi}_h \gamma_f \psi_{gh} \oplus 
  \bar{\psi}_i \gamma_{efg} \psi_{hi} \oplus
  \bar{\psi}_j \gamma_{defgh} \psi_{ij} \oplus \notag
  \\
  &\bar{\psi}_g \gamma_{efg} \psi_{hi} \oplus
  \bar{\psi}_h \gamma_{defgh} \psi_{ij} \oplus
  \bar{\psi}_i \gamma_{cdefghi} \psi_{jk} \oplus
  \\
  &\bar{\psi}_k \gamma_l \psi_{ij} \oplus
  \bar{\psi}_k \gamma_{lmn} \psi_{ij} \oplus
  \bar{\psi}_k \gamma_{lmnop} \psi_{ij}. \notag
\end{alignat}
In the first line or the second line the index of the gravitino is contracted with 
one of the indices of the gravitino field strength or the gamma matrix, 
and in the third line no indices are contracted.
There are several remarks at this stage.
First because of the cyclicity of the Riemann tensor, the number of the uncontracted
indices of the gamma matrix should be less than seven.
Second, as argued before, the terms of $\bar{\psi}_i \gamma_{j} \psi_{jk}$ are neglected 
because these are expressed by using field equations.
Third the term of $\bar{\psi}_i \gamma_{i} \psi_{jk}$ is dropped since 
there are no cubic terms of the Riemann tensor whose uncontracted two indices are antisymmetric.

The types of $[R^3]$ are classified by the positions of the contracted indices.
As an example, let us consider a cubic term $R_{\s\s\s c}R_{\s b \s d}R_{\s bcd}$
where $b$, $c$ and $d$ are the contracted indices and blanks are arbitrary. 
This term is classified by the positions of the contracted indices as $\{1,2,3\}\{1,2\}$.
The $\{1,2,3\}$ shows that the number of the contracted indices in each Riemann tensor.
That is, the first Riemann tensor contains one contracted index, the second does two
and the third does three.
The contracted index $c$ is contained in the first and the third Riemann tensor, so
the numbers $(1,3)$ are assigned for this index. Similarly for the indices $b$ and $d$, the numbers
$(2,3)$ are assigned, and totally this example has the numbers of $(1,3)^1(2,3)^2$.
The $\{1,2\}$ represents the numbers of the powers of $(1,3)^1$ and $(2,3)^2$.
The numbers are aligned in order of rising.
Thus the example $R_{\s\s\s c}R_{\s b \s d}R_{\s bcd}$ is classified by the numbers 
of $\{1,2,3\}\{1,2\}$ which are not affected by the properties of the Riemann tensor.
The types of $[R^3]$ are classified in this way and the complete list is given in the 
appendix \ref{BB}. The result there is checked both by hand and by the computer programming 
independently.

The independent terms of $\mathcal{L}[eR^3\bar{\psi}\psi_{(2)}]$ are obtained by 
inserting the uncontracted indices of $[\bar{\psi}\psi_{(2)}]$ into the blanks of $[R^3]$.
It is useful to classify these terms by the positions of the uncontracted indices of 
$[\bar{\psi}\psi_{(2)}]$ in the $[R^3]$.
For instance, the term $e R_{albi}R_{akcd}R_{bjcd} \bar{\psi}_k \gamma_l \psi_{ij}$
is assigned new numbers of $\{10,11,100\}$. The meaning of these numbers are as follows.
For each index of the Majorana gravitino, the gamma matrix and the gravitino field strength,
the number 100, 1 and 10 is assigned respectively. The first Riemann tensor of this example
has the number 11, the second does 100 and the third does 10, and these numbers are aligned
in order of rising. Thus the term $e R_{albi}R_{akcd}R_{bjcd} \bar{\psi}_k \gamma_l \psi_{ij}$ 
is classified by the numbers of $\{2,3,3\}\{1,1,2\}\{10,11,100\}$, 
which are not affected by the properties of the Riemann tensor, the gamma matrix and 
the gravitino field strength.

By using the numbers discussed above it is almost possible to classify the terms of 
$\mathcal{L}[eR^3\bar{\psi}\psi_{(2)}]$, and the explicit expression is given by
\begin{alignat}{3}
  &\mathcal{L}[eR^3\bar{\psi}\psi_{(2)}] = \notag 
\\[0.1cm]
  &+ ( f^1_1 R_{afbg}R_{acde}R_{bcde} + f^1_2 R_{afbc}R_{agde}R_{bcde} 
     + f^1_3 R_{bfad}R_{cgae}R_{bcde}) 
       e\bar{\psi}_h \gamma_f \psi_{gh} \notag 
\\[0.1cm]
  &+ ( f^1_4 R_{klij}R_{abcd}R_{abcd} + f^1_5 R_{klai}R_{bjcd}R_{abcd} 
     + f^1_6 R_{kial}R_{bjcd}R_{abcd} \notag\\
  &\;+ f^1_7 R_{ijak}R_{blcd}R_{abcd} + f^1_8 R_{ijal}R_{bkcd}R_{abcd} 
     + f^1_9 R_{klab}R_{ijcd}R_{abcd} \notag\\
  &\;+ f^1_{10} R_{kiab}R_{ljcd}R_{abcd} + f^1_{11} R_{klab}R_{aicd}R_{bjcd} 
     + f^1_{12} R_{kiab}R_{alcd}R_{bjcd} \notag \\
  &\;+ f^1_{13} R_{liab}R_{akcd}R_{bjcd} + f^1_{14} R_{ijab}R_{akcd}R_{blcd} 
     + f^1_{15} R_{klac}R_{biad}R_{bjcd} \notag \\
  &\;+ f^1_{16} R_{kiac}R_{blad}R_{bjcd} + f^1_{17} R_{liac}R_{bkad}R_{bjcd}
     + f^1_{18} R_{ijac}R_{bkad}R_{blcd} \notag \\
  &\;+ f^1_{19} R_{akci}R_{bldj}R_{abcd} + f^1_{20} R_{akci}R_{blad}R_{bjcd} 
     + f^1_{21} R_{alci}R_{bkad}R_{bjcd}\notag \\
  &\;+ f^1_{22} R_{akbi}R_{alcd}R_{bjcd} + f^1_{23} R_{albi}R_{akcd}R_{bjcd}) 
       e\bar{\psi}_k \gamma_l \psi_{ij} \notag 
\\[0.1cm]
  &+ ( f^1_{24} R_{efhi}R_{abcd}R_{abcd} + f^1_{25} R_{efah}R_{bicd}R_{abcd}
     + f^1_{26} R_{hiae}R_{bfcd}R_{abcd} \notag \\
  &\;+ f^1_{27} R_{efab}R_{hicd}R_{abcd} + f^1_{28} R_{ehab}R_{ficd}R_{abcd} 
     + f^1_{29} R_{efab}R_{ahcd}R_{bicd} \notag \\
  &\;+ f^1_{30} R_{ehab}R_{afcd}R_{bicd} + f^1_{31} R_{hiab}R_{aecd}R_{bfcd} 
     + f^1_{32} R_{efac}R_{bhad}R_{bicd} \notag \\
  &\;+ f^1_{33} R_{ehac}R_{bfad}R_{bicd} + f^1_{34} R_{hiac}R_{bead}R_{bfcd} 
     + f^1_{35} R_{aech}R_{bfdi}R_{abcd} \notag \\
  &\;+ f^1_{36} R_{aech}R_{bfad}R_{bicd} + f^1_{37} R_{aebh}R_{facd}R_{ibcd}) 
       e\bar{\psi}_g \gamma_{efg} \psi_{hi} \label{f1}
\\[0.1cm]
  &+ ( f^1_{38} R_{efah}R_{bgcd}R_{abcd} + f^1_{39} R_{efab}R_{ghcd}R_{abcd}
     + f^1_{40} R_{efab}R_{agcd}R_{bhcd} \notag \\
  &\;+ f^1_{41} R_{ehab}R_{afcd}R_{bgcd} + f^1_{42} R_{efac}R_{bgad}R_{bhcd} 
     + f^1_{43} R_{ehac}R_{bfad}R_{bgcd}) 
       e\bar{\psi}_i \gamma_{efg} \psi_{hi} \notag
\\[0.1cm]
  &+ ( f^1_{44} R_{kilm}R_{anbc}R_{ajbc} + f^1_{45} R_{lmij}R_{akbc}R_{anbc}
     + f^1_{46} R_{lmak}R_{ijbc}R_{anbc} \notag \\
  &\;+ f^1_{47} R_{lmak}R_{nibc}R_{ajbc} + f^1_{48} R_{lmai}R_{knbc}R_{ajbc} 
     + f^1_{49} R_{lmai}R_{njbc}R_{akbc} \notag \\
  &\;+ f^1_{50} R_{lmai}R_{kjbc}R_{anbc} + f^1_{51} R_{klai}R_{mnbc}R_{ajbc} 
     + f^1_{52} R_{klai}R_{mjbc}R_{anbc} \notag \\
  &\;+ f^1_{53} R_{liak}R_{mnbc}R_{ajbc} + f^1_{54} R_{liak}R_{mjbc}R_{anbc} 
     + f^1_{55} R_{ijal}R_{kmbc}R_{anbc} \notag \\
  &\;+ f^1_{56} R_{ijal}R_{mnbc}R_{akbc} + f^1_{57} R_{ijak}R_{lmbc}R_{anbc}
     + f^1_{58} R_{lmbk}R_{anci}R_{ajbc} \notag \\
  &\;+ f^1_{59} R_{lmbi}R_{akcn}R_{ajbc} + f^1_{60} R_{lmbi}R_{ancj}R_{akbc} 
     + f^1_{61} R_{lmbi}R_{akcj}R_{anbc} \notag \\
  &\;+ f^1_{62} R_{klbi}R_{ajcm}R_{anbc} + f^1_{63} R_{libk}R_{ajcm}R_{anbc} 
     + f^1_{64} R_{ijbl}R_{akcm}R_{anbc} \notag \\
  &\;+ f^1_{65} R_{klab}R_{mnac}R_{ijbc} + f^1_{66} R_{kiab}R_{lmac}R_{njbc} 
     + f^1_{67} R_{lmab}R_{ijac}R_{bkcn} \notag \\
  &\;+ f^1_{68} R_{liab}R_{mjac}R_{bkcn} + f^1_{69} R_{lmab}R_{niac}R_{bkcj}
     + f^1_{70} R_{lmab}R_{kiac}R_{bncj} \notag \\
  &\;+ f^1_{71} R_{klab}R_{miac}R_{bncj} + f^1_{72} R_{lmab}R_{ckai}R_{bncj} 
     + f^1_{73} R_{liab}R_{ckam}R_{bncj} \notag \\
  &\;+ f^1_{74} R_{akbl}R_{cmai}R_{bncj}) 
       e\bar{\psi}_k \gamma_{lmn} \psi_{ij} \notag 
\\[0.1cm]
  &+ ( f^1_{75} R_{deai}R_{fgbc}R_{ahbc} + f^1_{76} R_{deab}R_{fgac}R_{bhci}) 
       e\bar{\psi}_j \gamma_{defgh} \psi_{ij} \notag 
\end{alignat}
\begin{alignat}{3}
  &+ ( f^1_{77} R_{deai}R_{fgbc}R_{ajbc} + f^1_{78} R_{ijad}R_{efbc}R_{agbc}
     + f^1_{79} R_{deai}R_{fjbc}R_{agbc} \notag \\
  &\;+ f^1_{80} R_{debi}R_{afcj}R_{agbc} + f^1_{81} R_{deab}R_{fiac}R_{bgcj}) 
       e\bar{\psi}_h \gamma_{defgh} \psi_{ij} \notag 
\\[0.1cm]
  &+ ( f^1_{82} R_{lmki}R_{noab}R_{pjab} + f^1_{83} R_{klij}R_{mnab}R_{opab}
     + f^1_{84} R_{lmij}R_{knab}R_{opab} \notag \\
  &\;+ f^1_{85} R_{lmai}R_{nobj}R_{kpab} + f^1_{86} R_{lmak}R_{ijbn}R_{opab} 
     + f^1_{87} R_{liak}R_{mnbj}R_{opab} \notag \\
  &\;+ f^1_{88} R_{klai}R_{mnbj}R_{opab} + f^1_{89} R_{lmak}R_{nobi}R_{pjab} 
     + f^1_{90} R_{lmak}R_{nobi}R_{apbj}) 
       e\bar{\psi}_k \gamma_{lmnop} \psi_{ij} \notag 
\\[0.1cm]
  &+ ( f^1_{91} R_{cdjk}R_{efab}R_{ghab} + f^1_{92} R_{cdaj}R_{efbk}R_{ghab}) 
       e\bar{\psi}_i \gamma_{cdefghi} \psi_{jk}. \notag
\end{alignat}
Thus there are 92 terms for $\mathcal{L}[eR^3\bar{\psi}\psi_{(2)}]$.
This result is obtained both by hand and by the computer programming independently.
Note that this expression is also obtained in ref.~\cite{RSW2}.

\subsubsection{$\mathcal{L}[eR^2\bar{\psi}_{(2)}D\psi_{(2)}]$ terms}

We write down the bilinear terms of the Majorana gravitino
which are in the category of $\mathcal{L}[eR^2\bar{\psi}_{(2)}D\psi_{(2)}]$.

First of all let us classify the types of $[\bar{\psi}_{(2)}D\psi_{(2)}]$.
Since all indices are contracted, the total number of the indices is even
and the number of the indices for $[\bar{\psi}_{(2)}D\psi_{(2)}]$ is also even.
Then the number of the indices for the gamma matrix should be odd and
the types of $[\bar{\psi}_{(2)}D\psi_{(2)}]$ are classified as
\begin{alignat}{3}
  [\bar{\psi}_{(2)}D\psi_{(2)}] \sim \;
  &\bar{\psi}_{mn}\gamma_j D_i \psi_{mn} \oplus 
  \bar{\psi}_{mn} \gamma_{jkl} D_i \psi_{mn} \notag
  \\
  &\bar{\psi}_{mn} \gamma_j D_i \psi_{on} \oplus
  \bar{\psi}_{mp} \gamma_{jkl} D_i \psi_{op} 
  \\
  &\bar{\psi}_{mn} \gamma_j D_i \psi_{op} \oplus
  \bar{\psi}_{mn} \gamma_{jkl} D_i \psi_{op}. \notag
\end{alignat}
In the first line four indices of the gravitino field strengths are contracted.
In the second line two indices of the gravitino field strengths are contracted, and
in the third line no indices are contracted.
There are several remarks at this stage.
First the indices of the gamma matrix are all uncontracted with the indices of 
the gravitino field strength and the covariant derivative.
As discussed before, the terms whose indices of the gamma matrix are contracted
can be expressed by using field equations and neglected in the ansatz.
Second because of the cyclicity and the Bianchi identity of the Riemann tensor, 
the number of the indices for the gamma matrix should be less than five.
Third it is always possible to make the index of the covariant derivative uncontracted 
with the indices of the gravitino field strength by using the
relation of $D_{[c}\psi_{ab]}\sim \frac{1}{4}R_{de[ab}\gamma^{de}\psi_{c]}$.
For example, $\bar{\psi}_{ik} \gamma_j D_i \psi_{kl} = 
-\frac{1}{2} \bar{\psi}_{ik} \gamma_j D_l \psi_{ik} 
+ \frac{3}{8}R_{de[ik}\bar{\psi}^{ik} \gamma^j \gamma^{de}\psi_{l]}$, where the second
term is already included in the category of $\mathcal{L}[eR^3\bar{\psi}\psi_{(2)}]$.

The types of $[R^2]$ are classified by the positions of the contracted indices.
As an example, let us consider a quadratic term $R_{\s bcd}R_{\s bcd}$
where $b$, $c$ and $d$ are the contracted indices and blanks are arbitrary. 
This term is classified by the positions of the contracted indices as $\{3,3\}\{3\}$.
The $\{3,3\}$ shows that the number of the contracted indices in each Riemann tensor.
That is, the first and the second Riemann tensor contains three contracted indices, respectively.
The contracted index $b$ is contained in the first and the second Riemann tensor, so
the numbers $(1,2)$ are assigned for this index. Similarly for the indices $c$ and $d$, the numbers
$(1,2)$ are assigned, and totally this example has the numbers of $(1,2)^3$.
The $\{3\}$ represents the number of the power of $(1,2)^3$.
The numbers are aligned in order of rising.
Thus the example $R_{\s bcd}R_{\s bcd}$ is classified by the numbers 
of $\{3,3\}\{3\}$ which are not affected by the properties of the Riemann tensor.
The types of $[R^2]$ are classified in this way and the complete list is given in the 
appendix \ref{BB}. The result there is checked both by hand and by the computer programming 
independently.

The independent terms of $\mathcal{L}[eR^2\bar{\psi}_{(2)}D\psi_{(2)}]$ are obtained by 
inserting the uncontracted indices of $[\bar{\psi}_{(2)}D\psi_{(2)}]$ into the blanks of $[R^2]$.
It is useful to classify these terms by the positions of the uncontracted indices of 
$[\bar{\psi}_{(2)}D\psi_{(2)}]$ in the $[R^2]$.
For instance, the term $e R_{iaob}R_{majb} \bar{\psi}_{mn} \gamma_j D_i \psi_{on}$
is assigned new numbers of $\{20,101\}$. The meaning of these numbers are as follows.
For each index of the first gravitino field strength and the gamma matrix,
the number 100 and 1 is assigned respectively.
And for each index of the covariant derivative and the second gravitino field strength,
the number 10 is assigned. Note that the same number is assigned both for the covariant derivative
and the second gravitino field strength because of the relation 
$D_{[c}\psi_{ab]}\sim \frac{1}{4}R_{de[ab}\gamma^{de}\psi_{c]}$.
The first Riemann tensor of this example
has the number 20 and the second does 101, and these numbers are aligned
in order of rising. Thus the term $e R_{iaob}R_{majb} \bar{\psi}_{mn} \gamma_j D_i \psi_{on}$ 
is classified by the numbers of $\{2,2\}\{2\}\{20,101\}$, 
which are not affected by the properties of the Riemann tensor, the gamma matrix and 
the gravitino field strength.

By using the numbers discussed above it is almost possible to classify the terms of 
$\mathcal{L}[eR^2\bar{\psi}_{(2)}D\psi_{(2)}]$, and the explicit expression is given by
\begin{alignat}{3}
  &\mathcal{L}[eR^2\bar{\psi}_{(2)}D\psi_{(2)}] = \notag 
\\[0.1cm]
  &+ ( f^2_1 R_{ijma}R_{opna}+f^2_2 R_{ijoa}R_{mnpa}+f^2_3 R_{imja}R_{opna}
     + f^2_4 R_{ioja}R_{mnpa} \notag \\
  &\;+ f^2_5 R_{imoa}R_{jnpa}+f^2_6 R_{imoa}R_{jpna}+f^2_7 R_{iopa}R_{jmna})
      e\bar{\psi}_{mn} \gamma_j D_i \psi_{op} \notag
\\[0.1cm]
  &+ ( f^2_8 R_{iajb}R_{maob}+f^2_9 R_{iajb}R_{oamb}+f^2_{10} R_{iamb}R_{jaob}
     + f^2_{11} R_{iaob}R_{jamb} \notag \\
  &\;+ f^2_{12} R_{iamb}R_{oajb}+f^2_{13} R_{iaob}R_{majb} ) 
       e\bar{\psi}_{mn} \gamma_j D_i \psi_{on} \notag
\\[0.1cm]
  &\;+ f^2_{14} R_{iabc}R_{jabc} \, e\bar{\psi}_{mn}\gamma_j D_i \psi_{mn} \label{f2}
\\[0.1cm]
  &+ ( f^2_{15} R_{ijkm}R_{lnop}+f^2_{16} R_{ijko}R_{lpmn}
     + f^2_{17} R_{imjo}R_{klnp}+f^2_{18} R_{iojp}R_{klmn} )
       e\bar{\psi}_{mn} \gamma_{jkl} D_i \psi_{op} \notag
\\[0.1cm]
  &+ ( f^2_{19} R_{ijka}R_{lmoa}+f^2_{20} R_{ijka}R_{loma}+f^2_{21} R_{ijma}R_{kloa}
     + f^2_{22} R_{ijoa}R_{klma} \notag \\
  &\;+ f^2_{23} R_{imja}R_{kloa}+f^2_{24} R_{ioja}R_{klma}) 
       e\bar{\psi}_{mp} \gamma_{jkl} D_i \psi_{op} \notag
\\[0.1cm]
  &\;+ f^2_{25} R_{iajb}R_{kalb} \, e\bar{\psi}_{mn} \gamma_{jkl} D_i \psi_{mn} \notag
\end{alignat}
Thus there are 25 terms for $\mathcal{L}[eR^2\bar{\psi}_{(2)}D\psi_{(2)}]$.
This result is obtained both by hand and by the computer programming independently.

\subsection{The Bases for the Variations}

Now let us turn our attention to the variations of the ansatz.
In this subsection we write down all independent terms of the variations,
and in next subsection we will expand the variations of the ansatz in terms of these bases.

Since we are interested in the cancellation of the terms which are linear to the Majorana
gravitino and independent of the 3-form potential, the relevant supersymmetric transformation
rules of the fields are roughly expressed as
\begin{alignat}{3}
  &\delta_0 e \sim [\bar{\ep}\psi], \notag
  \\
  &\delta_0 R \sim [R\bar{\ep}\psi] \oplus [D(\bar{\ep}\psi_{(2)})], \notag
  \\
  &\delta_0 A \sim [\bar{\ep}\psi], 
  \\
  &\delta_0 \psi \sim [D\epsilon], \notag
  \\
  &\delta_0 \psi_{(2)} \sim [R \ep]. \notag
\end{alignat}
Then the variations of the ansatz $\dl\mathcal{L}[eR^4]$, $\dl\mathcal{L}[e\epsilon_{11}AR^4]$,
$\dl\mathcal{L}[eR^3\bar{\psi}\psi_{(2)}]$ and $\dl\mathcal{L}[e R^2 \bar{\psi}_{(2)} D \psi_{(2)}]$,
which are linear to the Majorana gravitino and independent of the 3-form potential,
are sketched as follows:
\begin{alignat}{5}
  &\delta \mathcal{L}[eR^4] &&\sim V[eR^4\bar{\epsilon}\psi] \oplus
  &&V[eR^2DR\bar{\epsilon}\psi_{(2)}], \notag
  \\
  &\delta \mathcal{L}[e\epsilon_{11}AR^4] &&\sim V[eR^4\bar{\epsilon}\psi], \notag
  \\
  &\delta \mathcal{L}[eR^3\bar{\psi}\psi_{(2)}] &&\sim V[eR^4\bar{\epsilon}\psi] \oplus
  &&V[eR^2DR\bar{\epsilon}\psi_{(2)}] \oplus &&V[eR^3\bar{\epsilon}D\psi_{(2)}], \label{eq:var}
  \\
  &\delta \mathcal{L}[e R^2 \bar{\psi}_{(2)} D \psi_{(2)}]
  &&\sim &&V[eR^2DR\bar{\epsilon}\psi_{(2)}] \oplus &&V[eR^3\bar{\epsilon}D\psi_{(2)}], \notag
  \\
  &0 &&\sim V[eR^4\bar{\epsilon}\psi] \oplus &&&& V[eR^3\bar{\epsilon}D\psi_{(2)}]. \notag
\end{alignat}
As we will see soon, there are 116 bases for $V[eR^4\bar{\epsilon}\psi]$, 88 bases for
$V[eR^2DR\bar{\epsilon}\psi_{(2)}]$ and 92 bases for $V[eR^3\bar{\epsilon}D\psi_{(2)}]$.
For the last, 60 bases are essentially required and the other 32 terms are rewritten
by the other bases with the aid of the field equations.
Note that the first terms $V[eR^4\bar{\epsilon}\psi]$ and
the third terms $V[eR^3\bar{\epsilon}D\psi_{(2)}]$ are not independent because of the identity,
$D_{[e} \psi_{cd]} \sim \tfrac{1}{4} \gamma^{ab} \psi_{[e} R_{cd] ab}$.
In fact, by using the computer program we find that there are 20 identities between the first terms
and the third terms,
\begin{alignat}{3}
  0 &= \sum_{i=n}^{20} i_n (V[R^4 \bar{\epsilon} \psi]
  + V[R^3 \bar{\epsilon} D \psi_{(2)}])_n,
\end{alignat}
where $i_n$ are arbitrary coefficients.
The last line of the eq.~(\ref{eq:var}) represents these identities.

Therefore under the local supersymmetric transformations, the variations of the ansatz 
are expanded by $264$ bases. Now we write down these 264 terms which are obtained
both by hand and by the computer programming independently.

\subsubsection{Bases for $V[eR^4\bar{\ep}\psi]$}

There are 116 bases for $V[eR^4\bar{\ep}\psi]$. Before giving the explicit 
expressions we clarify the algorithm to obtain the result.

First of all let us classify the types of $[\bar{\ep}\psi]$.
Since all indices are contracted, the total number of the indices is even
and the number of the indices for $[\bar{\ep}\psi]$ is also even.
Then the number of the indices for the gamma matrix should be odd and
the types of $[\bar{\ep}\psi]$ are classified as
\begin{alignat}{3}
  [\bar{\ep}\psi] \sim \;
  &\bar{\epsilon} \gamma^{z} \psi_z \oplus 
  \bar{\epsilon} \gamma^{ijz} \psi_z \oplus 
  \bar{\epsilon} \gamma^{ijklz} \psi_z \oplus 
  \bar{\epsilon} \gamma^{ijklmnopz} \psi_z \oplus \notag
  \\
  &\bar{\epsilon} \gamma^{i} \psi^z \oplus
  \bar{\epsilon} \gamma^{ijk} \psi^z \oplus
  \bar{\epsilon} \gamma^{ijklm} \psi^z \oplus
  \bar{\epsilon} \gamma^{ijklmno} \psi^z. 
\end{alignat}
In the first line the index of the gravitino is contracted with that of the gamma matrix, and
in the second line no indices are contracted.
There are two remarks at this stage.
First because of the cyclicity of the Riemann tensor, 
the number of the uncontracted indices for the gamma matrix should be less than nine.
Second the term $[\bar{\epsilon} \gamma^{ijklmnz} \psi_z]$ is dropped because
there are no quartic terms of the Riemann tensor whose uncontracted indices are
completely antisymmetric. (This is not obvious but can be checked by referring the
appendix \ref{BB}.)

The types of $[R^4]$ are classified by the positions of the contracted indices.
As an example, let us consider a quartic term $R \s _e \s _f R \s _{eaf} R \s _{bcd} R_{abcd}$
where $a$, $b$, $c$, $d$, $e$ and $f$ are contracted by the flat metric and blanks are arbitrary. 
This term is classified by the positions of the contracted indices as $\{2,3,3,4\}\{1,2,3\}$.
The $\{2,3,3,4\}$ shows that the number of the contracted indices in each Riemann tensor.
That is, the first Riemann tensor contains two contracted indices,
the second does three, the third does three and the fourth does four.
The contracted index $a$ is contained in the second and the fourth Riemann tensor, so
the numbers $(2,4)$ are assigned for this index. In a similar way the numbers
$(3,4)$, $(3,4)$, $(3,4)$, $(1,2)$ and $(1,2)$ are assigned for the 
indices $b$, $c$, $d$, $e$ and $f$, respectively, 
and totally this example has the numbers of $(2,4)^1(1,2)^2(3,4)^3$.
The $\{1,2,3\}$ represents the numbers of the powers of $(2,4)^1$, $(1,2)^2$ and $(3,4)^3$.
The numbers are aligned in order of rising.
Thus the example $R \s _e \s _f R \s _{eaf} R \s _{bcd} R_{abcd}$ is classified by the numbers 
of $\{2,3,3,4\}\{1,2,3\}$ which are not affected by the properties of the Riemann tensor.
The types of $[R^4]$ are classified in this way and the complete list is given in the 
appendix \ref{BB}. The result there is checked both by hand and by the computer programming 
independently.

The independent terms of $V[eR^4\bar{\ep}\psi]$ are obtained by 
inserting the uncontracted indices of $[\bar{\ep}\psi]$ into the blanks of $[R^4]$.
It is useful to classify these terms by the positions of the uncontracted indices of 
$[\bar{\ep}\psi]$ in the $[R^4]$.
For instance, the term $e R_{ijef} R_{kaef} R_{zbcd} R_{abcd} \bar{\epsilon} \gamma^{ijk} \psi^z$
is assigned new numbers of $\{0,1,2,10\}$. The meaning of these numbers are as follows.
For each index of the gamma matrix the number 1 is assigned,
and for each index of the Majorana gravitino the number 10 is assigned.
The first Riemann tensor of this example
has the number 2, the second does 1, the third does 10 and the fourth does 0, 
and these numbers are aligned in order of rising. 
Thus the term $e R_{ijef} R_{kaef} R_{zbcd} R_{abcd} \bar{\epsilon} \gamma^{ijk} \psi^z$ 
is classified by the numbers of $\{2,3,3,4\}\{1,2,3\}\{0,1,2,10\}$, 
which are not affected by the properties of the Riemann tensor and the gamma matrix.

By using the numbers discussed above it is almost possible to classify the terms of 
$V[eR^4\bar{\ep}\psi]$, and the explicit expressions for the bases are given by
\begin{alignat}{3}
V^1_{1} &= e R_{abcd} R_{abcd} R_{efgh} R_{efgh} \bar{\epsilon} \gamma^{z} \psi_z, \qquad&   
V^1_{2} &= e R_{abcd} R_{abce} R_{dfgh} R_{efgh} \bar{\epsilon} \gamma^{z} \psi_z, \notag \\
V^1_{3} &= e R_{abcd} R_{abef} R_{cdgh} R_{efgh} \bar{\epsilon} \gamma^{z} \psi_z, \qquad&
V^1_{4} &= e R_{abcd} R_{aecg} R_{bfdh} R_{efgh} \bar{\epsilon} \gamma^{z} \psi_z, \notag \\ 
V^1_{5} &= e R_{abce} R_{abdg} R_{cfdh} R_{efgh} \bar{\epsilon} \gamma^{z} \psi_z, \qquad&
V^1_{6} &= e R_{abce} R_{abdf} R_{cdgh} R_{efgh} \bar{\epsilon} \gamma^{z} \psi_z, \notag \\ 
V^1_{7} &= e R_{abce} R_{adcg} R_{bfdh} R_{efgh} \bar{\epsilon} \gamma^{z} \psi_z, \notag 
\\[0.1cm]
V^1_{8} &= e R_{iezf} R_{eafb} R_{agcd} R_{bgcd} \bar{\epsilon} \gamma^{i} \psi^z, \qquad&   
V^1_{9} &= e R_{iezf} R_{egab} R_{fgcd} R_{abcd} \bar{\epsilon} \gamma^{i} \psi^z, \notag \\
V^1_{10} &= e R_{iezf} R_{eagb} R_{fcgd} R_{acbd} \bar{\epsilon} \gamma^{i} \psi^z, \qquad&
V^1_{11} &= e R_{iefg} R_{zefg} R_{abcd} R_{abcd} \bar{\epsilon} \gamma^{i} \psi^z, \notag \\ 
V^1_{12} &= e R_{iafg} R_{zbcd} R_{eafg} R_{ebcd} \bar{\epsilon} \gamma^{i} \psi^z, \quad&   
V^1_{13} &= e R_{iaef} R_{zbef} R_{agcd} R_{bgcd} \bar{\epsilon} \gamma^{i} \psi^z, \notag \\ 
V^1_{14} &= e R_{ieaf} R_{zebf} R_{agcd} R_{bgcd} \bar{\epsilon} \gamma^{i} \psi^z, \quad&
V^1_{15} &= e R_{iefg} R_{zeab} R_{fgcd} R_{abcd} \bar{\epsilon} \gamma^{i} \psi^z, \notag \\     
V^1_{16} &= e R_{ifeg} R_{zaeb} R_{fcgd} R_{acbd} \bar{\epsilon} \gamma^{i} \psi^z, \quad&
V^1_{17} &= e R_{iagb} R_{zdfc} R_{eafb} R_{ecgd} \bar{\epsilon} \gamma^{i} \psi^z, \notag \\
V^1_{18} &= e R_{zagb} R_{idfc} R_{eafb} R_{ecgd} \bar{\epsilon} \gamma^{i} \psi^z, \quad&
V^1_{19} &= e R_{zagb} R_{icfd} R_{eafb} R_{ecgd} \bar{\epsilon} \gamma^{i} \psi^z, \notag \\
V^1_{20} &= e R_{zbga} R_{idfc} R_{eafb} R_{ecgd} \bar{\epsilon} \gamma^{i} \psi^z, \quad&           
V^1_{21} &= e R_{iaef} R_{zbeg} R_{fgcd} R_{abcd} \bar{\epsilon} \gamma^{i} \psi^z, \notag \\    
V^1_{22} &= e R_{iaef} R_{zgeb} R_{fgcd} R_{abcd} \bar{\epsilon} \gamma^{i} \psi^z, \qquad&   
V^1_{23} &= e R_{iaef} R_{zbeg} R_{fcgd} R_{acbd} \bar{\epsilon} \gamma^{i} \psi^z, \notag \\ 
V^1_{24} &= e R_{iaef} R_{zgeb} R_{fcgd} R_{acbd} \bar{\epsilon} \gamma^{i} \psi^z, \notag
\\[0.1cm]
V^1_{25} &= e R_{iafb} R_{jgcd} R_{eafg} R_{ebcd} \bar{\epsilon} \gamma^{ijz} \psi_z, \label{v1}
\\[0.1cm]
V^1_{26} &= e R_{ijze} R_{kaeb} R_{afcd} R_{bfcd} \bar{\epsilon} \gamma^{ijk} \psi^z, \qquad&
V^1_{27} &= e R_{ijze} R_{kfab} R_{efcd} R_{abcd} \bar{\epsilon} \gamma^{ijk} \psi^z, \notag \\
V^1_{28} &= e R_{ijze} R_{kafb} R_{ecfd} R_{acbd} \bar{\epsilon} \gamma^{ijk} \psi^z, \qquad&
V^1_{29} &= e R_{ijef} R_{zkef} R_{abcd} R_{abcd} \bar{\epsilon} \gamma^{ijk} \psi^z, \notag \\
V^1_{30} &= e R_{ijef} R_{zkab} R_{efcd} R_{abcd} \bar{\epsilon} \gamma^{ijk} \psi^z, \quad& 
V^1_{31} &= e R_{ijea} R_{kezb} R_{afcd} R_{bfcd} \bar{\epsilon} \gamma^{ijk} \psi^z, \notag \\        
V^1_{32} &= e R_{ijea} R_{zekb} R_{afcd} R_{bfcd} \bar{\epsilon} \gamma^{ijk} \psi^z, \quad&  
V^1_{33} &= e R_{ijea} R_{zfkb} R_{efcd} R_{abcd} \bar{\epsilon} \gamma^{ijk} \psi^z, \notag \\   
V^1_{34} &= e R_{ijea} R_{zfkb} R_{ecfd} R_{acbd} \bar{\epsilon} \gamma^{ijk} \psi^z, \qquad&
V^1_{35} &= e R_{ijef} R_{zaef} R_{kbcd} R_{abcd} \bar{\epsilon} \gamma^{ijk} \psi^z, \notag \\
V^1_{36} &= e R_{ijef} R_{kaef} R_{zbcd} R_{abcd} \bar{\epsilon} \gamma^{ijk} \psi^z, \qquad&  
V^1_{37} &= e R_{ijef} R_{zeab} R_{kfcd} R_{abcd} \bar{\epsilon} \gamma^{ijk} \psi^z, \notag \\
V^1_{38} &= e R_{ijef} R_{zaeb} R_{kcfd} R_{acbd} \bar{\epsilon} \gamma^{ijk} \psi^z, \qquad&
V^1_{39} &= e R_{ijab} R_{zcef} R_{kdef} R_{abcd} \bar{\epsilon} \gamma^{ijk} \psi^z, \notag \\
V^1_{40} &= e R_{ijab} R_{zecf} R_{kedf} R_{abcd} \bar{\epsilon} \gamma^{ijk} \psi^z, \qquad&   
V^1_{41} &= e R_{ijea} R_{zfbe} R_{kfcd} R_{abcd} \bar{\epsilon} \gamma^{ijk} \psi^z, \notag \\   
V^1_{42} &= e R_{ijea} R_{kfbe} R_{zfcd} R_{abcd} \bar{\epsilon} \gamma^{ijk} \psi^z, \qquad&
V^1_{43} &= e R_{ijea} R_{zebf} R_{kcfd} R_{acbd} \bar{\epsilon} \gamma^{ijk} \psi^z, \notag \\
V^1_{44} &= e R_{ijea} R_{kebf} R_{zcfd} R_{acbd} \bar{\epsilon} \gamma^{ijk} \psi^z, \qquad&
V^1_{45} &= e R_{ijea} R_{zfbe} R_{kcfd} R_{acbd} \bar{\epsilon} \gamma^{ijk} \psi^z, \notag \\  
V^1_{46} &= e R_{ijea} R_{kfbe} R_{zcfd} R_{acbd} \bar{\epsilon} \gamma^{ijk} \psi^z, \qquad&  
V^1_{47} &= e R_{ijea} R_{zebf} R_{kfcd} R_{abcd} \bar{\epsilon} \gamma^{ijk} \psi^z, \notag \\  
V^1_{48} &= e R_{ijea} R_{kebf} R_{zfcd} R_{abcd} \bar{\epsilon} \gamma^{ijk} \psi^z, \qquad&
V^1_{49} &= e R_{zeif} R_{jeaf} R_{kbcd} R_{abcd} \bar{\epsilon} \gamma^{ijk} \psi^z, \notag 
\end{alignat}
\begin{alignat}{3}
V^1_{50} &= e R_{iezf} R_{jeaf} R_{kbcd} R_{abcd} \bar{\epsilon} \gamma^{ijk} \psi^z, \qquad&
V^1_{51} &= e R_{iezf} R_{jeab} R_{kfcd} R_{abcd} \bar{\epsilon} \gamma^{ijk} \psi^z, \notag \\
V^1_{52} &= e R_{zeif} R_{jaeb} R_{kcfd} R_{acbd} \bar{\epsilon} \gamma^{ijk} \psi^z, \qquad&
V^1_{53} &= e R_{zaib} R_{jcef} R_{kdef} R_{acbd} \bar{\epsilon} \gamma^{ijk} \psi^z, \notag \\   
V^1_{54} &= e R_{zaib} R_{jecf} R_{kedf} R_{acbd} \bar{\epsilon} \gamma^{ijk} \psi^z, \qquad&   
V^1_{55} &= e R_{zeia} R_{jfbe} R_{kfcd} R_{abcd} \bar{\epsilon} \gamma^{ijk} \psi^z, \notag \\
V^1_{56} &= e R_{ieza} R_{jfbe} R_{kfcd} R_{abcd} \bar{\epsilon} \gamma^{ijk} \psi^z, \qquad&
V^1_{57} &= e R_{zeia} R_{jebf} R_{kcfd} R_{acbd} \bar{\epsilon} \gamma^{ijk} \psi^z, \notag \\
V^1_{58} &= e R_{ieza} R_{jebf} R_{kcfd} R_{acbd} \bar{\epsilon} \gamma^{ijk} \psi^z, \qquad&
V^1_{59} &= e R_{zeia} R_{jfbe} R_{kcfd} R_{acbd} \bar{\epsilon} \gamma^{ijk} \psi^z, \notag \\    
V^1_{60} &= e R_{ieza} R_{jfbe} R_{kcfd} R_{acbd} \bar{\epsilon} \gamma^{ijk} \psi^z, \qquad&   
V^1_{61} &= e R_{zeia} R_{jebf} R_{kfcd} R_{abcd} \bar{\epsilon} \gamma^{ijk} \psi^z, \notag \\
V^1_{62} &= e R_{ieza} R_{jebf} R_{kfcd} R_{abcd} \bar{\epsilon} \gamma^{ijk} \psi^z, \qquad&
V^1_{63} &= e R_{zeab} R_{ifab} R_{jecd} R_{kfcd} \bar{\epsilon} \gamma^{ijk} \psi^z, \notag \\
V^1_{64} &= e R_{zaeb} R_{iafb} R_{jecd} R_{kfcd} \bar{\epsilon} \gamma^{ijk} \psi^z, \qquad&
V^1_{65} &= e R_{iaeb} R_{jafb} R_{zecd} R_{kfcd} \bar{\epsilon} \gamma^{ijk} \psi^z, \notag \\
V^1_{66} &= e R_{zaeb} R_{iafb} R_{jced} R_{kcfd} \bar{\epsilon} \gamma^{ijk} \psi^z, \qquad& 
V^1_{67} &= e R_{zaec} R_{ibed} R_{jafb} R_{kcfd} \bar{\epsilon} \gamma^{ijk} \psi^z, \notag \\
V^1_{68} &= e R_{zaec} R_{ibed} R_{jdfa} R_{kbfc} \bar{\epsilon} \gamma^{ijk} \psi^z, \notag
\\[0.1cm]
V^1_{69} &= e R_{ijef} R_{klef} R_{abcd} R_{abcd} \bar{\epsilon} \gamma^{ijklz} \psi_z, \qquad& 
V^1_{70} &= e R_{ijef} R_{klab} R_{efcd} R_{abcd} \bar{\epsilon} \gamma^{ijklz} \psi_z, \notag \\    
V^1_{71} &= e R_{ijea} R_{kleb} R_{afcd} R_{bfcd} \bar{\epsilon} \gamma^{ijklz} \psi_z, \qquad&    
V^1_{72} &= e R_{ijea} R_{klfb} R_{efcd} R_{abcd} \bar{\epsilon} \gamma^{ijklz} \psi_z, \notag \\ 
V^1_{73} &= e R_{ijea} R_{klfb} R_{ecfd} R_{acbd} \bar{\epsilon} \gamma^{ijklz} \psi_z, \qquad&
V^1_{74} &= e R_{ijef} R_{kaef} R_{lbcd} R_{abcd} \bar{\epsilon} \gamma^{ijklz} \psi_z, \notag \\
V^1_{75} &= e R_{ijef} R_{keab} R_{lfcd} R_{abcd} \bar{\epsilon} \gamma^{ijklz} \psi_z, \qquad& 
V^1_{76} &= e R_{ijef} R_{kaeb} R_{lcfd} R_{acbd} \bar{\epsilon} \gamma^{ijklz} \psi_z, \notag\\
V^1_{77} &= e R_{ijab} R_{kcef} R_{ldef} R_{abcd} \bar{\epsilon} \gamma^{ijklz} \psi_z, \quad& 
V^1_{78} &= e R_{ijab} R_{kecf} R_{ledf} R_{abcd} \bar{\epsilon} \gamma^{ijklz} \psi_z, \notag \\ 
V^1_{79} &= e R_{ijea} R_{kebf} R_{lcfd} R_{acbd} \bar{\epsilon} \gamma^{ijklz} \psi_z, \qquad&    
V^1_{80} &= e R_{ijea} R_{kfbe} R_{lcfd} R_{acbd} \bar{\epsilon} \gamma^{ijklz} \psi_z, \notag \\   
V^1_{81} &= e R_{ijea} R_{kebf} R_{lfcd} R_{abcd} \bar{\epsilon} \gamma^{ijklz} \psi_z, \qquad&
V^1_{82} &= e R_{ijea} R_{kfbe} R_{lfcd} R_{abcd} \bar{\epsilon} \gamma^{ijklz} \psi_z, \notag \\   
V^1_{83} &= e R_{ieab} R_{jfab} R_{kecd} R_{lfcd} \bar{\epsilon} \gamma^{ijklz} \psi_z, \qquad&
V^1_{84} &= e R_{iaeb} R_{jafb} R_{kecd} R_{lfcd} \bar{\epsilon} \gamma^{ijklz} \psi_z, \notag \\          
V^1_{85} &= e R_{iaeb} R_{jafb} R_{kced} R_{lcfd} \bar{\epsilon} \gamma^{ijklz} \psi_z, \notag
\\[0.1cm]
  V^1_{86} &= e R_{ijze} R_{klea} R_{mbcd} R_{abcd} \bar{\epsilon} \gamma^{ijklm} \psi^z, \qquad&
  V^1_{87} &= e R_{ijze} R_{klab} R_{mecd} R_{abcd} \bar{\epsilon} \gamma^{ijklm} \psi^z, \notag \\
  V^1_{88} &= e R_{ijza} R_{kleb} R_{mecd} R_{abcd} \bar{\epsilon} \gamma^{ijklm} \psi^z, \qquad&
  V^1_{89} &= e R_{ijza} R_{kleb} R_{mced} R_{acbd} \bar{\epsilon} \gamma^{ijklm} \psi^z, \notag \\
  V^1_{90} &= e R_{zeia} R_{jkeb} R_{lmcd} R_{abcd} \bar{\epsilon} \gamma^{ijklm} \psi^z, \qquad&
  V^1_{91} &= e R_{ieza} R_{jkeb} R_{lmcd} R_{abcd} \bar{\epsilon} \gamma^{ijklm} \psi^z, \notag \\
  V^1_{92} &= e R_{ijea} R_{kleb} R_{mczd} R_{acbd} \bar{\epsilon} \gamma^{ijklm} \psi^z, \qquad&
  V^1_{93} &= e R_{ijze} R_{kaeb} R_{lacd} R_{mbcd} \bar{\epsilon} \gamma^{ijklm} \psi^z, \notag \\
  V^1_{94} &= e R_{ijze} R_{kaeb} R_{lcad} R_{mcbd} \bar{\epsilon} \gamma^{ijklm} \psi^z, \qquad&
  V^1_{95} &= e R_{zaib} R_{jkcd} R_{laeb} R_{mced} \bar{\epsilon} \gamma^{ijklm} \psi^z, \notag \\
  V^1_{96} &= e R_{iazb} R_{jkcd} R_{laeb} R_{mced} \bar{\epsilon} \gamma^{ijklm} \psi^z, \quad&
  V^1_{97} &= e R_{zeia} R_{jkeb} R_{lcad} R_{mcbd} \bar{\epsilon} \gamma^{ijklm} \psi^z, \notag \\
  V^1_{98} &= e R_{ieza} R_{jkeb} R_{lcad} R_{mcbd} \bar{\epsilon} \gamma^{ijklm} \psi^z, \quad&
  V^1_{99} &= e R_{zeia} R_{jkeb} R_{lacd} R_{mbcd} \bar{\epsilon} \gamma^{ijklm} \psi^z, \notag \\
  V^1_{100} &= e R_{ieza} R_{jkeb} R_{lacd} R_{mbcd} \bar{\epsilon} \gamma^{ijklm} \psi^z, \quad&
  V^1_{101} &= e R_{zaic} R_{jkbd} R_{laeb} R_{mced} \bar{\epsilon} \gamma^{ijklm} \psi^z, \notag \\
  V^1_{102} &= e R_{iajc} R_{zbkd} R_{laeb} R_{mced} \bar{\epsilon} \gamma^{ijklm} \psi^z, \quad&
  V^1_{103} &= e R_{zaid} R_{jkbc} R_{laeb} R_{mced} \bar{\epsilon} \gamma^{ijklm} \psi^z, \notag \\
  V^1_{104} &= e R_{iazd} R_{jkbc} R_{laeb} R_{mced} \bar{\epsilon} \gamma^{ijklm} \psi^z, \qquad&
  V^1_{105} &= e R_{ijea} R_{klea} R_{mbcd} R_{zbcd} \bar{\epsilon} \gamma^{ijklm} \psi^z, \notag \\
  V^1_{106} &= e R_{ijab} R_{klcd} R_{meab} R_{zced} \bar{\epsilon} \gamma^{ijklm} \psi^z, \quad&
  V^1_{107} &= e R_{ijea} R_{kleb} R_{mcad} R_{zcbd} \bar{\epsilon} \gamma^{ijklm} \psi^z, \notag \\
  V^1_{108} &= e R_{ijea} R_{kleb} R_{macd} R_{zbcd} \bar{\epsilon} \gamma^{ijklm} \psi^z, \quad&
  V^1_{109} &= e R_{ijac} R_{klbd} R_{maeb} R_{zced} \bar{\epsilon} \gamma^{ijklm} \psi^z, \notag \\
  V^1_{110} &= e R_{ijad} R_{klbc} R_{maeb} R_{zced} \bar{\epsilon} \gamma^{ijklm} \psi^z, \notag
\end{alignat}
\begin{alignat}{3}
  V^1_{111} &= e R_{ijza} R_{klac} R_{mnbd} R_{obcd} \bar{\epsilon} \gamma^{ijklmno} \psi^z, \qquad&
  V^1_{112} &= e R_{ijzc} R_{klab} R_{mnad} R_{obcd} \bar{\epsilon} \gamma^{ijklmno} \psi^z, \notag \\
  V^1_{113} &= e R_{ijab} R_{klab} R_{mncd} R_{ozcd} \bar{\epsilon} \gamma^{ijklmno} \psi^z, \qquad&
  V^1_{114} &= e R_{ijab} R_{klac} R_{mnbd} R_{oczd} \bar{\epsilon} \gamma^{ijklmno} \psi^z, \notag 
  \\[0.1cm]
  V^1_{115} &= e R_{ijab} R_{klab} R_{mncd} R_{opcd} \bar{\epsilon} \gamma^{ijklmnopz} \psi_z, \quad&
  V^1_{116} &= e R_{ijab} R_{klac} R_{mnbd} R_{opcd} \bar{\epsilon} \gamma^{ijklmnopz} \psi_z. \notag 
\end{alignat}
Thus there are 116 bases for $V[eR^4\bar{\ep}\psi]$.
This result is obtained both by hand and by the computer programming independently.

\subsubsection{Bases for $V[eR^2DR\bar{\ep}\psi_{(2)}]$}

There are 88 bases for $V[eR^2DR\bar{\ep}\psi_{(2)}]$. Before giving the explicit 
expressions we clarify the algorithm to obtain the result.

First of all let us classify the types of $[\bar{\ep}\psi_{(2)}]$.
Since all indices are contracted, the total number of the indices is even
and the number of the indices for $[\bar{\ep}\psi_{(2)}]$ is odd.
Then the number of the indices for the gamma matrix should be odd and
the types of $[\bar{\ep}\psi_{(2)}]$ are classified as
\begin{alignat}{3}
  [\bar{\ep}\psi_{(2)}] \sim \;
  &\bar{\epsilon} \gamma^{k} \psi^{ij} \oplus 
  \bar{\epsilon} \gamma^{klm} \psi^{ij} \oplus 
  \bar{\epsilon} \gamma^{klmno} \psi^{ij} . 
\end{alignat}
There are two remarks on this classification.
First of all since the ansatz does not contain the terms which depend on the
field equations, as a result the variations of $V[eR^2DR\bar{\ep}\psi_{(2)}]$ do
not contain the terms which depend on the field equations as well.
So we only take into account of the indices which are all uncontracted.
Second because of the cyclicity and the Bianchi identity of the Riemann tensor, 
the number of the uncontracted indices for the gamma matrix should be less than seven.

The types of $[R^2DR]$ are classified by the positions of the contracted indices.
As an example, let us consider a quartic term $R \s _c \s _d R \s _{cbd} D_b R \s\s\s\s$,
where $b$, $c$ and $d$ are contracted by the flat metric and blanks are arbitrary. 
This term is classified by the positions of the contracted indices as $\{2,3,1\}\{1,2\}$.
The numbers of 2 and 3 in $\{2,3,1\}$ represent the numbers of the contracted indices in 
the first and the second Riemann tensor, respectively.
The last number 1 in $\{2,3,1\}$ represents the number of the contracted indices in 
the covariant derivative and the third Riemann tensor,
because the indices of the covariant derivative and the third Riemann tensor
can be exchanged by using the Bianchi identity of the Riemann tensor.
Thus the index of the covariant derivative is grouped into the position of the third Riemann tensor.
The contracted index $b$ is contained in the second Riemann tensor and the third position, so
the numbers $(2,3)$ are assigned for this index. In a similar way the numbers
$(1,2)$ are assigned both for the indices $c$ and $d$, 
and totally this example has the numbers of $(2,3)^1(1,2)^2$.
The $\{1,2\}$ represents the numbers of the powers of $(2,3)^1$ and $(1,2)^2$,
where the numbers are aligned in order of rising.
Thus the example, $R \s _c \s _d R \s _{cbd} D_b R \s\s\s\s$, is classified by the numbers 
of $\{2,3,1\}\{1,2\}$ which are not affected by the properties of the Riemann tensor.
The types of $[R^2DR]$ are classified in this way and the complete list is given in the 
appendix \ref{BB}. The result there is checked both by hand and by the computer programming 
independently.

The independent terms of $V[eR^2DR\bar{\ep}\psi_{(2)}]$ are obtained by 
inserting the uncontracted indices of $[\bar{\ep}\psi_{(2)}]$ into the blanks of $[R^2DR]$.
Careful analysis shows that it is always possible to make the index of the covariant derivative 
filled by a contracted index by using the Bianchi identity of the Riemann tensor.
Therefore below we only consider the terms of $[R^2DR]$ in which the index of the covariant derivative
is already filled.

It is useful to classify the terms of $V[eR^2DR\bar{\ep}\psi_{(2)}]$ 
by the positions of the uncontracted indices of $[\bar{\ep}\psi_{(2)}]$ in the $[R^2DR]$.
For instance, the term $e R_{ijka} R_{ebcd} D_e R_{abcd} \bar{\epsilon} \gamma^{k} \psi^{ij}$
is assigned new numbers of $\{0,0,21\}$. The meaning of these numbers are as follows.
For each index of the gamma matrix the number 1 is assigned,
and for each index of the gravitino field strength the number 10 is assigned.
The first Riemann tensor of this example
has the number 21, the second does 0 and the third position does 0, 
and these numbers are aligned in order of rising. 
Thus the term $e R_{ijka} R_{ebcd} D_e R_{abcd} \bar{\epsilon} \gamma^{k} \psi^{ij}$ 
is classified by the numbers of $\{1,4,5\}\{1,4\}\{0,0,21\}$, 
which are not affected by the properties of the Riemann tensor and the gamma matrix.

By using the numbers discussed above it is almost possible to classify the terms of 
$V[eR^2DR\bar{\ep}\psi_{(2)}]$, and the explicit expressions for the bases are given by
\begin{alignat}{3}
V^2_{1} &= e R_{ijka} R_{ebcd} D_e R_{abcd} \bar{\epsilon} \gamma^{k} \psi^{ij}, \qquad&
V^2_{2} &= e R_{ijea} R_{kbcd} D_e R_{abcd} \bar{\epsilon} \gamma^{k} \psi^{ij}, \notag \\
V^2_{3} &= e R_{eaic} R_{jbkd} D_e R_{abcd} \bar{\epsilon} \gamma^{k} \psi^{ij}, \qquad&
V^2_{4} &= e R_{eaic} R_{kbjd} D_e R_{abcd} \bar{\epsilon} \gamma^{k} \psi^{ij}, \notag \\
V^2_{5} &= e R_{eaib} R_{jcad} D_e R_{kcbd} \bar{\epsilon} \gamma^{k} \psi^{ij}, \quad&
V^2_{6} &= e R_{eaib} R_{jdac} D_e R_{kcbd} \bar{\epsilon} \gamma^{k} \psi^{ij}, \notag \\
V^2_{7} &= e R_{eaib} R_{kcad} D_e R_{jcbd} \bar{\epsilon} \gamma^{k} \psi^{ij}, \quad&
V^2_{8} &= e R_{eakb} R_{icad} D_e R_{jcbd} \bar{\epsilon} \gamma^{k} \psi^{ij}, \notag \\
V^2_{9} &= e R_{eaib} R_{kdac} D_e R_{jcbd} \bar{\epsilon} \gamma^{k} \psi^{ij}, \qquad&
V^2_{10} &= e R_{kaeb} R_{idac} D_e R_{jcbd} \bar{\epsilon} \gamma^{k} \psi^{ij}, \notag \\
V^2_{11} &= e R_{eakb} R_{idac} D_e R_{jcbd} \bar{\epsilon} \gamma^{k} \psi^{ij}, \qquad& 
V^2_{12} &= e R_{ijab} R_{ecad} D_e R_{kcbd} \bar{\epsilon} \gamma^{k} \psi^{ij}, \notag \\
V^2_{13} &= e R_{ijab} R_{edac} D_e R_{kcbd} \bar{\epsilon} \gamma^{k} \psi^{ij}, \quad&
V^2_{14} &= e R_{iakb} R_{ecad} D_e R_{jcbd} \bar{\epsilon} \gamma^{k} \psi^{ij}, \notag \\
V^2_{15} &= e R_{kaib} R_{ecad} D_e R_{jcbd} \bar{\epsilon} \gamma^{k} \psi^{ij}, \qquad&
V^2_{16} &= e R_{iakb} R_{edac} D_e R_{jcbd} \bar{\epsilon} \gamma^{k} \psi^{ij}, \notag \\
V^2_{17} &= e R_{kaib} R_{edac} D_e R_{jcbd} \bar{\epsilon} \gamma^{k} \psi^{ij}, \quad&
V^2_{18} &= e R_{eaic} R_{abcd} D_e R_{jbkd} \bar{\epsilon} \gamma^{k} \psi^{ij}, \notag \\
V^2_{19} &= e R_{eaic} R_{abcd} D_e R_{kbjd} \bar{\epsilon} \gamma^{k} \psi^{ij}, \quad&
V^2_{20} &= e R_{ecad} R_{icbd} D_e R_{jakb} \bar{\epsilon} \gamma^{k} \psi^{ij}, \notag \\
V^2_{21} &= e R_{ecad} R_{icbd} D_e R_{kajb} \bar{\epsilon} \gamma^{k} \psi^{ij}, \qquad&
V^2_{22} &= e R_{edac} R_{icbd} D_e R_{kajb} \bar{\epsilon} \gamma^{k} \psi^{ij}, \notag \\
V^2_{23} &= e R_{ecad} R_{kcbd} D_e R_{ijab} \bar{\epsilon} \gamma^{k} \psi^{ij}, \quad& 
V^2_{24} &= e R_{edac} R_{kcbd} D_e R_{ijab} \bar{\epsilon} \gamma^{k} \psi^{ij}, \notag \\
V^2_{25} &= e R_{ebcd} R_{abcd} D_e R_{ijka} \bar{\epsilon} \gamma^{k} \psi^{ij}, \label{v2}
\\[0.1cm]
V^2_{26} &= e R_{ijkl} R_{dabc} D_d R_{mabc} \bar{\epsilon} \gamma^{klm} \psi^{ij}, \qquad&
V^2_{27} &= e R_{ijka} R_{dblc} D_d R_{mbac} \bar{\epsilon} \gamma^{klm} \psi^{ij}, \notag \\ 
V^2_{28} &= e R_{ijka} R_{lbdc} D_d R_{mbac} \bar{\epsilon} \gamma^{klm} \psi^{ij}, \qquad&
V^2_{29} &= e R_{ijda} R_{klbc} D_d R_{mabc} \bar{\epsilon} \gamma^{klm} \psi^{ij}, \notag \\
V^2_{30} &= e R_{ijac} R_{kadb} D_d R_{lmbc} \bar{\epsilon} \gamma^{klm} \psi^{ij}, \quad&
V^2_{31} &= e R_{ijac} R_{dakb} D_d R_{lmbc} \bar{\epsilon} \gamma^{klm} \psi^{ij}, \notag \\
V^2_{32} &= e R_{ijka} R_{dabc} D_d R_{lmbc} \bar{\epsilon} \gamma^{klm} \psi^{ij}, \qquad&
V^2_{33} &= e R_{iakl} R_{jbdc} D_d R_{mbac} \bar{\epsilon} \gamma^{klm} \psi^{ij}, \notag \\
V^2_{34} &= e R_{iakl} R_{jcdb} D_d R_{mbac} \bar{\epsilon} \gamma^{klm} \psi^{ij}, \quad& 
V^2_{35} &= e R_{iakc} R_{jbda} D_d R_{lmbc} \bar{\epsilon} \gamma^{klm} \psi^{ij}, \notag 
\end{alignat}
\begin{alignat}{3}
V^2_{36} &= e R_{icka} R_{jbda} D_d R_{lmbc} \bar{\epsilon} \gamma^{klm} \psi^{ij}, \quad&
V^2_{37} &= e R_{iakc} R_{jadb} D_d R_{lmbc} \bar{\epsilon} \gamma^{klm} \psi^{ij}, \notag \\
V^2_{38} &= e R_{icka} R_{jadb} D_d R_{lmbc} \bar{\epsilon} \gamma^{klm} \psi^{ij}, \qquad&
V^2_{39} &= e R_{iakd} R_{jblc} D_d R_{mbac} \bar{\epsilon} \gamma^{klm} \psi^{ij}, \notag \\
V^2_{40} &= e R_{iakd} R_{jclb} D_d R_{mbac} \bar{\epsilon} \gamma^{klm} \psi^{ij}, \quad&
V^2_{41} &= e R_{idka} R_{jclb} D_d R_{mbac} \bar{\epsilon} \gamma^{klm} \psi^{ij}, \notag \\
V^2_{42} &= e R_{iakl} R_{dbmc} D_d R_{jbac} \bar{\epsilon} \gamma^{klm} \psi^{ij}, \quad&
V^2_{43} &= e R_{iakl} R_{mbdc} D_d R_{jbac} \bar{\epsilon} \gamma^{klm} \psi^{ij}, \notag \\
V^2_{44} &= e R_{iakl} R_{dbac} D_d R_{jcmb} \bar{\epsilon} \gamma^{klm} \psi^{ij}, \qquad& 
V^2_{45} &= e R_{iakl} R_{dbac} D_d R_{jbmc} \bar{\epsilon} \gamma^{klm} \psi^{ij}, \notag \\
V^2_{46} &= e R_{ibda} R_{klac} D_d R_{jcmb} \bar{\epsilon} \gamma^{klm} \psi^{ij}, \quad&
V^2_{47} &= e R_{ibda} R_{klac} D_d R_{jbmc} \bar{\epsilon} \gamma^{klm} \psi^{ij}, \notag \\
V^2_{48} &= e R_{iadb} R_{klac} D_d R_{jcmb} \bar{\epsilon} \gamma^{klm} \psi^{ij}, \quad&
V^2_{49} &= e R_{iadb} R_{klac} D_d R_{jbmc} \bar{\epsilon} \gamma^{klm} \psi^{ij}, \notag \\
V^2_{50} &= e R_{ibac} R_{klda} D_d R_{jcmb} \bar{\epsilon} \gamma^{klm} \psi^{ij}, \qquad&
V^2_{51} &= e R_{ickb} R_{lmda} D_d R_{jbac} \bar{\epsilon} \gamma^{klm} \psi^{ij}, \notag \\
V^2_{52} &= e R_{ibkc} R_{lmda} D_d R_{jbac} \bar{\epsilon} \gamma^{klm} \psi^{ij}, \quad&
V^2_{53} &= e R_{ibkc} R_{dbac} D_d R_{jalm} \bar{\epsilon} \gamma^{klm} \psi^{ij}, \notag \\
V^2_{54} &= e R_{ickb} R_{dbac} D_d R_{jalm} \bar{\epsilon} \gamma^{klm} \psi^{ij}, \quad&
V^2_{55} &= e R_{iakd} R_{lbac} D_d R_{jbmc} \bar{\epsilon} \gamma^{klm} \psi^{ij}, \notag \\
V^2_{56} &= e R_{idka} R_{lbac} D_d R_{jbmc} \bar{\epsilon} \gamma^{klm} \psi^{ij}, \qquad&
V^2_{57} &= e R_{iakc} R_{dalb} D_d R_{jcmb} \bar{\epsilon} \gamma^{klm} \psi^{ij}, \notag \\
V^2_{58} &= e R_{icka} R_{dalb} D_d R_{jcmb} \bar{\epsilon} \gamma^{klm} \psi^{ij}, \quad& 
V^2_{59} &= e R_{iakc} R_{dalb} D_d R_{jbmc} \bar{\epsilon} \gamma^{klm} \psi^{ij}, \notag \\
V^2_{60} &= e R_{icka} R_{dalb} D_d R_{jbmc} \bar{\epsilon} \gamma^{klm} \psi^{ij}, \quad&
V^2_{61} &= e R_{iakc} R_{ladb} D_d R_{jcmb} \bar{\epsilon} \gamma^{klm} \psi^{ij}, \notag \\
V^2_{62} &= e R_{icka} R_{ladb} D_d R_{jcmb} \bar{\epsilon} \gamma^{klm} \psi^{ij}, \qquad&
V^2_{63} &= e R_{iakc} R_{ladb} D_d R_{jbmc} \bar{\epsilon} \gamma^{klm} \psi^{ij}, \notag \\
V^2_{64} &= e R_{icka} R_{ladb} D_d R_{jbmc} \bar{\epsilon} \gamma^{klm} \psi^{ij}, \quad&
V^2_{65} &= e R_{icdb} R_{kbac} D_d R_{jalm} \bar{\epsilon} \gamma^{klm} \psi^{ij}, \notag \\
V^2_{66} &= e R_{icab} R_{kbdc} D_d R_{jalm} \bar{\epsilon} \gamma^{klm} \psi^{ij}, \quad&
V^2_{67} &= e R_{ibdc} R_{kbac} D_d R_{jalm} \bar{\epsilon} \gamma^{klm} \psi^{ij}, \notag \\
V^2_{68} &= e R_{ibac} R_{kbdc} D_d R_{jalm} \bar{\epsilon} \gamma^{klm} \psi^{ij}, \qquad&
V^2_{69} &= e R_{kcdb} R_{lbac} D_d R_{ijma} \bar{\epsilon} \gamma^{klm} \psi^{ij}, \notag \\
V^2_{70} &= e R_{kbdc} R_{lbac} D_d R_{ijma} \bar{\epsilon} \gamma^{klm} \psi^{ij}, \quad&
V^2_{71} &= e R_{kbda} R_{lmac} D_d R_{ijbc} \bar{\epsilon} \gamma^{klm} \psi^{ij}, \notag \\
V^2_{72} &= e R_{kadb} R_{lmac} D_d R_{ijbc} \bar{\epsilon} \gamma^{klm} \psi^{ij}, \quad&
V^2_{73} &= e R_{klbc} R_{dbac} D_d R_{ijma} \bar{\epsilon} \gamma^{klm} \psi^{ij}, \notag \\
V^2_{74} &= e R_{kabc} R_{dabc} D_d R_{ijlm} \bar{\epsilon} \gamma^{klm} \psi^{ij}, \notag
\\[0.1cm]
V^2_{75} &= e R_{ijka} R_{lmcb} D_c R_{noab} \bar{\epsilon} \gamma^{klmno} \psi^{ij}, \qquad&
V^2_{76} &= e R_{klia} R_{jcmb} D_c R_{noab} \bar{\epsilon} \gamma^{klmno} \psi^{ij}, \notag \\
V^2_{77} &= e R_{klia} R_{mcjb} D_c R_{noab} \bar{\epsilon} \gamma^{klmno} \psi^{ij}, \qquad&
V^2_{78} &= e R_{klia} R_{mncb} D_c R_{jaob} \bar{\epsilon} \gamma^{klmno} \psi^{ij}, \notag \\
V^2_{79} &= e R_{klia} R_{mncb} D_c R_{oajb} \bar{\epsilon} \gamma^{klmno} \psi^{ij}, \quad&
V^2_{80} &= e R_{ijkl} R_{cmab} D_c R_{noab} \bar{\epsilon} \gamma^{klmno} \psi^{ij}, \notag \\
V^2_{81} &= e R_{klca} R_{mnab} D_c R_{ijob} \bar{\epsilon} \gamma^{klmno} \psi^{ij}, \quad&
V^2_{82} &= e R_{icka} R_{lmab} D_c R_{nojb} \bar{\epsilon} \gamma^{klmno} \psi^{ij}, \notag \\
V^2_{83} &= e R_{kcia} R_{lmab} D_c R_{nojb} \bar{\epsilon} \gamma^{klmno} \psi^{ij}, \qquad&
V^2_{84} &= e R_{iakb} R_{lmca} D_c R_{nojb} \bar{\epsilon} \gamma^{klmno} \psi^{ij}, \notag \\
V^2_{85} &= e R_{kaib} R_{lmca} D_c R_{nojb} \bar{\epsilon} \gamma^{klmno} \psi^{ij}, \quad&
V^2_{86} &= e R_{klia} R_{camb} D_c R_{nojb} \bar{\epsilon} \gamma^{klmno} \psi^{ij}, \notag \\
V^2_{87} &= e R_{klia} R_{macb} D_c R_{nojb} \bar{\epsilon} \gamma^{klmno} \psi^{ij}, \quad&
V^2_{88} &= e R_{ckab} R_{lmab} D_c R_{ijno} \bar{\epsilon} \gamma^{klmno} \psi^{ij}, \notag 
\end{alignat}
Thus there are 88 bases for $V[eR^2DR\bar{\ep}\psi_{(2)}]$.
This result is obtained both by hand and by the computer programming independently.

\subsubsection{Bases for $V[eR^3\bar{\ep}D\psi_{(2)}]$} \label{sec:V3}

The bases for $V[eR^3\bar{\ep}D\psi_{(2)}]$ are obtained as in the case
of deriving the ansatz for $\mathcal{L}[eR^3\bar{\psi}\psi_{(2)}]$.
That is, by replacing the Majorana gravitino in the $\mathcal{L}[eR^3\bar{\psi}\psi_{(2)}]$ 
with the covariant derivative, we obtain 92 bases for the $V[eR^3\bar{\ep}D\psi_{(2)}]$.
The explicit expressions for the bases are given by
\begin{alignat}{3}
  \big( V^3_1 &= e R_{afbg}R_{acde}R_{bcde} \bar{\ep} \gamma_f D_h \psi_{gh}, \qquad&
  V^3_2 &= e R_{afbc}R_{agde}R_{bcde} \bar{\ep} \gamma_f D_h \psi_{gh}, \notag \\
  V^3_3 &= e R_{bfad}R_{cgae}R_{bcde} \bar{\ep} \gamma_f D_h \psi_{gh} \big)_2 , \notag
  \\[0.1cm]
V^3_4 &= e R_{klij}R_{abcd}R_{abcd} \bar{\epsilon} \gamma^l D^k \psi^{ij}, \qquad&
V^3_5 &= e R_{klai}R_{bjcd}R_{abcd} \bar{\epsilon} \gamma^l D^k \psi^{ij}, \notag \\
V^3_6 &= e R_{kial}R_{bjcd}R_{abcd} \bar{\epsilon} \gamma^l D^k \psi^{ij}, \qquad&
V^3_7 &= e R_{ijak}R_{blcd}R_{abcd} \bar{\epsilon} \gamma^l D^k \psi^{ij}, \notag \\
V^3_8 &= e R_{ijal}R_{bkcd}R_{abcd} \bar{\epsilon} \gamma^l D^k \psi^{ij}, \quad&
V^3_9 &= e R_{klab}R_{ijcd}R_{abcd} \bar{\epsilon} \gamma^l D^k \psi^{ij}, \notag \\
V^3_{10} &= e R_{kiab}R_{ljcd}R_{abcd} \bar{\epsilon} \gamma^l D^k \psi^{ij}, \quad&
V^3_{11} &= e R_{klab}R_{aicd}R_{bjcd} \bar{\epsilon} \gamma^l D^k \psi^{ij}, \notag \\
V^3_{12} &= e R_{kiab}R_{alcd}R_{bjcd} \bar{\epsilon} \gamma^l D^k \psi^{ij}, \qquad&
V^3_{13} &= e R_{liab}R_{akcd}R_{bjcd} \bar{\epsilon} \gamma^l D^k \psi^{ij}, \notag \\
V^3_{14} &= e R_{ijab}R_{akcd}R_{blcd} \bar{\epsilon} \gamma^l D^k \psi^{ij}, \quad&
V^3_{15} &= e R_{klac}R_{biad}R_{bjcd} \bar{\epsilon} \gamma^l D^k \psi^{ij}, \notag \\
V^3_{16} &= e R_{kiac}R_{blad}R_{bjcd} \bar{\epsilon} \gamma^l D^k \psi^{ij}, \quad&
V^3_{17} &= e R_{liac}R_{bkad}R_{bjcd} \bar{\epsilon} \gamma^l D^k \psi^{ij}, \notag \\
V^3_{18} &= e R_{ijac}R_{bkad}R_{blcd} \bar{\epsilon} \gamma^l D^k \psi^{ij}, \qquad&
V^3_{19} &= e R_{akci}R_{bldj}R_{abcd} \bar{\epsilon} \gamma^l D^k \psi^{ij}, \notag \\
V^3_{20} &= e R_{akci}R_{blad}R_{bjcd} \bar{\epsilon} \gamma^l D^k \psi^{ij}, \qquad&
V^3_{21} &= e R_{alci}R_{bkad}R_{bjcd} \bar{\epsilon} \gamma^l D^k \psi^{ij}, \notag \\
V^3_{22} &= e R_{akbi}R_{alcd}R_{bjcd} \bar{\epsilon} \gamma^l D^k \psi^{ij}, \quad&
V^3_{23} &= e R_{albi}R_{akcd}R_{bjcd} \bar{\epsilon} \gamma^l D^k \psi^{ij}, \label{v3}
\\[0.1cm]
  \big(V^3_{24} &= e R_{efhi}R_{abcd}R_{abcd} \bar{\ep} \gamma_{efg} D_g \psi_{hi}, \qquad&
  V^3_{25} &= e R_{efah}R_{bicd}R_{abcd} \bar{\ep} \gamma_{efg} D_g \psi_{hi}, \notag \\
  V^3_{26} &= e R_{hiae}R_{bfcd}R_{abcd} \bar{\ep} \gamma_{efg} D_g \psi_{hi}, \qquad&
  V^3_{27} &= e R_{efab}R_{hicd}R_{abcd} \bar{\ep} \gamma_{efg} D_g \psi_{hi}, \notag \\
  V^3_{28} &= e R_{ehab}R_{ficd}R_{abcd} \bar{\ep} \gamma_{efg} D_g \psi_{hi}, \qquad&
  V^3_{29} &= e R_{efab}R_{ahcd}R_{bicd} \bar{\ep} \gamma_{efg} D_g \psi_{hi}, \notag \\
  V^3_{30} &= e R_{ehab}R_{afcd}R_{bicd} \bar{\ep} \gamma_{efg} D_g \psi_{hi}, \qquad&
  V^3_{31} &= e R_{hiab}R_{aecd}R_{bfcd} \bar{\ep} \gamma_{efg} D_g \psi_{hi}, \notag \\
  V^3_{32} &= e R_{efac}R_{bhad}R_{bicd} \bar{\ep} \gamma_{efg} D_g \psi_{hi}, \qquad&
  V^3_{33} &= e R_{ehac}R_{bfad}R_{bicd} \bar{\ep} \gamma_{efg} D_g \psi_{hi}, \notag \\
  V^3_{34} &= e R_{hiac}R_{bead}R_{bfcd} \bar{\ep} \gamma_{efg} D_g \psi_{hi}, \qquad&
  V^3_{35} &= e R_{aech}R_{bfdi}R_{abcd} \bar{\ep} \gamma_{efg} D_g \psi_{hi}, \notag \\
  V^3_{36} &= e R_{aech}R_{bfad}R_{bicd} \bar{\ep} \gamma_{efg} D_g \psi_{hi}, \qquad&
  V^3_{37} &= e R_{aebh}R_{facd}R_{ibcd} \bar{\ep} \gamma_{efg} D_g \psi_{hi}\big)_1, \notag
\\[0.1cm]
  \big( V^3_{38} &= e R_{efah}R_{bgcd}R_{abcd} \bar{\ep} \gamma_{efg} D_i \psi_{hi}, \qquad&
  V^3_{39} &= e R_{efab}R_{ghcd}R_{abcd} \bar{\ep} \gamma_{efg} D_i \psi_{hi}, \notag \\
  V^3_{40} &= e R_{efab}R_{agcd}R_{bhcd} \bar{\ep} \gamma_{efg} D_i \psi_{hi}, \qquad&
  V^3_{41} &= e R_{ehab}R_{afcd}R_{bgcd} \bar{\ep} \gamma_{efg} D_i \psi_{hi}, \notag \\
  V^3_{42} &= e R_{efac}R_{bgad}R_{bhcd} \bar{\ep} \gamma_{efg} D_i \psi_{hi}, \qquad&
  V^3_{43} &= e R_{ehac}R_{bfad}R_{bgcd} \bar{\ep} \gamma_{efg} D_i \psi_{hi} \big)_2, \notag
  \\[0.1cm]
V^3_{44} &= e R_{kilm}R_{anbc}R_{ajbc} \bar{\epsilon} \gamma^{lmn} D^k \psi^{ij}, \qquad&
V^3_{45} &= e R_{lmij}R_{akbc}R_{anbc} \bar{\epsilon} \gamma^{lmn} D^k \psi^{ij}, \notag \\
V^3_{46} &= e R_{lmak}R_{ijbc}R_{anbc} \bar{\epsilon} \gamma^{lmn} D^k \psi^{ij}, \qquad&
V^3_{47} &= e R_{lmak}R_{nibc}R_{ajbc} \bar{\epsilon} \gamma^{lmn} D^k \psi^{ij}, \notag \\
V^3_{48} &= e R_{lmai}R_{knbc}R_{ajbc} \bar{\epsilon} \gamma^{lmn} D^k \psi^{ij}, \quad&
V^3_{49} &= e R_{lmai}R_{njbc}R_{akbc} \bar{\epsilon} \gamma^{lmn} D^k \psi^{ij}, \notag \\
V^3_{50} &= e R_{lmai}R_{kjbc}R_{anbc} \bar{\epsilon} \gamma^{lmn} D^k \psi^{ij}, \quad&
V^3_{51} &= e R_{klai}R_{mnbc}R_{ajbc} \bar{\epsilon} \gamma^{lmn} D^k \psi^{ij}, \notag \\
V^3_{52} &= e R_{klai}R_{mjbc}R_{anbc} \bar{\epsilon} \gamma^{lmn} D^k \psi^{ij}, \qquad&
V^3_{53} &= e R_{liak}R_{mnbc}R_{ajbc} \bar{\epsilon} \gamma^{lmn} D^k \psi^{ij}, \notag 
\end{alignat}
\begin{alignat}{3}
V^3_{54} &= e R_{liak}R_{mjbc}R_{anbc} \bar{\epsilon} \gamma^{lmn} D^k \psi^{ij}, \quad&
V^3_{55} &= e R_{ijal}R_{kmbc}R_{anbc} \bar{\epsilon} \gamma^{lmn} D^k \psi^{ij}, \notag \\
V^3_{56} &= e R_{ijal}R_{mnbc}R_{akbc} \bar{\epsilon} \gamma^{lmn} D^k \psi^{ij}, \quad&
V^3_{57} &= e R_{ijak}R_{lmbc}R_{anbc} \bar{\epsilon} \gamma^{lmn} D^k \psi^{ij}, \notag \\
V^3_{58} &= e R_{lmbk}R_{anci}R_{ajbc} \bar{\epsilon} \gamma^{lmn} D^k \psi^{ij}, \qquad&
V^3_{59} &= e R_{lmbi}R_{akcn}R_{ajbc} \bar{\epsilon} \gamma^{lmn} D^k \psi^{ij}, \notag \\
V^3_{60} &= e R_{lmbi}R_{ancj}R_{akbc} \bar{\epsilon} \gamma^{lmn} D^k \psi^{ij}, \quad&
V^3_{61} &= e R_{lmbi}R_{akcj}R_{anbc} \bar{\epsilon} \gamma^{lmn} D^k \psi^{ij}, \notag \\
V^3_{62} &= e R_{klbi}R_{ajcm}R_{anbc} \bar{\epsilon} \gamma^{lmn} D^k \psi^{ij}, \quad&
V^3_{63} &= e R_{libk}R_{ajcm}R_{anbc} \bar{\epsilon} \gamma^{lmn} D^k \psi^{ij}, \notag \\
V^3_{64} &= e R_{ijbl}R_{akcm}R_{anbc} \bar{\epsilon} \gamma^{lmn} D^k \psi^{ij}, \qquad&
V^3_{65} &= e R_{klab}R_{mnac}R_{ijbc} \bar{\epsilon} \gamma^{lmn} D^k \psi^{ij}, \notag \\
V^3_{66} &= e R_{kiab}R_{lmac}R_{njbc} \bar{\epsilon} \gamma^{lmn} D^k \psi^{ij}, \quad&
V^3_{67} &= e R_{lmab}R_{ijac}R_{bkcn} \bar{\epsilon} \gamma^{lmn} D^k \psi^{ij}, \notag \\
V^3_{68} &= e R_{liab}R_{mjac}R_{bkcn} \bar{\epsilon} \gamma^{lmn} D^k \psi^{ij}, \quad&
V^3_{69} &= e R_{lmab}R_{niac}R_{bkcj} \bar{\epsilon} \gamma^{lmn} D^k \psi^{ij}, \notag \\
V^3_{70} &= e R_{lmab}R_{kiac}R_{bncj} \bar{\epsilon} \gamma^{lmn} D^k \psi^{ij}, \qquad&
V^3_{71} &= e R_{klab}R_{miac}R_{bncj} \bar{\epsilon} \gamma^{lmn} D^k \psi^{ij}, \notag \\
V^3_{72} &= e R_{lmab}R_{ckai}R_{bncj} \bar{\epsilon} \gamma^{lmn} D^k \psi^{ij}, \quad&
V^3_{73} &= e R_{liab}R_{ckam}R_{bncj} \bar{\epsilon} \gamma^{lmn} D^k \psi^{ij}, \notag \\
V^3_{74} &= e R_{akbl}R_{cmai}R_{bncj} \bar{\epsilon} \gamma^{lmn} D^k \psi^{ij}, \notag
\\[0,1cm]
  \big( V^3_{75} &= e R_{deai}R_{fgbc}R_{ahbc} \bar{\ep} \gamma_{defgh} D_j \psi_{ij}, \qquad&
  V^3_{76} &= e R_{deab}R_{fgac}R_{bhci} \bar{\ep} \gamma_{defgh} D_j \psi_{ij} \big)_2, \notag 
\\[0.1cm]
  \big(V^3_{77} &= e R_{deai}R_{fgbc}R_{ajbc} \bar{\ep} \gamma_{defgh} D_h \psi_{ij}, \qquad&
  V^3_{78} &= e R_{ijad}R_{efbc}R_{agbc} \bar{\ep} \gamma_{defgh} D_h \psi_{ij}, \notag \\
  V^3_{79} &= e R_{deai}R_{fjbc}R_{agbc} \bar{\ep} \gamma_{defgh} D_h \psi_{ij}, \qquad&
  V^3_{80} &= e R_{debi}R_{afcj}R_{agbc} \bar{\ep} \gamma_{defgh} D_h \psi_{ij}, \notag \\
  V^3_{81} &= e R_{deab}R_{fiac}R_{bgcj} \bar{\ep} \gamma_{defgh} D_h \psi_{ij}\big)_1, \notag
  \\[0.1cm]
V^3_{82} &= e R_{lmki}R_{noab}R_{pjab} \bar{\epsilon} \gamma^{lmnop} D^k \psi^{ij}, \qquad&
V^3_{83} &= e R_{klij}R_{mnab}R_{opab} \bar{\epsilon} \gamma^{lmnop} D^k \psi^{ij}, \notag \\
V^3_{84} &= e R_{lmij}R_{knab}R_{opab} \bar{\epsilon} \gamma^{lmnop} D^k \psi^{ij}, \qquad&
V^3_{85} &= e R_{lmai}R_{nobj}R_{kpab} \bar{\epsilon} \gamma^{lmnop} D^k \psi^{ij}, \notag \\
V^3_{86} &= e R_{lmak}R_{ijbn}R_{opab} \bar{\epsilon} \gamma^{lmnop} D^k \psi^{ij}, \quad&
V^3_{87} &= e R_{liak}R_{mnbj}R_{opab} \bar{\epsilon} \gamma^{lmnop} D^k \psi^{ij}, \notag \\
V^3_{88} &= e R_{klai}R_{mnbj}R_{opab} \bar{\epsilon} \gamma^{lmnop} D^k \psi^{ij}, \quad&
V^3_{89} &= e R_{lmak}R_{nobi}R_{pjab} \bar{\epsilon} \gamma^{lmnop} D^k \psi^{ij}, \notag \\
V^3_{90} &= e R_{lmak}R_{nobi}R_{apbj} \bar{\epsilon} \gamma^{lmnop} D^k \psi^{ij}, \notag
\\[0.1cm]
  \big(V^3_{91} &= e R_{cdjk}R_{efab}R_{ghab} \bar{\ep} \gamma_{cdefghi} D_i \psi_{jk}, \qquad&
  V^3_{92} &= e R_{cdaj}R_{efbk}R_{ghab} \bar{\ep} \gamma_{cdefghi} D_i \psi_{jk}\big)_1. \notag       
\end{alignat}
Note that these terms are classified into 3 types.
The first one is $V^3_{24\sim 37}$, $V^3_{77\sim 81}$ and $V^3_{91,92}$
where the index of the covariant derivative is contracted with the index of the gamma matrix.
By applying the first equation in (\ref{feq}), these terms are expanded by
$V^3_{4\sim 23}$, $V^3_{44\sim 74}$, $V^3_{82\sim 90}$ and $V^1_{1\sim 116}$.
The second one is $V^3_{1\sim 3}$, $V^3_{38\sim 43}$ and $V^3_{75,76}$
where the index of the covariant derivative is contracted with the index of the gravitino field strength.
By applying the second equation in (\ref{feq}), these terms are expanded by $V^1_{1\sim 116}$.
The third one is $V^3_{4\sim 23}$, $V^3_{44\sim 74}$ and $V^3_{82\sim 90}$, which
are the 60 bases for the $V[eR^3\bar{\ep}D\psi_{(2)}]$.

\subsubsection{The Identities}

As mentioned before the bases for $V[eR^4\bar{\epsilon}\psi]$ and
for $V[eR^3\bar{\epsilon}D\psi_{(2)}]$ are not independent because of the identity,
$D_{[e} \psi_{cd]} \sim \tfrac{1}{4} \gamma^{ab} \psi_{[e} R_{cd] ab}$.
By using the computer programming we find 20 identities among these bases.
The result is as follows.
\begin{alignat}{3}
  I_1 &= 2 V^1_{12} + 2 V^1_{13} - 4 V^1_{14} + 2 V^1_{32} + V^1_{36} 
  - 8 V^3_{6} - 4 V^3_{8} = 0, \notag \\
  I_{2}&=V^1_{30}-2 V^1_{37}-4 V^3_{9}+8 V^3_{10}=0, \notag \\
  I_{3}&=4 V^1_{17}+4 V^1_{18}-4 V^1_{19}-4 V^1_{20}+2 V^1_{21}-2 V^1_{22}-V^1_{33}+V^1_{48}
        +2 V^3_{11}+4 V^3_{13}=0, \notag \\
  I_{4}&=8 V^1_{17}+8 V^1_{18}-8 V^1_{19}-8 V^1_{20}+4 V^1_{21}-4 V^1_{22}+V^1_{39}+2 V^1_{47}
        +8 V^3_{12}-4 V^3_{14}=0, \notag \\
  I_{5}&=2 V^1_{17}-2 V^1_{20}+2 V^1_{23}-2 V^1_{24}-V^1_{34}+V^1_{44}
        +2 V^3_{15}+4 V^3_{17}=0, \notag \\
  I_{6}&=4 V^1_{17}-4 V^1_{20}+4 V^1_{23}-4 V^1_{24}+V^1_{40}+2 V^1_{43}
        +8 V^3_{16}-4 V^3_{18}=0, \notag \\
  I_{7}&=8 V^1_{8}+2 V^1_{11}-8 V^1_{12}+4 V^1_{26}+2 V^1_{36}-8 V^1_{50}-2 V^1_{86}-V^1_{105}
        +8 V^3_{44}+4 V^3_{45}=0, \notag \\
  I_{8}&=2 V^1_{9}+2 V^1_{13}-V^1_{15}+8 V^1_{17}-8 V^1_{19}+V^1_{27}
          -V^1_{37}+2 V^1_{42}-16 V^1_{54}+4 V^1_{62}+4 V^1_{65} \notag \\
       &\quad+V^1_{88}-V^1_{108}+4 V^3_{47}+4 V^3_{48}+4 V^3_{49}=0, \notag \\
  I_{9}&=4 V^1_{9}+4 V^1_{13}-2 V^1_{15}+16 V^1_{17}-16 V^1_{19}+2 V^1_{27}
        +4 V^1_{42}-4 V^1_{48}-4 V^1_{51}-8 V^1_{65} \notag \\
       &\quad+V^1_{87}-2 V^1_{108}-4 V^3_{46}+8 V^3_{50}=0, \notag \\
  I_{10}&=2 V^1_{9}+2 V^1_{13}-V^1_{15}+8 V^1_{18}-8 V^1_{19}-4 V^1_{49}+4 V^1_{50}+2 V^1_{51}
         +8 V^1_{53}+4 V^1_{55}-4 V^1_{61} \notag \\
        &\quad-4 V^1_{63}+4 V^1_{64}-4 V^1_{95}+4 V^1_{96}-2 V^1_{99}
         -8 V^3_{52}-8 V^3_{54}+4 V^3_{55}=0, \notag \\
  I_{11}&=2 V^1_{9}+2 V^1_{13}-V^1_{15}+8 V^1_{18}-8 V^1_{19}+V^1_{35}-V^1_{39}+2 V^1_{40} \label{id}
         +4 V^1_{55}+2 V^1_{63}-V^1_{90} \\
        &\quad+V^1_{106}-4 V^3_{51}-4 V^3_{53}-2 V^3_{56}=0, \notag \\
  I_{12}&=4 V^1_{8}-2 V^1_{9}+8 V^1_{10}-4 V^1_{14}+8 V^1_{16}-8 V^1_{17}+8 V^1_{19}
         -8 V^1_{20}-V^1_{27}+4 V^1_{28} \notag \\
        &\quad+4 V^1_{38}-4 V^1_{46}+8 V^1_{54}+8 V^1_{58}-4 V^1_{62}-8 V^1_{66}-V^1_{88}+2 V^1_{89}
         -2 V^1_{92}+2 V^1_{107} \notag \\
        &\quad-4 V^3_{48}-8 V^3_{58}+8 V^3_{59}-8 V^3_{60}=0, \notag \\
  I_{13}&=2 V^1_{8}-V^1_{9}+4 V^1_{10}-2 V^1_{14}+4 V^1_{16}-4 V^1_{18}+4 V^1_{19}-4 V^1_{20}
         +2 V^1_{49}-V^1_{51} \notag \\
        &\quad-4 V^1_{52}-2 V^1_{53}+4 V^1_{57}-4 V^1_{59}+V^1_{63}-2 V^1_{65}+4 V^1_{66}
         +2 V^1_{95}-2 V^1_{97}+V^1_{99} \notag \\
        &\quad-8 V^3_{62}-8 V^3_{63}-4 V^3_{64}=0, \notag \\
  I_{14}&=12 V^1_{21}-4 V^1_{41}+4 V^1_{55}-4 V^1_{56}+8 V^1_{67}+8 V^1_{68}+V^1_{90}
            -V^1_{91}-2 V^1_{109}-4 V^3_{65}-8 V^3_{66}=0, \notag \\
  I_{15}&=2 V^1_{9}+8 V^1_{22}+8 V^1_{23}-2 V^1_{41}-4 V^1_{45}+2 V^1_{47}-4 V^1_{56}-8 V^1_{67}
           -V^1_{91}-2 V^1_{110} \notag \\
        &\quad-4 V^3_{65}+4 V^3_{67}+8 V^3_{70}=0, \notag \\
  I_{16}&=V^1_{9}-3 V^1_{21}+4 V^1_{22}+4 V^1_{23}-V^1_{55}-V^1_{56}+4 V^1_{59}-4 V^1_{60}
       -2 V^1_{61}+2 V^1_{62} \notag \\
        &\quad-2 V^1_{67}-2 V^1_{68}-2 V^1_{102}-2 V^1_{103}+2 V^1_{104}
         +4 V^3_{68}-8 V^3_{71}=0, \notag \\
  I_{17}&=V^1_{9}-12 V^1_{10}-6 V^1_{21}+4 V^1_{22}+28 V^1_{23}-24 V^1_{24}+2 V^1_{55}
       +2 V^1_{56}+12 V^1_{57} \notag \\
        &\quad+12 V^1_{58}-8 V^1_{59}-4 V^1_{60}-2 V^1_{61}-4 V^1_{62}+16 V^1_{67}+4 V^1_{68}
        +6 V^1_{101}+4 V^1_{102} \notag \\
        &\quad-2 V^1_{103}-4 V^1_{104}-8 V^3_{68}-8 V^3_{71}-24 V^3_{74}=0, \notag 
\end{alignat}
\begin{alignat}{3}
  I_{18}&=2 V^1_{29}+4 V^1_{30}-4 V^1_{35}+4 V^1_{36}-4 V^1_{39}-16 V^1_{49}+16 V^1_{50}
        -16 V^1_{53}-2 V^1_{87} \notag \\
        &\quad+8 V^1_{93}+16 V^1_{95}+4 V^1_{106}+4 V^1_{111}+V^1_{113}
        -8 V^3_{82}+4 V^3_{84}=0, \notag \\
  I_{19}&=2 V^1_{26}-V^1_{27}-2 V^1_{31}+2 V^1_{32}+2 V^1_{33}-4 V^1_{38}+4 V^1_{44}-4 V^1_{46}
         -4 V^1_{55}+4 V^1_{56}+8 V^1_{58} \notag \\
        &\quad-8 V^1_{60}+4 V^1_{61}-4 V^1_{62}+V^1_{86}-2 V^1_{93}
         +4 V^1_{98}-2 V^1_{99}-4 V^1_{101}-8 V^1_{102}+4 V^1_{103} \notag \\
        &\quad+4 V^1_{104}+2 V^1_{107}-V^1_{108}+V^1_{111}+V^1_{114}
         -2 V^3_{86}+4 V^3_{87}+4 V^3_{88}=0, \notag \\
  I_{20}&=2 V^1_{26}-V^1_{27}+2 V^1_{31}-2 V^1_{32}-2 V^1_{33}+2 V^1_{41}-2 V^1_{42}+4 V^1_{43}
         -4 V^1_{45}-2 V^1_{47}+2 V^1_{48} \notag \\
        &\quad+8 V^1_{52}+8 V^1_{57}-8 V^1_{59}+V^1_{88}-4 V^1_{94}-4 V^1_{97}+4 V^1_{98}
        -4 V^1_{101}+4 V^1_{104}-2 V^1_{109} \notag \\
        &\quad+2 V^1_{110}-V^1_{112}-V^1_{114}-2 V^3_{85}+4 V^3_{89}=0. \notag
\end{alignat}
Now we are ready to examine the variations of the ansatz under the local supersymmetry
transformations.

\subsection{The Variations of the Ansatz}

\subsubsection{$ \delta \mc{L}[eR^4]$ terms}

In the case of the supergravity, the variation of the spin connection is automatically
cancelled by solving (\ref{spin}).
For the higher derivative corrections, however, the variations for the spin connection 
do not cancel automatically. As discussed before the Riemann tensors in the ansatz of $\mc{L}[eR^4]$
are expressed by the supercovariant spin connection, and the supersymmetric transformation 
for this field is given by
\begin{alignat}{3}
  \delta \hat{\omega}_\mu{}^{ab} &= - \tfrac{1}{2} \bar{\epsilon} \gamma_\mu \psi^{ab}
  + \tfrac{1}{2} \bar{\epsilon} \gamma^a \psi^b{}_{\mu}
  - \tfrac{1}{2} \bar{\epsilon} \gamma^b \psi^a{}_{\mu}.
\end{alignat}
By definition the above transformation does not include derivatives of the supersymmetric 
parameter $\epsilon$.

Now let us consider the variation of the ansatz $\mc{L}[eR^4]$ under the local supersymmetry.
In order to execute this, it is convenient to use the variation of the Riemann tensor,
\begin{alignat}{3}
  \delta R_{abcd} &= e^\mu{}_c e^\nu{}_d \delta R_{ab\mu\nu}(\hat{\omega})
  - R_{abd\mu} \delta e^\mu{}_c + R_{abc\nu} \delta e^\nu{}_d \notag
  \\
  &= D_c \delta \hat{\omega}_{dab} - D_d \delta \hat{\omega}_{cab}
  + R_{abde} \bar{\epsilon} \gamma^e \psi_c - R_{abce} \bar{\epsilon} \gamma^e \psi_d
  + \mathcal{O}(\psi^3).
\end{alignat}
The covariant derivative acts on all local Lorentz indices.
Since we focus on the cancellation of the terms which linearly depend on the gravitino,
we neglect the torsion part in the second line.
Let us vary $b^1_1$ term of $\mathcal{L}[eR^4]$ in the eq.(\ref{b1}) as an example.
\begin{alignat}{3}
  \delta(b^1_1 e R_{abcd}R_{abcd}R_{efgh}R_{efgh}) &\sim
  + b^1_1 e R_{abcd}R_{abcd}R_{efgh}R_{efgh} \bar{\epsilon} \gamma^z \psi_z \notag
  \\
  &\quad -8 b^1_1 e R_{iefg}R_{zefg}R_{abcd}R_{abcd} \bar{\epsilon} \gamma^i \psi_z
  \\
  &\quad -32 b^1_1 e R_{ijka}R_{ebcd}D_eR_{abcd} \bar{\epsilon} \gamma^k \psi^{ij}. \notag
\end{alignat}
Here the partial integral and the Bianchi identity of the Riemann tensor are used to
derive the above expression.
We also dropped the terms which are proportional to the field equations since
those terms are canceled by modifying the supersymmetry transformation of $\dl e^a{}_\mu$.

In a similar way, it is basically possible to vary the ansatz $\mc{L}[eR^4]$
under the local supersymmetry. Since this operation is systematic, 
we employ the computer programming by the mathematica to calculate the variations.
The result is given as follows.
\begin{alignat}{3}
  & \delta \mc{L}[eR^4] = \notag \\
  &+ b^1_{1} (V^1_{1} - 8 V^1_{11} - 32 V^2_{1}) \notag \\
  &+ b^1_{2} (-\tfrac{1}{2}V^1_{2} + 4 V^1_{14} - V^2_{2} - 4 V^2_{3} - 4 V^2_{4} + 
          2 V^2_{13} + 4 V^2_{16} + 4 V^2_{17}) \notag \\
  &+ b^1_{3} (V^1_{2} - 4 V^1_{12} - 4 V^1_{13} - 4 V^2_{1} + 2 V^2_{2} - 
          8 V^2_{3} + 8 V^2_{4} - 4 V^2_{13} - 8 V^2_{16} + 8 V^2_{17} - 
          4 V^2_{25}) \notag \\
  &+ b^1_{4} (-\tfrac{1}{8}V^1_{3} + V^1_{15} - \tfrac{7}{4} V^2_{2} + V^2_{3} - V^2_{4} + 
          V^2_{18} - V^2_{19} - \tfrac{7}{2} V^2_{23} + \tfrac{7}{2} V^2_{24}) \notag \\
  &+ b^1_{5} (-\tfrac{1}{4}V^1_{3} + 2 V^1_{15} - 3 V^2_{2} + 4 V^2_{3} - 
          4 V^2_{4} + 4 V^2_{18} - 4 V^2_{19} - 6 V^2_{23} + 6 V^2_{24}) \notag \\
  &+ b^1_{6} (\tfrac{1}{8}V^1_{3} - V^1_{4} - V^1_{15} + 8 V^1_{16} + V^2_{2} - 4 V^2_{3} - 
          4 V^2_{4} - 4 V^2_{18} - 4 V^2_{19} + 2 V^2_{23} - 2 V^2_{24}) \notag \\
  &+ b^1_{7} (V^1_{5} + 4 V^1_{18} - 4 V^1_{20} - 2 V^1_{21} - 8 V^1_{23} + 
          8 V^1_{24} - 6 V^2_{5} + 2 V^2_{6} + 2 V^2_{7} + 2 V^2_{8} + 2 V^2_{9} \notag \\
          &\qquad+ 2 V^2_{11} + V^2_{12} - V^2_{13} - 2 V^2_{14} - 2 V^2_{15} + 2 V^2_{16} + 
          2 V^2_{17} + 2 V^2_{18} - 2 V^2_{19} + 4 V^2_{21} - 4 V^2_{22} \notag \\
          &\qquad+ V^2_{23} - V^2_{24}) \notag \\
  &+ b^1_{8} (V^1_{5} + 4 V^1_{17} + 4 V^1_{18} - 8 V^1_{20} - 2 V^1_{21} - 
          4 V^1_{23} + 4 V^1_{24} - 2 V^2_{5} + 2 V^2_{7} + 2 V^2_{8} + 
          2 V^2_{12} \notag \\
          &\qquad- V^2_{13} - 4 V^2_{15} + 2 V^2_{16} + 2 V^2_{17} + 
          2 V^2_{18} - 2 V^2_{19} - 2 V^2_{20} + 6 V^2_{21} - 4 V^2_{22} + \label{vb1}
          2 V^2_{23} - V^2_{24}) \\ 
  &+ b^1_{9} (\tfrac{1}{2}V^1_{5} - V^1_{7} + 2 V^1_{17} + 2 V^1_{18} - V^1_{21} + 2 V^1_{23} + 
          2 V^1_{24} + V^2_{5} - 3 V^2_{6} - V^2_{7} - V^2_{8} - V^2_{9} \notag \\
          &\qquad+ V^2_{11} + \tfrac{1}{2}V^2_{12} - \tfrac{1}{2}V^2_{13} + V^2_{14} + V^2_{15} + V^2_{16} + V^2_{17} + 
          V^2_{18} - V^2_{19} - 2 V^2_{20} - 2 V^2_{22} \notag \\
          &\qquad+ \tfrac{1}{2}V^2_{23} - \tfrac{1}{2}V^2_{24}) \notag \\ 
  &+ b^1_{10} (\tfrac{1}{4}V^1_{6} + 6 V^1_{17} + 6 V^1_{18} - 6 V^1_{19} - 6 V^1_{20} - 
          V^1_{21} + V^1_{22} - 2 V^2_{7} + 2 V^2_{8} + 2 V^2_{9} + 2 V^2_{10} \notag \\
          &\qquad- 2 V^2_{11} + 2 V^2_{12} - 2 V^2_{13} + 2 V^2_{14} - 2 V^2_{15} - 
          2 V^2_{16} + 2 V^2_{17} - 2 V^2_{18} + 2 V^2_{19} + 2 V^2_{23} - 
          2 V^2_{24}) \notag \\ 
  &+ b^1_{11} (-\tfrac{1}{2}V^1_{6} - 14 V^1_{17} - 14 V^1_{18} + 14 V^1_{19} + 
          14 V^1_{20} + V^1_{21} - V^1_{22} + 2 V^2_{7} - 2 V^2_{8} - 2 V^2_{9} \notag \\
          &\qquad- 2 V^2_{10} + 2 V^2_{11} - 4 V^2_{12} + 4 V^2_{13} - 6 V^2_{14} + 
          6 V^2_{15} + 6 V^2_{16} - 6 V^2_{17} + 6 V^2_{18} - 6 V^2_{19} \notag \\
          &\qquad- 4 V^2_{23} + 4 V^2_{24}) \notag \\
  &+ b^1_{12} (-V^1_{6} + 16 V^1_{21} - 16 V^1_{22} + 16 V^2_{5} - 16 V^2_{6} + 
          16 V^2_{7} - 16 V^2_{8} - 16 V^2_{9} - 32 V^2_{10} + 16 V^2_{11}) \notag \\
  &+ b^1_{13} (-V^1_{7} + 4 V^1_{20} + 4 V^1_{23} - 2 V^2_{6} - 4 V^2_{8} - 
          2 V^2_{9} + 2 V^2_{11} - V^2_{12} + 2 V^2_{14} + 2 V^2_{15} - 2 V^2_{20} \notag \\
          &\qquad- 2 V^2_{21} - V^2_{23}). \notag
\end{alignat}
The variations which depend on the Ricci tensor or the scalar curvature are neglected.
As explained in the sec.~\ref{sec:ov}, those terms are cancelled by modifying the 
supersymmetric transformation rule of the vielbein.

\subsubsection{$ \delta \mc{L}[e\ep_{11}AR^4]$ terms}

Let us consider the variations of $\mc{L}[e\ep_{11}AR^4]$ under the local
supersymmetry. Since we are interested in the variations which are independent of 
the 3-form potential, we only need to consider the variation of the 3-form potential
in this ansatz. The calculation is straightforward and the result is written as
\begin{alignat}{3}
  &\delta \mc{L}[e\ep_{11}AR^4] = b^2_{1} V^1_{115} + b^2_{2} V^1_{116}. \label{vb2}
\end{alignat}
Note that, because of the presence of the $\ep_{11}$ tensor, the gamma matrix with 2 indices
is mapped to the one with 9 indices.

\subsubsection{$ \delta \mc{L}[eR^3\bar{\psi}\psi_{(2)}]$ terms}

As discussed in the sec.~\ref{sec:V3}, by applying the field equations (\ref{feq}),
the variations of the ansatz $\mc{L}[eR^3\bar{\psi}\psi_{(2)}]$ are
classified into three types.

For the terms with the coefficients $f^1_{24\sim 37}$, $f^1_{77\sim 81}$ and $f^1_{91,92}$
in the $\mc{L}[eR^3\bar{\psi}\psi_{(2)}]$, the variations of them are evaluated as
\begin{alignat}{3}
  &\delta (e R^3_{i_1 \cdots i_n ij} \bar{\psi}_z \gamma^{i_1 \cdots i_n z} \psi^{ij}) \notag
  \\[0.1cm]
  &= 2 e R^3_{i_1 \cdots i_n ij} \overline{D_z \epsilon} \gamma^{i_1 \cdots i_n z} \psi^{ij} 
  + \tfrac{1}{2} e R^3_{i_1 \cdots i_n ij}R_{xy}{}^{ij} \bar{\psi}_z 
  \gamma^{i_1 \cdots i_n z} \gamma^{xy} \epsilon 
  \notag
  \\
  &= - 2 e \bar{\epsilon} (D_z R^3_{i_1 \cdots i_n ij}) \gamma^{i_1 \cdots i_n z} \psi^{ij} 
  + 2n e \bar{\epsilon} R^3_{i_1 \cdots i_n ij} \gamma^{[i_1 \cdots i_{n-1}} D^{i_n]} \psi^{ij} 
  \\
  &\quad\,
  - \tfrac{1}{2} e R^3_{i_1 \cdots i_n ij}R_{xy}{}^{ij} \bar{\epsilon} 
  \big\{ \gamma^{i_1 \cdots i_n} \gamma^z \gamma^{xy} 
  + C_{n+1} \gamma^{xy} \gamma^{i_1 \cdots i_n z} \big\} \psi_z, \notag
\end{alignat}
where $C_n = (-1)^{\frac{n(n+1)}{2}}$.
We used $\gamma^{i_1 \cdots i_n z} = \gamma^{i_1 \cdots i_n}\gamma^z - n \gamma^{[i_1 \cdots 
i_{n-1}} \eta^{i_n] z}$ and the first equation in (\ref{feq}).

For the terms with the coefficients $f^1_{1\sim 3}$, $f^1_{38\sim 43}$ and $f^1_{75,76}$
in the $\mc{L}[eR^3\bar{\psi}\psi_{(2)}]$, the variations of them are evaluated as
\begin{alignat}{3}
  &\delta (e R^3_{i_1 \cdots i_n i} \bar{\psi}_z \gamma^{i_1 \cdots i_n} \psi^{iz}) \notag
  \\[0.1cm]
  &= 2 e R^3_{i_1 \cdots i_n i} \overline{D_z \epsilon} \gamma^{i_1 \cdots i_n} \psi^{iz} 
  + \tfrac{1}{2} e R^3_{i_1 \cdots i_n i}R_{xy}{}^{iz} \bar{\psi}_z \gamma^{i_1 \cdots i_n} 
  \gamma^{xy} \epsilon 
  \\
  &= - 2 e \bar{\epsilon} (D_z R^3_{i_1 \cdots i_n i}) \gamma^{i_1 \cdots i_n} \psi^{iz}
  - \tfrac{1}{2} e R^3_{i_1 \cdots i_n i}R_{xy}{}^{iz} \bar{\epsilon} \big\{ \gamma^{i_1 \cdots i_n} \gamma^{xy}
  + C_n \gamma^{xy} \gamma^{i_1 \cdots i_n} \big\} \psi_z. \notag
\end{alignat}
Here we used the second equation in (\ref{feq}).

For the terms with the coefficients $f^1_{4\sim 23}$, $f^1_{44\sim 74}$ and $f^1_{82\sim 90}$
in the $\mc{L}[eR^3\bar{\psi}\psi_{(2)}]$, the variations of them are evaluated as
\begin{alignat}{3}
  &\delta (e R^3_{i_1 \cdots i_n ijz} \bar{\psi}^z \gamma^{i_1 \cdots i_n} \psi^{ij}) 
  \notag
  \\[0.1cm]
  &= 2 e R^3_{i_1 \cdots i_n ij}{}^z \overline{D_z \epsilon} \gamma^{i_1 \cdots i_n} \psi^{ij} 
  + \tfrac{1}{2} e R^3_{i_1 \cdots i_n ijz}R_{xy}{}^{ij} \bar{\psi}^z \gamma^{i_1 \cdots i_n} \gamma^{xy} 
  \epsilon \notag
  \\
  &= - 2 e \bar{\epsilon} (D_z R^3_{i_1 \cdots i_n ij}{}^z) \gamma^{i_1 \cdots i_n} \psi^{ij}
  - 2 e \bar{\epsilon} R^3_{i_1 \cdots i_n ij}{}^z \gamma^{i_1 \cdots i_n} D_z \psi^{ij}
  \\
  &\quad\,- \tfrac{1}{2} C_n e R^3_{i_1 \cdots i_n ijz}R_{xy}{}^{ij} \bar{\epsilon} 
  \gamma^{xy} \gamma^{i_1 \cdots i_n} \psi_z. \notag
\end{alignat}
These variations are quite systematic and it is practical to calculate them 
with the aid of the computer programming\footnote{We partially made use of the mathematica package 
GAMMA by U. Gran\cite{Gr}.}. The result is expressed as follows.
\begin{alignat}{3}
  &\delta \mc{L}[eR^3\bar{\psi}\psi_{(2)}] = \notag \\
  &+f^1_{1} (-2 V^1_{8} - 4 V^2_{14} - 4 V^2_{15} - V^2_{25}) \notag \\ 
  &+f^1_{2} (-2 V^1_{9} - 8 V^2_{7} + 8 V^2_{9} + 8 V^2_{10} - 8 V^2_{18} + 
          8 V^2_{19} - 8 V^2_{21} + 8 V^2_{22} + 4 V^2_{23} - 4 V^2_{24}) \notag \\ 
  &+f^1_{3} (-2 V^1_{10} + 2 V^2_{9} + 2 V^2_{11} - 2 V^2_{18} + 2 V^2_{22} + 
          V^2_{23}) \notag \\ 
  &+f^1_{4} (V^1_{11} + V^1_{29}/2 + 8 V^2_{1} - 2 V^3_{4}) \notag \\ 
  &+f^1_{5} (V^1_{14} - V^1_{31}/2 + V^1_{32}/2 - 4 V^2_{4} + 4 V^2_{16} - 
          2 V^3_{5}) \notag \\ 
  &+f^1_{6} (-V^1_{13}/2 + V^1_{14} + V^1_{32}/2 - 4 V^2_{3} + 
          4 V^2_{17} - 2 V^3_{6}) \notag \\ 
  &+f^1_{7} (V^1_{13} + V^1_{35}/2 - 2 V^2_{2} + 4 V^2_{13} - 2 V^3_{7}) \notag \\ 
  &+f^1_{8} (-V^1_{12} + V^1_{36}/2 - 2 V^2_{1} - 2 V^2_{25} - 2 V^3_{8}) \notag \\ 
  &+f^1_{9} (V^1_{15} + V^1_{30}/2 - 4 V^2_{2} - 8 V^2_{23} + 8 V^2_{24} - 
          2 V^3_{9}) \notag \\ 
  &+f^1_{10} (V^1_{15}/2 + V^1_{37}/2 + 8 V^2_{3} - 8 V^2_{4} + 8 V^2_{18} - 
          8 V^2_{19} - 2 V^3_{10}) \notag \\ 
  &+f^1_{11} (2 V^1_{21} - 2 V^1_{22} + V^1_{33} + 8 V^2_{7} - 8 V^2_{8} - 
          8 V^2_{9} - 8 V^2_{10} + 8 V^2_{11} - 2 V^3_{11}) \notag \\ 
  &+f^1_{12} (V^1_{21} - V^1_{22} - V^1_{47}/2 + 4 V^2_{5} - 4 V^2_{6} - 
          4 V^2_{10} - 2 V^3_{12}) \notag \\ 
  &+f^1_{13} (2 V^1_{17} + 2 V^1_{18} - 2 V^1_{19} - 2 V^1_{20} - V^1_{48}/2 + 
          4 V^2_{14} - 4 V^2_{15} - 4 V^2_{16} + 4 V^2_{17} - 4 V^2_{18} \notag \\
         &\qquad\;\;+ 4 V^2_{19} - 2 V^3_{13}) \label{vf1} \\ 
  &+f^1_{14} (-4 V^1_{17} - 4 V^1_{18} + 4 V^1_{19} + 4 V^1_{20} + 
          V^1_{39}/2 - 4 V^2_{12} + 4 V^2_{13} - 4 V^2_{23} + 4 V^2_{24} - 
          2 V^3_{14}) \notag \\ 
  &+f^1_{15} (2 V^1_{23} - 2 V^1_{24} + V^1_{34} + 4 V^2_{7} - 4 V^2_{8} - 
          4 V^2_{9} - 2 V^3_{15}) \notag \\ 
  &+f^1_{16} (V^1_{23} - V^1_{24} - V^1_{43}/2 + 4 V^2_{5} - 2 V^2_{6} - 2 V^2_{7} + 
          2 V^2_{8} - 2 V^2_{11} - 2 V^3_{16}) \notag \\ 
  &+f^1_{17} (V^1_{17} - V^1_{20} - V^1_{44}/2 + 2 V^2_{14} - 2 V^2_{15} - 
          2 V^2_{20} + 2 V^2_{21} - 2 V^3_{17}) \notag \\ 
  &+f^1_{18} (-2 V^1_{17} + 2 V^1_{20} + V^1_{40}/2 - 2 V^2_{12} - 
          2 V^2_{23} - 2 V^3_{18}) \notag \\ 
  &+f^1_{19} (V^1_{16} + V^1_{38}/2 - 2 V^2_{4} - 2 V^2_{19} - 2 V^3_{19}) \notag \\ 
  &+f^1_{20} (V^1_{23} - V^1_{45}/2 + 2 V^2_{5} - 2 V^2_{6} - 2 V^2_{7} - 
          2 V^3_{20}) \notag \\ 
  &+f^1_{21} (-V^1_{20} - V^1_{46}/2 - 2 V^2_{15} + 2 V^2_{21} - 
          2 V^3_{21}) \notag \\ 
  &+f^1_{22} (V^1_{21} - V^1_{41}/2 - 4 V^2_{7} + 4 V^2_{9} - 2 V^3_{22}) \notag \\ 
  &+f^1_{23} (2 V^1_{18} - 2 V^1_{20} - V^1_{42}/2 - 4 V^2_{15} + 4 V^2_{17} + 
          4 V^2_{21} - 4 V^2_{22} - 2 V^3_{23}) \notag \\ 
  &+f^1_{24} (2 V^1_{1} - 6 V^1_{11} + V^1_{29} - V^1_{69} - 8 V^2_{26} - 
          4 V^3_{4}) \notag \\ 
  &+f^1_{25} (2 V^1_{2} - 6 V^1_{14} - V^1_{31} + V^1_{32} - V^1_{71} - 4 V^2_{34} - 
          4 V^2_{44} - 4 V^3_{5}) \notag 
\end{alignat}
\begin{alignat}{3}
  &+f^1_{26} (2 V^1_{2} - 3 V^1_{12} - 3 V^1_{13} + V^1_{35}/2 - V^1_{36}/2 - 
          V^1_{74} - 4 V^2_{27} + V^2_{32} + V^2_{74} - 2 V^3_{7} + 2 V^3_{8}) \notag \\
  &+f^1_{27} (2 V^1_{3} - 6 V^1_{15} + V^1_{30} - V^1_{70} - 4 V^2_{29} - 
          8 V^2_{73} - 4 V^3_{9}) \notag \\
  &+f^1_{28} (V^1_{3} - 3 V^1_{15} + V^1_{37} - V^1_{75} + 8 V^2_{39} - 16 V^2_{40} + 
          8 V^2_{41} + 8 V^2_{53} - 8 V^2_{54} - 4 V^3_{10}) \notag \\
  &+f^1_{29} (2 V^1_{6} - 12 V^1_{21} + 12 V^1_{22} + 2 V^1_{33} - V^1_{72} - 
          8 V^2_{48} - 4 V^3_{11}) \notag \\ 
  &+f^1_{30} (V^1_{6} + 6 V^1_{17} + 6 V^1_{18} - 6 V^1_{19} - 6 V^1_{20} - 
          3 V^1_{21} + 3 V^1_{22} - V^1_{47}/2 + V^1_{48}/2 + V^1_{81} + 2 V^2_{37} \notag \\
          &\qquad\;\;- 2 V^2_{38} + 4 V^2_{61} - 4 V^2_{62} - 2 V^2_{65} + 
          2 V^2_{66} + 2 V^2_{67} - 2 V^2_{68} - 2 V^3_{12} + 2 V^3_{13}) \notag \\
  &+f^1_{31} (2 V^1_{6} + 24 V^1_{17} + 24 V^1_{18} - 24 V^1_{19} - 24 V^1_{20} + 
          V^1_{39} - V^1_{77} + 4 V^2_{30} + 8 V^2_{69} - 8 V^2_{70} - 4 V^3_{14}) \notag \\ 
  &+f^1_{32} (2 V^1_{5} - 12 V^1_{23} + 12 V^1_{24} + 2 V^1_{34} - V^1_{73} - 
          4 V^2_{47} - 4 V^2_{48} + 4 V^2_{49} + 4 V^2_{50} - 4 V^3_{15}) \notag \\ 
  &+f^1_{33} (V^1_{5} + V^1_{17} + 2 V^1_{18} - 3 V^1_{20} - 3 V^1_{23} + 
          3 V^1_{24} - V^1_{43}/2 + V^1_{44}/2 + V^1_{79} - V^2_{35} + V^2_{36} + 
          V^2_{37} \notag \\
          &\qquad\;\;- V^2_{38} + 2 V^2_{55} - 2 V^2_{56} + 2 V^2_{59} - 2 V^2_{60} + 
          V^2_{67} - V^2_{68} - 2 V^3_{16} + 2 V^3_{17}) \notag \\ 
  &+f^1_{34} (2 V^1_{5} + 4 V^1_{17} + 8 V^1_{18} - 12 V^1_{20} + V^1_{40} - 
          V^1_{78} + 2 V^2_{30} - 2 V^2_{31} - 4 V^2_{70} - 4 V^3_{18}) \notag \\ 
  &+f^1_{35} (2 V^1_{4} - 6 V^1_{16} + V^1_{38} - V^1_{76} - 4 V^2_{40} - 
          2 V^2_{54} - 4 V^3_{19}) \notag \\ 
  &+f^1_{36} (2 V^1_{7} - 3 V^1_{20} - 3 V^1_{23} - V^1_{45}/2 + V^1_{46}/2 + 
          V^1_{80} - V^2_{35} + V^2_{37} - 2 V^2_{56} - 2 V^2_{60} - V^2_{68} \notag \\
          &\qquad\;\;- 2 V^3_{20} + 2 V^3_{21}) \notag \\ 
  &+f^1_{37} (2 V^1_{5} + 4 V^1_{17} + 2 V^1_{18} - 6 V^1_{20} - 3 V^1_{21} - 
          V^1_{41}/2 + V^1_{42}/2 + V^1_{82} + 2 V^2_{37} - 4 V^2_{62} + 
          2 V^2_{66} \notag \\
          &\qquad\;\;- 2 V^2_{68} - 2 V^3_{22} + 2 V^3_{23}) \notag \\ 
  &+f^1_{38} (2 V^1_{12} + 2 V^1_{13} - 4 V^1_{14} + V^1_{86} - 4 V^2_{42} - 
          4 V^2_{45} - V^2_{74}) \notag \\ 
  &+f^1_{39} (V^1_{87} - 8 V^2_{51} + 8 V^2_{52} + 8 V^2_{53} - 8 V^2_{54} + 
          4 V^2_{73}) \notag \\ 
  &+f^1_{40} (8 V^1_{17} + 8 V^1_{18} - 8 V^1_{19} - 8 V^1_{20} + 4 V^1_{21} - 
          4 V^1_{22} + V^1_{88} + 4 V^2_{49} - 4 V^2_{65} + 4 V^2_{66} + 
          4 V^2_{67} \notag \\
          &\qquad\;\;- 4 V^2_{68} + 2 V^2_{72}) \notag \\ 
  &+f^1_{41} (-8 V^1_{17} - 8 V^1_{18} + 8 V^1_{19} + 8 V^1_{20} - 
          4 V^1_{21} + 4 V^1_{22} + 2 V^1_{93} + 8 V^2_{63} - 8 V^2_{64} - 
          4 V^2_{69} + 4 V^2_{70}) \notag \\ 
  &+f^1_{42} (4 V^1_{18} - 4 V^1_{20} + 4 V^1_{23} - 4 V^1_{24} + V^1_{89} + 
          2 V^2_{46} - 2 V^2_{50} + 2 V^2_{67} - 2 V^2_{68} - V^2_{71} + V^2_{72}) \notag \\ 
  &+f^1_{43} (-4 V^1_{18} + 4 V^1_{20} - 4 V^1_{23} + 4 V^1_{24} + 
          2 V^1_{94} + 4 V^2_{55} - 4 V^2_{56} + 4 V^2_{57} - 4 V^2_{58} - 
          4 V^2_{61} + 4 V^2_{62} \notag \\
          &\qquad\;\;+ 4 V^2_{63} - 4 V^2_{64} + 2 V^2_{70}) \notag \\ 
  &+f^1_{44} (2 V^1_{8} - V^1_{12} - V^1_{26} + 2 V^1_{50} - V^1_{86}/2 + 
          4 V^2_{33} + 4 V^2_{43} - 2 V^3_{44}) \notag \\ 
  &+f^1_{45} (V^1_{11} - 2 V^1_{12} - V^1_{36} - V^1_{105}/2 - 2 V^2_{26} - 
          2 V^2_{74} - 2 V^3_{45}) \notag \\ 
  &+f^1_{46} (-2 V^1_{9} + V^1_{15} + V^1_{27} - 2 V^1_{51} - V^1_{87}/2 - 
          2 V^2_{29} - 4 V^2_{72} - 2 V^3_{46}) \notag \\ 
  &+f^1_{47} (V^1_{9} + 4 V^1_{17} + 4 V^1_{18} - 4 V^1_{19} - 4 V^1_{20} - 
          V^1_{27}/2 - 2 V^1_{62} + V^1_{88}/2 + 4 V^2_{48} - 4 V^2_{49} - 
          4 V^2_{51} \notag \\
          &\qquad\;\;+ 4 V^2_{52} - 2 V^3_{47}) \notag \\ 
  &+f^1_{48} (-4 V^1_{18} + 4 V^1_{20} - V^1_{42} + 8 V^1_{54} - 4 V^2_{42} + 
          4 V^2_{43} + 4 V^2_{66} - 4 V^2_{68} - 2 V^3_{48}) \notag \\
  &+f^1_{49} (V^1_{13} - V^1_{15}/2 + V^1_{37}/2 - 2 V^1_{65} - V^1_{108}/2 - 
          4 V^2_{44} + 4 V^2_{45} - 4 V^2_{53} + 4 V^2_{54} - 2 V^3_{49}) \notag
\end{alignat}
\begin{alignat}{3}
  &+f^1_{50} (V^1_{13} + 4 V^1_{17} - 4 V^1_{19} - V^1_{42} + V^1_{48} + 2 V^1_{65} - 
          V^1_{108}/2 + 4 V^2_{33} - 4 V^2_{34} + 4 V^2_{65} - 4 V^2_{67} - 
          2 V^3_{50}) \notag \\ 
  &+f^1_{51} (4 V^1_{17} - 4 V^1_{20} + V^1_{40} + 2 V^1_{55} - 2 V^1_{56} + 
          V^1_{90}/2 - V^1_{91}/2 - 4 V^2_{37} - 4 V^2_{52} - 2 V^3_{51}) \notag \\
  &+f^1_{52} (-V^1_{15}/4 - 2 V^1_{18} + 2 V^1_{19} + V^1_{51} + V^1_{55} - 
          V^1_{56} - V^1_{61} + V^1_{62} + V^1_{64} + V^1_{99}/2 - V^1_{100}/2 - 
          4 V^2_{39} \notag \\
          &\qquad\;\;+ 4 V^2_{40} + 4 V^2_{61} - 4 V^2_{63} - 2 V^3_{52}) \notag \\
  &+f^1_{53} (-V^1_{9} - 4 V^1_{17} - 4 V^1_{18} + 4 V^1_{19} + 4 V^1_{20} - 
          V^1_{39}/2 + 2 V^1_{56} + V^1_{91}/2 + 4 V^2_{37} - 4 V^2_{38} - 
          4 V^2_{51} \notag \\
          &\qquad\;\;+ 4 V^2_{52} - 2 V^3_{53}) \notag \\ 
  &+f^1_{54} (-V^1_{9}/2 + V^1_{15}/2 + 2 V^1_{17} + 2 V^1_{18} - 
          2 V^1_{19} - 2 V^1_{20} - V^1_{51}/2 + V^1_{56} - V^1_{62} - V^1_{63}/2 + 
          V^1_{100}/2 \notag \\
          &\qquad\;\;+ 4 V^2_{39} - 8 V^2_{40} + 4 V^2_{41} - 4 V^2_{61} + 
          4 V^2_{62} + 4 V^2_{63} - 4 V^2_{64} - 2 V^3_{54}) \notag \\ 
  &+f^1_{55} (V^1_{13} + 4 V^1_{17} + 4 V^1_{18} - 4 V^1_{19} - 4 V^1_{20} + 
          2 V^1_{49} - 2 V^1_{50} - 4 V^1_{53} + V^1_{63} - 2 V^1_{95} + 
          2 V^1_{96} - 4 V^2_{27} \notag \\
          &\qquad\;\;+ 4 V^2_{28} - 4 V^2_{69} + 4 V^2_{70} - 
          2 V^3_{55}) \notag \\ 
  &+f^1_{56} (-2 V^1_{13} + V^1_{15} + V^1_{35} + 2 V^1_{63} - V^1_{106} - 
          2 V^2_{32} - 4 V^2_{73} - 2 V^3_{56}) \notag \\ 
  &+f^1_{57} (V^1_{15} + 8 V^1_{17} + 8 V^1_{18} - 8 V^1_{19} - 8 V^1_{20} + 
          V^1_{39} - 2 V^1_{63} - V^1_{106} - 2 V^2_{29} - 4 V^2_{30} - 
          2 V^3_{57}) \notag \\ 
  &+f^1_{58} (V^1_{9}/2 - 2 V^1_{10} + 2 V^1_{17} - 2 V^1_{19} - V^1_{27}/4 + 
          V^1_{28} + 2 V^1_{58} - V^1_{62} + V^1_{88}/4 - V^1_{89}/2 + 2 V^2_{50} \notag \\
          &\qquad\;\;- 2 V^2_{51} - 2 V^3_{58}) \notag \\ 
  &+f^1_{59} (V^1_{8} - 2 V^1_{18} - V^1_{42}/2 + V^1_{46} + 2 V^1_{54} - 
          V^1_{92}/2 - 2 V^2_{42} + 2 V^2_{66} - 2 V^3_{59}) \notag \\ 
  &+f^1_{60} (V^1_{14} - 2 V^1_{16} + V^1_{38} - 2 V^1_{66} - V^1_{107}/2 - 
          2 V^2_{44} + 2 V^2_{54} - 2 V^3_{60}) \notag \\ 
  &+f^1_{61} (V^1_{14} - 2 V^1_{19} - V^1_{42}/2 - V^1_{44} + V^1_{46} + V^1_{48}/2 + 
          2 V^1_{66} - V^1_{107}/2 - 2 V^2_{34} + 2 V^2_{65} - 2 V^3_{61}) \notag \\ 
  &+f^1_{62} (V^1_{15}/8 - V^1_{16} + V^1_{18} - V^1_{51}/2 - 2 V^1_{52} + V^1_{57} - 
          V^1_{58} - V^1_{59} + V^1_{60} - V^1_{64}/2 + V^1_{66} + V^1_{97}/2 \notag \\
          &\qquad\;\;- V^1_{98}/2 - V^1_{99}/4 + V^1_{100}/4 + 2 V^2_{39} - 2 V^2_{55} - 
          2 V^3_{62}) \notag \\ 
  &+f^1_{63} (V^1_{9}/4 - V^1_{10} - V^1_{15}/8 - V^1_{18} + V^1_{20} + V^1_{51}/4 + 
          V^1_{52} + V^1_{58} - V^1_{60} + V^1_{63}/4 - V^1_{65}/2 \notag \\
          &\qquad\;\;+ V^1_{98}/2 - V^1_{100}/4 - 2 V^2_{39} + 2 V^2_{40} + 2 V^2_{55} - 2 V^2_{56} - 
          2 V^3_{63}) \notag \\ 
  &+f^1_{64} (-V^1_{8} + V^1_{14} + 2 V^1_{18} - 2 V^1_{19} + V^1_{49} - 
          V^1_{53} + V^1_{64} - V^1_{95} - 2 V^2_{27} - 2 V^2_{69} - 2 V^3_{64}) \notag \\ 
  &+f^1_{65} (-2 V^1_{21} - V^1_{41} + 2 V^1_{55} - 2 V^1_{56} - V^1_{90}/2 + 
          V^1_{91}/2 - 2 V^2_{30} + 2 V^2_{31} - 2 V^2_{71} + 2 V^2_{72} - 
          2 V^3_{65}) \notag \\ 
  &+f^1_{66} (-2 V^1_{21} - V^1_{41}/2 + 2 V^1_{67} + 2 V^1_{68} + 
          V^1_{109}/2 + 2 V^2_{35} - 2 V^2_{36} - 2 V^2_{37} + 2 V^2_{38} - 
          2 V^2_{46} + 2 V^2_{47} \notag \\
          &\qquad\;\;+ 2 V^2_{48} - 2 V^2_{49} - 2 V^3_{66}) \notag \\ 
  &+f^1_{67} (V^1_{9} - 2 V^1_{21} + 2 V^1_{22} - V^1_{47} + 2 V^1_{55} - 
          V^1_{90}/2 - 2 V^2_{30} - 2 V^2_{71} - 2 V^3_{67}) \notag \\ 
  &+f^1_{68} (V^1_{9}/2 - V^1_{21}/2 + V^1_{22} + V^1_{55}/2 + V^1_{56}/2 - 
          V^1_{67} + V^1_{68} - V^1_{102} - 2 V^2_{57} + 2 V^2_{58} + 2 V^2_{59} \notag \\
          &\qquad\;\;- 2 V^2_{60} - 2 V^2_{61} + 2 V^2_{62} + 2 V^2_{63} - 2 V^2_{64} - 
          2 V^3_{68}) \notag \\
  &+f^1_{69} (V^1_{21} + V^1_{47}/2 - 2 V^1_{67} - V^1_{110}/2 + 2 V^2_{37} - 
          2 V^2_{38} + 2 V^2_{46} - 2 V^2_{47} - 2 V^3_{69}) \notag \\ 
  &+f^1_{70} (V^1_{22} + 2 V^1_{23} + V^1_{45} + 2 V^1_{67} - V^1_{110}/2 - 
          2 V^2_{35} + 2 V^2_{37} - 2 V^2_{47} + 2 V^2_{49} - 2 V^3_{70}) \notag \\
  &+f^1_{71} (V^1_{21}/2 - V^1_{22}/2 - V^1_{23} + V^1_{59} - V^1_{60} - V^1_{61}/2 + 
          V^1_{62}/2 - V^1_{67} + V^1_{103}/2 - V^1_{104}/2 \notag \\
          &\qquad\;\;- 2 V^2_{57} + 
          2 V^2_{60} + 2 V^2_{61} - 2 V^2_{64} - 2 V^3_{71}) \notag 
\end{alignat}
\begin{alignat}{3}
  &+f^1_{72} (V^1_{21} - V^1_{22} - 2 V^1_{24} + V^1_{43} - V^1_{45} + 2 V^1_{68} - 
          V^1_{109}/2 + V^1_{110}/2 + 2 V^2_{35} - 2 V^2_{49} - 2 V^3_{72}) \notag \\ 
  &+f^1_{73} (V^1_{9}/4 + V^1_{21}/2 - V^1_{24} + V^1_{55}/2 + V^1_{56}/2 - 
          V^1_{60} - V^1_{61}/2 + V^1_{68} - V^1_{102} + V^1_{103}/2 \notag \\
          &\qquad\;\;- 2 V^2_{57} + 
          2 V^2_{59} + 2 V^2_{63} - 2 V^2_{64} - 2 V^3_{73}) \notag \\
  &+f^1_{74} (-V^1_{9}/4 + V^1_{10} + V^1_{21}/2 - V^1_{22}/2 - 
          2 V^1_{23} + 2 V^1_{24} + V^1_{57} + V^1_{58} - V^1_{59} - V^1_{62}/2 + 
          2 V^1_{67} \notag \\
          &\qquad\;\;- V^1_{101}/2 + V^1_{104}/2 + 2 V^2_{57} + 2 V^2_{64} - 
          2 V^3_{74}) \notag \\
  &+f^1_{75} (-2 V^1_{90} - 8 V^1_{96} + 4 V^1_{99} - 4 V^2_{79} - 
          4 V^2_{87} + V^2_{88}) \notag \\ 
  &+f^1_{76} (-2 V^1_{92} - 4 V^1_{101} + 8 V^1_{104} + V^2_{81} - 
          2 V^2_{82} - 2 V^2_{83} - 2 V^2_{84} - 2 V^2_{85}) \notag \\ 
  &+f^1_{77} (-8 V^1_{17} + 8 V^1_{18} + 2 V^1_{40} - 2 V^1_{42} + 
          16 V^1_{54} + 4 V^1_{55} - 4 V^1_{56} - 4 V^1_{78} - 4 V^1_{82} - 
          3 V^1_{90} + 3 V^1_{91} \notag \\
          &\qquad\;\;- 4 V^2_{78} - 4 V^3_{48} - 4 V^3_{51}) \notag \\ 
  &+f^1_{78} (-V^1_{35} + V^1_{39} - 4 V^1_{49} + 4 V^1_{50} + 8 V^1_{53} - 
          6 V^1_{63} + 2 V^1_{74} - 2 V^1_{77} + 4 V^1_{83} - 12 V^1_{95} + 
          12 V^1_{96} \notag \\
          &\qquad\;\;- 2 V^2_{75} - V^2_{88} + 4 V^3_{55} + 2 V^3_{56} - 
          2 V^3_{57}) \notag \\ 
  &+f^1_{79} (-4 V^1_{17} + 4 V^1_{18} - V^1_{37}/2 - V^1_{42} + V^1_{48} + 
          2 V^1_{51} + 2 V^1_{55} - 2 V^1_{56} - 2 V^1_{61} + 2 V^1_{62} + 
          2 V^1_{64} \notag \\
          &\qquad\;\;+ 4 V^1_{65} + V^1_{75} + 2 V^1_{81} - 2 V^1_{82} - 
          4 V^1_{84} - 3 V^1_{99} + 3 V^1_{100} + 2 V^2_{76} - 2 V^2_{77} + 
          2 V^2_{87} \notag \\
          &\qquad\;\;+ 2 V^3_{49} - 2 V^3_{50} - 4 V^3_{52}) \notag \\ 
  &+f^1_{80} (-V^1_{38} - V^1_{42}/2 - V^1_{44} + V^1_{46} + V^1_{48}/2 - 
          4 V^1_{52} + V^1_{55} - V^1_{56} + 2 V^1_{57} - 2 V^1_{58} - 2 V^1_{59} \notag \\
          &\qquad\;\;+ 2 V^1_{60} - V^1_{61} + V^1_{62} + 6 V^1_{66} + 2 V^1_{76} - 2 V^1_{79} + 
          2 V^1_{80} + V^1_{81} - V^1_{82} - 4 V^1_{85} - 3 V^1_{97} \notag \\
          &\qquad\;\;+ 3 V^1_{98} - 
          V^2_{77} - V^2_{86} + V^2_{87} - 2 V^3_{52} + 2 V^3_{60} - 2 V^3_{61} - 
          4 V^3_{62}) \notag \\ 
  &+f^1_{81} (V^1_{45} - V^1_{47}/2 + 2 V^1_{59} - 2 V^1_{60} - V^1_{61} + V^1_{62} + 
          2 V^1_{67} - 2 V^1_{80} + V^1_{81} - 3 V^1_{103} + 3 V^1_{104} \notag \\
          &\qquad\;\;- V^2_{82} - 
          V^2_{84} + 2 V^3_{69} - 2 V^3_{70} - 4 V^3_{71}) \notag \\ 
  &+f^1_{82} (V^1_{30}/2 + V^1_{36} - 4 V^1_{53} + V^1_{87}/2 - 2 V^1_{93} - 
          4 V^1_{96} + V^1_{111} - 4 V^2_{76} + 4 V^2_{77} - 4 V^2_{78} + 
          4 V^2_{79} \notag \\
          &\qquad\;\;- 2 V^3_{82}) \notag \\ 
  &+f^1_{83} (-2 V^1_{30} - 4 V^1_{36} + 16 V^1_{53} - 16 V^1_{95} + 
          16 V^1_{96} + V^1_{105} - V^1_{113}/2 + 8 V^2_{75} - 2 V^3_{83}) \notag \\ 
  &+f^1_{84} (-V^1_{29} - V^1_{30} + 2 V^1_{35} + 2 V^1_{39} + 8 V^1_{49} - 
          8 V^1_{50} + 8 V^1_{95} - 8 V^1_{96} + 2 V^1_{106} - V^1_{113}/2 - 
          2 V^2_{80} \notag \\
          &\qquad\;\;- 2 V^2_{88} - 2 V^3_{84}) \notag \\
  &+f^1_{85} (2 V^1_{31} - 2 V^1_{32} + 2 V^1_{41} + 4 V^1_{43} - 4 V^1_{45} - 
          2 V^1_{47} + 8 V^1_{52} + 4 V^1_{97} - 4 V^1_{98} + 2 V^1_{109} - 
          2 V^1_{110} \notag \\
          &\qquad\;\;- V^1_{114} - 4 V^2_{86} + 4 V^2_{87} - 2 V^3_{85}) \notag \\ 
  &+f^1_{86} (2 V^1_{26} - V^1_{27} + 2 V^1_{32} - V^1_{37} - 4 V^1_{55} + 
          4 V^1_{61} - V^1_{86} + 2 V^1_{93} + 2 V^1_{99} \notag \\
          &\qquad\;\;+ V^1_{111} + 2 V^2_{75} - 
          2 V^2_{81} - 2 V^3_{86}) \notag \\ 
  &+f^1_{87} (V^1_{32} + V^1_{33} - V^1_{37}/2 + V^1_{42} - V^1_{48} - 2 V^1_{55} - 
          4 V^1_{57} + 4 V^1_{59} + 2 V^1_{61} + 2 V^1_{97} - 2 V^1_{101} \notag \\
          &\qquad\;\;- 4 V^1_{102} + 2 V^1_{103} + 2 V^1_{104} - V^1_{108}/2 + V^1_{114}/2 + 
          2 V^2_{76} - 2 V^2_{77} + 2 V^2_{82} - 2 V^2_{83} - 2 V^3_{87}) \notag \\ 
  &+f^1_{88} (V^1_{31} - V^1_{32} - 2 V^1_{33} + 2 V^1_{38} - V^1_{42} - 2 V^1_{44} + 
          2 V^1_{46} + V^1_{48} + 2 V^1_{55} - 2 V^1_{56} + 4 V^1_{57} - 4 V^1_{58} \notag \\
          &\qquad\;\;- 4 V^1_{59} + 4 V^1_{60} - 2 V^1_{61} + 2 V^1_{62} - 
          2 V^1_{97} + 2 V^1_{98} + V^1_{107} - V^1_{114} + 2 V^2_{77} + 2 V^2_{83} - 
          2 V^3_{88}) \notag
\end{alignat}
\begin{alignat}{3}
  &+f^1_{89} (-V^1_{26} + V^1_{27}/2 + V^1_{33} + V^1_{42} - V^1_{48} - 
          4 V^1_{57} + 4 V^1_{59} + V^1_{88}/2 - 2 V^1_{94} - 2 V^1_{101} + 
          2 V^1_{104} \notag \\
          &\qquad\;\;+ V^1_{112}/2 - 2 V^2_{78} + 2 V^2_{79} - 2 V^2_{84} + 
          2 V^2_{85} - 2 V^3_{89}) \notag \\
  &+f^1_{90} (2 V^1_{28} + 2 V^1_{34} + V^1_{42} - 4 V^1_{57} + V^1_{89} - 
          4 V^1_{94} - 2 V^1_{101} - 2 V^2_{78} + 2 V^2_{85} - 2 V^3_{90}) \notag \\
  &+f^1_{91} (-4 V^1_{29} - 8 V^1_{30} + 8 V^1_{35} - 8 V^1_{36} + 
          8 V^1_{39} + 32 V^1_{49} - 32 V^1_{50} + 32 V^1_{53} + 2 V^1_{69} + 
          4 V^1_{70} \notag \\
          &\qquad\;\;- 16 V^1_{74} - 8 V^1_{77} - 6 V^1_{105} - 24 V^1_{106} - 
          3 V^1_{113} - V^1_{115} - 4 V^3_{83} - 8 V^3_{84}) \notag \\ 
  &+f^1_{92} (8 V^1_{31} - 8 V^1_{32} - 8 V^1_{33} + 8 V^1_{38} + 4 V^1_{41} - 
          4 V^1_{42} + 8 V^1_{43} - 8 V^1_{44} - 8 V^1_{45} + 8 V^1_{46} - 
          4 V^1_{47} \notag \\
          &\qquad\;\;+ 4 V^1_{48} + 16 V^1_{52} + 8 V^1_{55} - 8 V^1_{56} + 
          16 V^1_{57} - 16 V^1_{58} - 16 V^1_{59} + 16 V^1_{60} - 8 V^1_{61} + 
          8 V^1_{62} \notag \\
          &\qquad\;\;+ 4 V^1_{71} + 2 V^1_{72} - 8 V^1_{76} - 16 V^1_{79} + 
          16 V^1_{80} + 8 V^1_{81} - 8 V^1_{82} - 12 V^1_{107} - 12 V^1_{109} + 
          12 V^1_{110} \notag \\
          &\qquad\;\;- 6 V^1_{114} - V^1_{116} - 4 V^3_{85} - 8 V^3_{88}). \notag
\end{alignat}
The variations which depend on the Ricci tensor, the scalar curvature and 
the field equations of the Majorana gravitino are neglected.
As explained in the sec.~\ref{sec:ov}, these terms are cancelled by modifying the 
supersymmetric transformation rules of the vielbein and the Majorana gravitino.

\subsubsection{$ \delta \mc{L}[eR^2\bar{\psi}_{(2)}D\psi_{(2)}]$ terms}

Finally let us evaluate the variations of $\mc{L}[eR^2\bar{\psi}_{(2)}D\psi_{(2)}]$.
The transformations of the terms in this ansatz are expressed as
\begin{alignat}{3}
  \delta_1 (e R^2_{ijkl i_1\cdots i_n z} \bar{\psi}^{kl} \gamma^{i_1\cdots i_n} D_z \psi^{ij}) &=
  - \tfrac{1}{2} e R^2_{ijkl i_1\cdots i_n z} R_{xy}{}^{kl} 
  \bar{\epsilon} \gamma^{xy} \gamma^{i_1\cdots i_n} D_z \psi^{ij} \notag
  \\
  &\quad
  + \tfrac{1}{2} (-1)^{\frac{n(n+1)}{2}} e R^2_{ijkl i_1\cdots i_n z} R_{xy}{}^{ij} 
  \bar{\epsilon} \gamma^{xy} \gamma^{i_1\cdots i_n} D_z \psi^{kl} 
  \\
  &\quad
  + \tfrac{1}{2} (-1)^{\frac{n(n+1)}{2}} e (D_z R^2_{ijkl i_1\cdots i_n z}) R_{xy}{}^{ij} 
  \bar{\epsilon} \gamma^{xy} \gamma^{i_1\cdots i_n} \psi^{kl}. \notag
\end{alignat}
And it is practical to obtain all variations by employing the computer programming.
The result becomes as follows.
\begin{alignat}{3}
  &\delta \mc{L}[eR^2\bar{\psi}_{(2)}D\psi_{(2)}] = \notag \\
  &+f^2_{1} (-2 V^2_{7} + 2 V^2_{9} - V^2_{52} + V^3_{18} - V^3_{22} - 
          V^3_{51}/2 + V^3_{65}/2) \notag \\ 
  &+f^2_{2} (V^2_{23} + V^2_{71}/2 - V^2_{72}/2 + V^3_{18} - V^3_{22} - V^3_{51}/2 + 
          V^3_{65}/2) \notag \\ 
  &+f^2_{3} (-2 V^2_{10} - V^2_{51} - V^3_{12} - V^3_{14}/2 + V^3_{18} - 
          V^3_{51}/2 - V^3_{53}/2 + V^3_{67}/2) \notag \\ 
  &+f^2_{4} (V^2_{24} + V^2_{71}/2 - V^3_{12} - V^3_{14}/2 + V^3_{18} - V^3_{51}/2 - 
          V^3_{53}/2 + V^3_{67}/2) \notag \\ 
  &+f^2_{5} (-V^2_{7} + V^2_{8} - V^2_{11} + V^2_{46}/2 - V^2_{47}/2 - 
          V^2_{48}/2 + V^2_{49}/2 \notag \\
          &\qquad\;- V^3_{16} - V^3_{22}/2 + V^3_{66}/2 - 
          V^3_{70}/2 - V^3_{72}/2) \label{vf2} \\ 
  &+f^2_{6} (-V^2_{10} - V^2_{47}/2 + V^2_{49}/2 - V^3_{12}/2 - V^3_{20} - 
          V^3_{69}/2 - V^3_{70}/2) \notag \\ 
  &+f^2_{7} (V^2_{23}/2 - V^2_{24}/2 - V^2_{73}/4 + V^3_{7}/4 + V^3_{14}/4 - 
          V^3_{56}/8 - V^3_{57}/8) \notag \\ 
  &+f^2_{8} (-V^2_{15} - V^2_{42}/2 - 2 V^3_{21} + V^3_{23}/2 + V^3_{48}/4 - 
          V^3_{59}) \notag \\ 
  &+f^2_{9} (-V^2_{17} - V^2_{43}/2 - 2 V^3_{21} + V^3_{23}/2 + V^3_{48}/4 - 
          V^3_{59}) \notag 
\end{alignat}
\begin{alignat}{3}
  &+f^2_{10} (V^2_{4} - V^2_{34}/2 - V^3_{13}/2 + V^3_{17} + V^3_{19} - V^3_{21} + 
          V^3_{23}/2 - V^3_{60}/2 - V^3_{61}/2) \notag \\
  &+f^2_{11} (-V^2_{16} - V^2_{44}/2 - V^3_{13}/2 + V^3_{17} + V^3_{19} - 
          V^3_{21} + V^3_{23}/2 - V^3_{60}/2 - V^3_{61}/2) \notag \\
  &+f^2_{12} (V^2_{3} - V^2_{33}/2 - V^3_{10}/4 + V^3_{17} + V^3_{19} - V^3_{21} + 
          V^3_{49}/4 + V^3_{50}/4 - V^3_{60}/2 - V^3_{61}/2) \notag \\
  &+f^2_{13} (-V^2_{14} - V^2_{45}/2 - V^3_{10}/4 + V^3_{17} + V^3_{19} - 
          V^3_{21} + V^3_{49}/4 + V^3_{50}/4 - V^3_{60}/2 - V^3_{61}/2) \notag \\
  &+f^2_{14} (-V^2_{1} - V^2_{26}/2 - 2 V^3_{8} - V^3_{45}) \notag \\ 
  &+f^2_{15} (2 V^2_{17} - 2 V^2_{18} + 2 V^2_{19} - V^2_{43} - 2 V^2_{62} + 
          2 V^2_{64} - V^2_{78}/2 + V^2_{79}/2 - V^3_{6} + V^3_{8}/2 \notag \\
          &\qquad\;+ V^3_{9}/4 + 
          V^3_{10}/2 + V^3_{44}/2 + V^3_{46}/4 + V^3_{52} + V^3_{54} + V^3_{55}/2 - 
          V^3_{64} - V^3_{82}/4 + V^3_{86}/4) \notag \\
  &+f^2_{16} (-V^2_{23} + V^2_{24} - V^2_{25}/2 + V^2_{70} - V^2_{72}/2 - 
          V^2_{81}/4 + V^3_{6} - V^3_{8}/2 - V^3_{9}/4 - V^3_{10}/2 \notag \\
          &\qquad\;\;- V^3_{44}/2 - 
          V^3_{46}/4 - V^3_{52} - V^3_{54} - V^3_{55}/2 + V^3_{64} + V^3_{82}/4 - 
          V^3_{86}/4) \notag \\ 
  &+f^2_{17} (2 V^2_{17} - 2 V^2_{18} + 2 V^2_{19} + 2 V^2_{56} - 2 V^2_{62} + 
          2 V^2_{64} - V^2_{65} + V^2_{67} - V^2_{82}/2 + V^3_{5} - 2 V^3_{6} \notag \\
          &\qquad\;\;+ V^3_{10} - 2 V^3_{19} + V^3_{50} + V^3_{52} + 2 V^3_{54} - V^3_{61} + 
          2 V^3_{62} + 4 V^3_{63} + V^3_{87} + V^3_{88}/2) \notag \\ 
  &+f^2_{18} (-2 V^2_{23} + 2 V^2_{24} - V^2_{25} - 2 V^2_{69} + 2 V^2_{70} + 
          V^2_{74}/2 + V^2_{88}/4 + V^3_{4}/2 - V^3_{7} - V^3_{8} \notag \\
          &\qquad\;\;- V^3_{9}/2 + 
          V^3_{45}/2 - V^3_{55} - V^3_{57}/2 - V^3_{83}/4 + V^3_{84}/4) \notag \\ 
  &+f^2_{19} (V^2_{10} + V^2_{18} - V^2_{19} - V^2_{51}/2 - V^2_{62} + V^2_{64} + 
          V^2_{78}/4 - V^2_{79}/4 - V^3_{15}/2 + V^3_{23}/2 + V^3_{47}/4 \notag \\
          &\qquad\;\;- V^3_{58}/2 + V^3_{68} + V^3_{71} - V^3_{73} + V^3_{74} + V^3_{90}/4) \notag \\ 
  &+f^2_{20} (-V^2_{7} + V^2_{9} + V^2_{10} + V^2_{18} - V^2_{19} + V^2_{21} - 
          V^2_{22} - V^2_{51}/2 + V^2_{52}/2 + V^2_{64} \notag \\
          &\qquad\;\;- V^2_{79}/4 + V^3_{15}/2 - 
          V^3_{23}/2 - V^3_{47}/4 + V^3_{58}/2 - V^3_{68} - V^3_{71} + V^3_{73} - 
          V^3_{74} - V^3_{90}/4) \notag \\ 
  &+f^2_{21} (-2 V^2_{4} + 2 V^2_{6} - V^2_{36} + 2 V^2_{40} - V^2_{77}/2 - 
          V^3_{5} + V^3_{11} + V^3_{12} + V^3_{13} - 2 V^3_{16} \notag \\
          &\qquad\;\;- 2 V^3_{17} + 
          2 V^3_{20} + 2 V^3_{21} - V^3_{22} - V^3_{23} - V^3_{52} - V^3_{60} - 
          2 V^3_{62} - V^3_{66} - V^3_{69} \notag \\
          &\qquad\;\;+ V^3_{72} - V^3_{85}/2 + V^3_{88}/2) \notag \\ 
  &+f^2_{22} (2 V^2_{7} - 2 V^2_{8} - 2 V^2_{9} - 2 V^2_{10} + 2 V^2_{11} - 
          2 V^2_{18} + 2 V^2_{19} + 2 V^2_{20} - 2 V^2_{21} + 2 V^2_{22} - 
          V^2_{54} \notag \\
          &\qquad\;\;+ 2 V^2_{59} - 2 V^2_{63} + V^2_{86}/2 - V^2_{87}/2 + V^3_{5} - 
          V^3_{11} - V^3_{12} - V^3_{13} + 2 V^3_{16} + 2 V^3_{17} - 2 V^3_{20} \notag \\
          &\qquad\;\;- 2 V^3_{21} + V^3_{22} + V^3_{23} + V^3_{52} + V^3_{60} + 2 V^3_{62} + 
          V^3_{66} + V^3_{69} - V^3_{72} + V^3_{85}/2 - V^3_{88}/2) \notag \\ 
  &+f^2_{23} (-2 V^2_{3} + 2 V^2_{5} - V^2_{36} + V^2_{38} + 2 V^2_{41} - 
          V^2_{76}/2 - V^3_{6} + V^3_{11}/2 + V^3_{12} - 2 V^3_{16} - 2 V^3_{17} \notag \\
          &\qquad\;\;+ 2 V^3_{21} + V^3_{49}/2 - V^3_{52} - V^3_{54} - V^3_{60} - 2 V^3_{62} - 
          2 V^3_{63} - V^3_{66} - V^3_{70} + 2 V^3_{73} + 2 V^3_{74} \notag \\
          &\qquad\;\;- V^3_{85}/4 + 
          V^3_{87}/2 + V^3_{88}/2) \notag \\ 
  &+f^2_{24} (-2 V^2_{8} + 2 V^2_{11} + 2 V^2_{20} - V^2_{53} + 2 V^2_{59} + 
          V^2_{86}/2 + V^3_{6} - V^3_{11}/2 - V^3_{12} + 2 V^3_{16} + 2 V^3_{17} \notag \\
          &\qquad\;\;- 2 V^3_{21} - V^3_{49}/2 + V^3_{52} + V^3_{54} + V^3_{60} + 2 V^3_{62} + 
          2 V^3_{63} + V^3_{66} + V^3_{70} - 2 V^3_{73} - 2 V^3_{74} \notag \\
          &\qquad\;\;+ V^3_{85}/4 - 
          V^3_{87}/2 - V^3_{88}/2) \notag \\ 
  &+f^2_{25} (-V^2_{2}/2 + V^2_{12} - V^2_{13} + V^2_{27} - V^2_{28} + 
          V^2_{32}/4 + V^2_{80}/8). \notag
\end{alignat}
Now we obtained all the variations of the ansatz.

\subsection{Summary}

As this section is quite long, let us briefly summarize the results obtained so far.
The ansatz for the higher derivative effective action is given by the sum of
the eqs.~(\ref{b1}), (\ref{b2}), (\ref{f1}) and (\ref{f2}), 
\begin{alignat}{3}
  \mc{L}_1 = \mc{L}[eR^4] + \mc{L}[e\ep_{11}AR^4] + \mc{L}[eR^3\bar{\psi}\psi_{(2)}]
  + \mc{L}[eR^2\bar{\psi}_{(2)}D\psi_{(2)}] ,
\end{alignat}
which contains totally 132 terms.
The variations of the ansatz are expanded by the 264 bases of 
$V[eR^4\bar{\ep}\psi]$, $V[eR^2DR\bar{\ep}\psi_{(2)}]$ and $V[eR^3\bar{\ep}D\psi_{(2)}]$,
which are given by the eqs.~(\ref{v1}), (\ref{v2}) and (\ref{v3}) respectively.
The terms of $V[eR^4\bar{\ep}\psi]$ and $V[eR^3\bar{\ep}D\psi_{(2)}]$ 
are related by the 20 identities (\ref{id}).
The results of the variations of the ansatz are expanded by the 264 bases as 
the eqs.~(\ref{vb1}), (\ref{vb2}), (\ref{vf1}) and (\ref{vf2}). 
Notice that these variations do not contain terms which are proportional to the 
field equations. Thus the sum of them is just the variations $V$ defined in the eq.~(\ref{V}).
\begin{alignat}{3}
  V = \delta \mc{L}[eR^4] + \delta \mc{L}[e\ep_{11}AR^4]
  + \delta \mc{L}[eR^3\bar{\psi}\psi_{(2)}]
  + \delta \mc{L}[eR^2\bar{\psi}_{(2)}D\psi_{(2)}] + \sum_{n=1}^{20} i_n I_n.
\end{alignat}
The 20 identities are also included.

The requirement of the local supersymmetry insists that all coefficients of the bases
should vanish, which gives 264 simultaneous equations.
For this purpose, we arrange the variations as
\begin{alignat}{3}
  &V = \notag
  \\
  &+ (b^1_{1} + 2 f^1_{24}) V^1_{1} \notag 
  \\
  &+ (-b^1_{2}/2 + b^1_{3} + 2 f^1_{25} + 2 f^1_{26}) V^1_{2} \notag 
  \\
  &+ (-b^1_{4}/8 - b^1_{5}/4 + b^1_{6}/8 + 2 f^1_{27} + 
          f^1_{28}) V^1_{3} \notag 
  \\
  &+ (-b^1_{6} + 2 f^1_{35}) V^1_{4} \notag 
  \\
  &+ (b^1_{7} + b^1_{8} + b^1_{9}/2 + 2 f^1_{32} + f^1_{33} + 2 f^1_{34} + 
          2 f^1_{37}) V^1_{5} \notag 
  \\
  &+ (b^1_{10}/4 - b^1_{11}/2 - b^1_{12} + 2 f^1_{29} + f^1_{30} + 
          2 f^1_{31}) V^1_{6} \notag 
  \\
  &+ (-b^1_{9} - b^1_{13} + 2 f^1_{36}) V^1_{7} \label{var}
  \\
  &+ (-2 f^1_{1} + 2 f^1_{44} + f^1_{59} - f^1_{64} + 8 i_{7} + 
          4 i_{12} + 2 i_{13}) V^1_{8} \notag
  \\
  &+ (-2 f^1_{2} - 2 f^1_{46} + f^1_{47} - f^1_{53} - f^1_{54}/2 + 
          f^1_{58}/2 + f^1_{63}/4 + f^1_{67} + f^1_{68}/2 + f^1_{73}/4 \notag \\
          &\quad\;- f^1_{74}/4 + 2 i_{8} + 4 i_{9} + 2 i_{10} + 2 i_{11} - 
          2 i_{12} - i_{13} + 2 i_{15} + i_{16} + i_{17}) V^1_{9} \notag
  \\
  &+ (-2 f^1_{3} - 2 f^1_{58} - f^1_{63} + f^1_{74} + 8 i_{12} + 
          4 i_{13} - 12 i_{17}) V^1_{10} \notag
  \\
  &+ (-8 b^1_{1} + f^1_{4} - 6 f^1_{24} + f^1_{45} + 2 i_{7}) V^1_{11} \notag
  \\
  &+ (-4 b^1_{3} - f^1_{8} - 3 f^1_{26} + 2 f^1_{38} - f^1_{44} - 
          2 f^1_{45} + 2 i_{1} - 8 i_{7}) V^1_{12} \notag 
  \\
  &+ (-4 b^1_{3} - f^1_{6}/2 + f^1_{7} - 3 f^1_{26} + 2 f^1_{38} + 
          f^1_{49} + f^1_{50} + f^1_{55} - 2 f^1_{56} \notag \\
          &\quad\;+ 2 i_{1} + 2 i_{8} + 
          4 i_{9} + 2 i_{10} + 2 i_{11}) V^1_{13} \notag
\end{alignat}
\begin{alignat}{3}
  &+ (4 b^1_{2} + f^1_{5} + f^1_{6} - 6 f^1_{25} - 4 f^1_{38} + f^1_{60} + 
          f^1_{61} + f^1_{64} - 4 i_{1} - 4 i_{12} - 2 i_{13}) V^1_{14} \notag \\
  &+ (b^1_{4} + 2 b^1_{5} - b^1_{6} + f^1_{9} + f^1_{10}/2 - 6 f^1_{27} - 
          3 f^1_{28} + f^1_{46} - f^1_{49}/2 - f^1_{52}/4 + f^1_{54}/2 \notag \\
          &\quad\;+ f^1_{56} + 
          f^1_{57} + f^1_{62}/8 - f^1_{63}/8 - i_{8} - 2 i_{9} - i_{10} - 
          i_{11}) V^1_{15} \notag \\
  &+ (8 b^1_{6} + f^1_{19} - 6 f^1_{35} - 2 f^1_{60} - f^1_{62} + 
          8 i_{12} + 4 i_{13}) V^1_{16} \notag \\
  &+ (4 b^1_{8} + 2 b^1_{9} + 6 b^1_{10} - 14 b^1_{11} + 2 f^1_{13} - 
          4 f^1_{14} + f^1_{17} - 2 f^1_{18} + 6 f^1_{30} + 24 f^1_{31} + f^1_{33} \notag \\
          &\quad\;+ 4 f^1_{34} + 4 f^1_{37} + 8 f^1_{40} - 8 f^1_{41} + 4 f^1_{47} + 
          4 f^1_{50} + 4 f^1_{51} - 4 f^1_{53} + 2 f^1_{54} + 4 f^1_{55} + 
          8 f^1_{57} \notag \\
          &\quad\;+ 2 f^1_{58} - 8 f^1_{77} - 4 f^1_{79} + 4 i_{3} + 
          8 i_{4} + 2 i_{5} + 4 i_{6} + 8 i_{8} + 16 i_{9} - 
          8 i_{12}) V^1_{17} \notag \\
  &+ (4 b^1_{7} + 4 b^1_{8} + 2 b^1_{9} + 6 b^1_{10} - 14 b^1_{11} + 
          2 f^1_{13} - 4 f^1_{14} + 2 f^1_{23} + 6 f^1_{30} + 24 f^1_{31} + 
          2 f^1_{33} + 8 f^1_{34} \notag \\
          &\quad\;+ 2 f^1_{37} + 8 f^1_{40} - 8 f^1_{41} + 
          4 f^1_{42} - 4 f^1_{43} + 4 f^1_{47} - 4 f^1_{48} - 2 f^1_{52} - 
          4 f^1_{53} + 2 f^1_{54} + 4 f^1_{55} \notag \\
          &\quad\;+ 8 f^1_{57} - 2 f^1_{59} + 
          f^1_{62} - f^1_{63} + 2 f^1_{64} + 8 f^1_{77} + 4 f^1_{79} + 4 i_{3} + 
          8 i_{4} + 8 i_{10} + 8 i_{11} - 4 i_{13}) V^1_{18} \notag \\
  &+ (-6 b^1_{10} + 14 b^1_{11} - 2 f^1_{13} + 4 f^1_{14} - 
          6 f^1_{30} - 24 f^1_{31} - 8 f^1_{40} + 8 f^1_{41} - 4 f^1_{47} - 
          4 f^1_{50} + 2 f^1_{52} \notag \\
          &\quad\;+ 4 f^1_{53} - 2 f^1_{54} - 4 f^1_{55} - 
          8 f^1_{57} - 2 f^1_{58} - 2 f^1_{61} - 2 f^1_{64} - 4 i_{3} - 
          8 i_{4} - 8 i_{8} - 16 i_{9} - 8 i_{10} \notag \\
          &\quad\;- 8 i_{11} + 
          8 i_{12} + 4 i_{13}) V^1_{19} \notag \\
  &+ (-4 b^1_{7} - 8 b^1_{8} - 6 b^1_{10} + 14 b^1_{11} + 
          4 b^1_{13} - 2 f^1_{13} + 4 f^1_{14} - f^1_{17} + 2 f^1_{18} - f^1_{21} - 
          2 f^1_{23} \notag \\
          &\quad\;- 6 f^1_{30} - 24 f^1_{31} - 3 f^1_{33} - 12 f^1_{34} - 
          3 f^1_{36} - 6 f^1_{37} - 8 f^1_{40} + 8 f^1_{41} - 4 f^1_{42} + 
          4 f^1_{43} \notag \\
          &\quad\;- 4 f^1_{47} + 4 f^1_{48} - 4 f^1_{51} + 4 f^1_{53} - 
          2 f^1_{54} - 4 f^1_{55} - 8 f^1_{57} + f^1_{63} - 4 i_{3} - 8 i_{4} - 
          2 i_{5} \notag \\
          &\quad\;- 4 i_{6} - 8 i_{12} - 4 i_{13}) V^1_{20} \notag
  \\
  &+ (-2 b^1_{7} - 2 b^1_{8} - b^1_{9} - b^1_{10} + b^1_{11} + 
          16 b^1_{12} + 2 f^1_{11} + f^1_{12} + f^1_{22} - 12 f^1_{29} - 3 f^1_{30} - 
          3 f^1_{37} \notag \\
          &\quad\;+ 4 f^1_{40} - 4 f^1_{41} - 2 f^1_{65} - 2 f^1_{66} - 
          2 f^1_{67} - f^1_{68}/2 + f^1_{69} + f^1_{71}/2 + f^1_{72} + f^1_{73}/2 + 
          f^1_{74}/2 \notag \\
          &\quad\;+ 2 i_{3} + 4 i_{4} + 12 i_{14} - 3 i_{16} - 
          6 i_{17}) V^1_{21} \notag
  \\
  &+ (b^1_{10} - b^1_{11} - 16 b^1_{12} - 2 f^1_{11} - f^1_{12} + 
          12 f^1_{29} + 3 f^1_{30} - 4 f^1_{40} + 4 f^1_{41} + 2 f^1_{67} + 
          f^1_{68} + f^1_{70} \notag \\
          &\quad\;- f^1_{71}/2 - f^1_{72} - f^1_{74}/2 - 2 i_{3} - 
          4 i_{4} + 8 i_{15} + 4 i_{16} + 4 i_{17}) V^1_{22} \notag
  \\
  &+ (-8 b^1_{7} - 4 b^1_{8} + 2 b^1_{9} + 4 b^1_{13} + 
          2 f^1_{15} + f^1_{16} + f^1_{20} - 12 f^1_{32} - 3 f^1_{33} - 3 f^1_{36} + 
          4 f^1_{42} - 4 f^1_{43} \notag \\
          &\quad\;+ 2 f^1_{70} - f^1_{71} - 2 f^1_{74} + 2 i_{5} + 
          4 i_{6} + 8 i_{15} + 4 i_{16} + 28 i_{17}) V^1_{23} \notag
  \\
  &+ (8 b^1_{7} + 4 b^1_{8} + 2 b^1_{9} - 2 f^1_{15} - f^1_{16} + 
          12 f^1_{32} + 3 f^1_{33} - 4 f^1_{42} + 4 f^1_{43} - 2 f^1_{72} - 
          f^1_{73} + 2 f^1_{74} \notag \\
          &\quad\;- 2 i_{5} - 4 i_{6} - 24 i_{17}) V^1_{24} \notag
  \\
  &+ (-f^1_{44} + 2 f^1_{86} - f^1_{89} + 4 i_{7} + 2 i_{19} + 
          2 i_{20}) V^1_{26} \notag
  \\
  &+ (f^1_{46} - f^1_{47}/2 - f^1_{58}/4 - f^1_{86} + f^1_{89}/2 + 
          i_{8} + 2 i_{9} - i_{12} - i_{19} - i_{20}) V^1_{27}\notag
  \\
  & + (f^1_{58} + 2 f^1_{90} + 4 i_{12}) V^1_{28} \notag
  \\
  &+ (f^1_{4}/2 + f^1_{24} - f^1_{84} - 4 f^1_{91} + 2 i_{18}) V^1_{29} \notag
  \\
  &+ (f^1_{9}/2 + f^1_{27} + f^1_{82}/2 - 2 f^1_{83} - f^1_{84} - 
          8 f^1_{91} + i_{2} + 4 i_{18}) V^1_{30} \notag \\
  &+ (-f^1_{5}/2 - f^1_{25} + 2 f^1_{85} + f^1_{88} + 
          8 f^1_{92} - 2 i_{19} + 2 i_{20}) V^1_{31} \notag
\end{alignat}
\begin{alignat}{3}
  &+ (f^1_{5}/2 + f^1_{6}/2 + f^1_{25} - 2 f^1_{85} + 2 f^1_{86} + 
          f^1_{87} - f^1_{88} - 8 f^1_{92} + 2 i_{1} + 2 i_{19} - 
          2 i_{20}) V^1_{32} \notag
  \\
  &+ (f^1_{11} + 2 f^1_{29} + f^1_{87} - 2 f^1_{88} + f^1_{89} - 
          8 f^1_{92} - i_{3} + 2 i_{19} - 2 i_{20}) V^1_{33} \notag
  \\
  &+ (f^1_{15} + 2 f^1_{32} + 2 f^1_{90} - i_{5}) V^1_{34} \notag \\
  &+ (f^1_{7}/2 + f^1_{26}/2 + f^1_{56} - f^1_{78} + 2 f^1_{84} + 
          8 f^1_{91} + i_{11} - 4 i_{18}) V^1_{35} \notag \\
  &+ (f^1_{8}/2 - f^1_{26}/2 - f^1_{45} + f^1_{82} - 4 f^1_{83} - 
          8 f^1_{91} + i_{1} + 2 i_{7} + 4 i_{18}) V^1_{36} \notag
  \\
  &+ (f^1_{10}/2 + f^1_{28} + f^1_{49}/2 - f^1_{79}/2 - f^1_{86} - 
          f^1_{87}/2 - 2 i_{2} - i_{8}) V^1_{37} \notag
  \\
  &+ (f^1_{19}/2 + f^1_{35} + f^1_{60} - f^1_{80} + 2 f^1_{88} + 
          8 f^1_{92} + 4 i_{12} - 4 i_{19}) V^1_{38} \notag
  \\
  &+ (f^1_{14}/2 + f^1_{31} - f^1_{53}/2 + f^1_{57} + f^1_{78} + 
          2 f^1_{84} + 8 f^1_{91} + i_{4} - i_{11} - 4 i_{18}) V^1_{39} \notag
  \\
  &+ (f^1_{18}/2 + f^1_{34} + f^1_{51} + 2 f^1_{77} + i_{6} + 
          2 i_{11}) V^1_{40} \notag \\
  &+ (-f^1_{22}/2 - f^1_{37}/2 - f^1_{65} - f^1_{66}/2 + 
          2 f^1_{85} + 4 f^1_{92} - 4 i_{14} - 2 i_{15} + 2 i_{20}) V^1_{41} \notag
  \\
  &+ (-f^1_{23}/2 + f^1_{37}/2 - f^1_{48} - f^1_{50} - 
          f^1_{59}/2 - f^1_{61}/2 - 2 f^1_{77} - f^1_{79} - f^1_{80}/2 + f^1_{87} \notag \\
          &\quad\;- f^1_{88} + f^1_{89} + f^1_{90} - 4 f^1_{92} + 2 i_{8} + 4 i_{9} - 
          2 i_{20}) V^1_{42} \notag \\
  &+ (-f^1_{16}/2 - f^1_{33}/2 + f^1_{72} + 4 f^1_{85} + 
          8 f^1_{92} + 2 i_{6} + 4 i_{20}) V^1_{43} \notag
  \\
  &+ (-f^1_{17}/2 + f^1_{33}/2 - f^1_{61} - f^1_{80} - 
          2 f^1_{88} - 8 f^1_{92} + i_{5} + 4 i_{19}) V^1_{44} \notag
  \\
  &+ (-f^1_{20}/2 - f^1_{36}/2 + f^1_{70} - f^1_{72} + f^1_{81} - 
          4 f^1_{85} - 8 f^1_{92} - 4 i_{15} - 4 i_{20}) V^1_{45} \notag
  \\
  &+ (-f^1_{21}/2 + f^1_{36}/2 + f^1_{59} + f^1_{61} + f^1_{80} + 
          2 f^1_{88} + 8 f^1_{92} - 4 i_{12} - 4 i_{19}) V^1_{46} \notag
  \\
  &+ (-f^1_{12}/2 - f^1_{30}/2 - f^1_{67} + f^1_{69}/2 - 
          f^1_{81}/2 - 2 f^1_{85} - 4 f^1_{92} + 2 i_{4} + 2 i_{15} - 
          2 i_{20}) V^1_{47} \notag
  \\
  &+ (-f^1_{13}/2 + f^1_{30}/2 + f^1_{50} + f^1_{61}/2 + 
          f^1_{79} + f^1_{80}/2 - f^1_{87} + f^1_{88} - f^1_{89} + 4 f^1_{92} \notag \\
          &\quad\;+ i_{3} - 
          4 i_{9} + 2 i_{20}) V^1_{48} \notag
  \\
  &+ (2 f^1_{55} + f^1_{64} - 4 f^1_{78} + 8 f^1_{84} + 32 f^1_{91} - 
          4 i_{10} + 2 i_{13} - 16 i_{18}) V^1_{49} \notag
  \\
  &+ (2 f^1_{44} - 2 f^1_{55} + 4 f^1_{78} - 8 f^1_{84} - 32 f^1_{91} - 
          8 i_{7} + 4 i_{10} + 16 i_{18}) V^1_{50} \notag
  \\
  &+ (-2 f^1_{46} + f^1_{52} - f^1_{54}/2 - f^1_{62}/2 + 
          f^1_{63}/4 + 2 f^1_{79} - 4 i_{9} + 2 i_{10} - i_{13}) V^1_{51} \notag
  \\
  &+ (-2 f^1_{62} + f^1_{63} - 4 f^1_{80} + 8 f^1_{85} + 
          16 f^1_{92} - 4 i_{13} + 8 i_{20}) V^1_{52} \notag
  \\
  &+ (-4 f^1_{55} - f^1_{64} + 8 f^1_{78} - 4 f^1_{82} + 
          16 f^1_{83} + 32 f^1_{91} + 8 i_{10} - 2 i_{13} - 16 i_{18}) V^1_{53} \notag
  \\
  &+ (8 f^1_{48} + 2 f^1_{59} + 16 f^1_{77} - 16 i_{8} + 
          8 i_{12}) V^1_{54} \notag
  \\
  &+ (2 f^1_{51} + f^1_{52} + 2 f^1_{65} + 2 f^1_{67} + f^1_{68}/2 + 
          f^1_{73}/2 + 4 f^1_{77} + 2 f^1_{79} + f^1_{80} - 4 f^1_{86} - 
          2 f^1_{87} \notag \\
          &\quad\;+ 2 f^1_{88} + 8 f^1_{92} + 4 i_{10} + 4 i_{11} + 
          4 i_{14} - i_{16} + 2 i_{17} - 4 i_{19}) V^1_{55} \notag
  \\
  &+ (-2 f^1_{51} - f^1_{52} + 2 f^1_{53} + f^1_{54} - 2 f^1_{65} + 
          f^1_{68}/2 + f^1_{73}/2 - 4 f^1_{77} - 2 f^1_{79} - f^1_{80} - 
          2 f^1_{88} \notag \\
          &\quad\;- 8 f^1_{92} - 4 i_{14} - 4 i_{15} - i_{16} + 
          2 i_{17} + 4 i_{19}) V^1_{56} \notag
  \\
  &+ (f^1_{62} + f^1_{74} + 2 f^1_{80} - 4 f^1_{87} + 4 f^1_{88} - 
          4 f^1_{89} - 4 f^1_{90} + 16 f^1_{92} + 4 i_{13} + 12 i_{17} + 
          8 i_{20}) V^1_{57} \notag
  \\
  &+ (2 f^1_{58} - f^1_{62} + f^1_{63} + f^1_{74} - 2 f^1_{80} - 
          4 f^1_{88} - 16 f^1_{92} + 8 i_{12} + 12 i_{17} + 8 i_{19}) V^1_{58} \notag
\end{alignat}
\begin{alignat}{3}
  &+ (-f^1_{62} + f^1_{71} - f^1_{74} - 2 f^1_{80} + 2 f^1_{81} + 
          4 f^1_{87} - 4 f^1_{88} + 4 f^1_{89} - 16 f^1_{92} \notag \\
          &\quad\;- 4 i_{13} + 
          4 i_{16} - 8 i_{17} - 8 i_{20}) V^1_{59} \notag
  \\
  &+ (f^1_{62} - f^1_{63} - f^1_{71} - f^1_{73} + 2 f^1_{80} - 2 f^1_{81} + 
          4 f^1_{88} + 16 f^1_{92} - 4 i_{16} - 4 i_{17} - 8 i_{19}) V^1_{60} \notag \\
  &+ (-f^1_{52} - f^1_{71}/2 - f^1_{73}/2 - 2 f^1_{79} - f^1_{80} - 
          f^1_{81} + 4 f^1_{86} + 2 f^1_{87} - 2 f^1_{88} - 8 f^1_{92} \notag \\
          &\quad\;- 4 i_{10} - 2 i_{16} - 2 i_{17} + 4 i_{19}) V^1_{61} \notag
  \\
  &+ (-2 f^1_{47} + f^1_{52} - f^1_{54} - f^1_{58} + f^1_{71}/2 - 
          f^1_{74}/2 + 2 f^1_{79} + f^1_{80} + f^1_{81} + 2 f^1_{88} + 8 f^1_{92} \notag \\
          &\quad\;+ 4 i_{8} - 4 i_{12} + 2 i_{16} - 4 i_{17} - 4 i_{19}) V^1_{62} \notag
  \\
  &+ (-f^1_{54}/2 + f^1_{55} + 2 f^1_{56} - 2 f^1_{57} + 
          f^1_{63}/4 - 6 f^1_{78} - 4 i_{10} + 2 i_{11} + i_{13}) V^1_{63} \notag
  \\
  &+ (f^1_{52} - f^1_{62}/2 + f^1_{64} + 2 f^1_{79} + 4 i_{10}) V^1_{64} \notag 
  \\
  &+ (-2 f^1_{49} + 2 f^1_{50} - f^1_{63}/2 + 4 f^1_{79} + 
          4 i_{8} - 8 i_{9} - 2 i_{13}) V^1_{65} \notag 
  \\
  &+ (-2 f^1_{60} + 2 f^1_{61} + f^1_{62} + 6 f^1_{80} - 
          8 i_{12} + 4 i_{13}) V^1_{66} \notag
  \\
  &+ (2 f^1_{66} - f^1_{68} - 2 f^1_{69} + 2 f^1_{70} - f^1_{71} + 
          2 f^1_{74} + 2 f^1_{81} + 8 i_{14} - 8 i_{15} - 2 i_{16} + 
          16 i_{17}) V^1_{67} \notag
  \\
  &+ (2 f^1_{66} + f^1_{68} + 2 f^1_{72} + f^1_{73} + 8 i_{14} - 
          2 i_{16} + 4 i_{17}) V^1_{68} \notag
  \\
  &+ (-f^1_{24} + 2 f^1_{91}) V^1_{69} \notag
  \\
  &+ (-f^1_{27} + 4 f^1_{91}) V^1_{70} \notag
  \\
  &+ (-f^1_{25} + 4 f^1_{92}) V^1_{71} \notag
  \\
  &+ (-f^1_{29} + 2 f^1_{92}) V^1_{72} \notag
  \\
  &- f^1_{32} V^1_{73} \notag
  \\
  &+ (-f^1_{26} + 2 f^1_{78} - 16 f^1_{91}) V^1_{74} \notag
  \\
  &+ (-f^1_{28} + f^1_{79}) V^1_{75} \notag
  \\
  &+ (-f^1_{35} + 2 f^1_{80} - 8 f^1_{92}) V^1_{76} \notag
  \\
  &+ (-f^1_{31} - 2 f^1_{78} - 8 f^1_{91}) V^1_{77} \notag
  \\
  &+ (-f^1_{34} - 4 f^1_{77}) V^1_{78} \notag
  \\
  &+ (f^1_{33} - 2 f^1_{80} - 16 f^1_{92}) V^1_{79} \notag
  \\
  &+ (f^1_{36} + 2 f^1_{80} - 2 f^1_{81} + 16 f^1_{92}) V^1_{80} \notag 
  \\
  &+ (f^1_{30} + 2 f^1_{79} + f^1_{80} + f^1_{81} + 8 f^1_{92}) V^1_{81} \notag
  \\
  &+ (f^1_{37} - 4 f^1_{77} - 2 f^1_{79} - f^1_{80} - 8 f^1_{92}) V^1_{82} \notag
  \\
  &+ 4 f^1_{78} V^1_{83} \notag
  \\
  &- 4 f^1_{79} V^1_{84} \notag
  \\
  &- 4 f^1_{80} V^1_{85} \notag
  \\
  &+ (f^1_{38} - f^1_{44}/2 - f^1_{86} - 2 i_{7} + i_{19}) V^1_{86} \notag
  \\
  &+ (f^1_{39} - f^1_{46}/2 + f^1_{82}/2 + i_{9} - 2 i_{18}) V^1_{87} \notag
\end{alignat}
\begin{alignat}{3}
  &+ (f^1_{40} + f^1_{47}/2 + f^1_{58}/4 + f^1_{89}/2 + i_{8} - 
          i_{12} + i_{20}) V^1_{88} \notag
  \\
  &+ (f^1_{42} - f^1_{58}/2 + f^1_{90} + 2 i_{12}) V^1_{89} \notag
  \\
  &+ (f^1_{51}/2 - f^1_{65}/2 - f^1_{67}/2 - 2 f^1_{75} - 3 f^1_{77} - 
          i_{11} + i_{14}) V^1_{90} \notag \\
  &+ (-f^1_{51}/2 + f^1_{53}/2 + f^1_{65}/2 + 3 f^1_{77} - 
          i_{14} - i_{15}) V^1_{91} \notag
  \\
  &+ (-f^1_{59}/2 - 2 f^1_{76} - 2 i_{12}) V^1_{92} \notag
  \\
  &+ (2 f^1_{41} - 2 f^1_{82} + 2 f^1_{86} + 8 i_{18} - 
          2 i_{19}) V^1_{93} \notag
  \\
  &+ (2 f^1_{43} - 2 f^1_{89} - 4 f^1_{90} - 4 i_{20}) V^1_{94} \notag
  \\
  &+ (-2 f^1_{55} - f^1_{64} - 12 f^1_{78} - 16 f^1_{83} + 
          8 f^1_{84} - 4 i_{10} + 2 i_{13} + 16 i_{18}) V^1_{95} \notag 
  \\
  &+ (2 f^1_{55} - 8 f^1_{75} + 12 f^1_{78} - 4 f^1_{82} + 
          16 f^1_{83} - 8 f^1_{84} + 4 i_{10}) V^1_{96} \notag
  \\
  &+ (f^1_{62}/2 - 3 f^1_{80} + 4 f^1_{85} + 2 f^1_{87} - 2 f^1_{88} - 
          2 i_{13} - 4 i_{20}) V^1_{97} \notag
  \\
  &+ (-f^1_{62}/2 + f^1_{63}/2 + 3 f^1_{80} - 4 f^1_{85} + 
          2 f^1_{88} + 4 i_{19} + 4 i_{20}) V^1_{98} \notag
  \\
  &+ (f^1_{52}/2 - f^1_{62}/4 + 4 f^1_{75} - 3 f^1_{79} + 2 f^1_{86} - 
          2 i_{10} + i_{13} - 2 i_{19}) V^1_{99} \notag
  \\
  &+ (-f^1_{52}/2 + f^1_{54}/2 + f^1_{62}/4 - f^1_{63}/4 + 
          3 f^1_{79}) V^1_{100} \notag
  \\
  &+ (-f^1_{74}/2 - 4 f^1_{76} - 2 f^1_{87} - 2 f^1_{89} - 
          2 f^1_{90} + 6 i_{17} - 4 i_{19} - 4 i_{20}) V^1_{101} \notag
  \\
  &+ (-f^1_{68} - f^1_{73} - 4 f^1_{87} - 2 i_{16} + 4 i_{17} - 
          8 i_{19}) V^1_{102} \notag
  \\
  &+ (f^1_{71}/2 + f^1_{73}/2 - 3 f^1_{81} + 2 f^1_{87} - 2 i_{16} - 
          2 i_{17} + 4 i_{19}) V^1_{103} \notag
  \\
  &+ (-f^1_{71}/2 + f^1_{74}/2 + 8 f^1_{76} + 3 f^1_{81} + 
          2 f^1_{87} + 2 f^1_{89} + 2 i_{16} - 4 i_{17} + 4 i_{19} + 
          4 i_{20}) V^1_{104} \notag
  \\
  &+ (-f^1_{45}/2 + f^1_{83} - 6 f^1_{91} - i_{7}) V^1_{105} \notag
  \\
  &+ (-f^1_{56} - f^1_{57} + 2 f^1_{84} - 24 f^1_{91} + i_{11} + 
          4 i_{18}) V^1_{106} \notag
  \\
  &+ (-f^1_{60}/2 - f^1_{61}/2 + f^1_{88} - 12 f^1_{92} + 
          2 i_{12} + 2 i_{19}) V^1_{107} \notag
  \\
  &+ (-f^1_{49}/2 - f^1_{50}/2 - f^1_{87}/2 - i_{8} - 
          2 i_{9} - i_{19}) V^1_{108} \notag
  \\
  &+ (f^1_{66}/2 - f^1_{72}/2 + 2 f^1_{85} - 12 f^1_{92} - 
          2 i_{14} - 2 i_{20}) V^1_{109} \notag
  \\
  &+ (-f^1_{69}/2 - f^1_{70}/2 + f^1_{72}/2 - 2 f^1_{85} + 
          12 f^1_{92} - 2 i_{15} + 2 i_{20}) V^1_{110} \notag
  \\
  &+ (f^1_{82} + f^1_{86} + 4 i_{18} + i_{19}) V^1_{111} \notag
  \\
  &+ (f^1_{89}/2 - i_{20}) V^1_{112} \notag
  \\
  &+ (-f^1_{83}/2 - f^1_{84}/2 - 3 f^1_{91} + i_{18}) V^1_{113} \notag
  \\
  &+ (-f^1_{85} + f^1_{87}/2 - f^1_{88} - 6 f^1_{92} + i_{19} - 
          i_{20}) V^1_{114} \notag
  \\
  &+ (b^2_{1} - f^1_{91}) V^1_{115} \notag
  \\
  &+ (b^2_{2} - f^1_{92}) V^1_{116} \notag
  \\
  &+ (-32 b^1_{1} - 4 b^1_{3} + 8 f^1_{4} - 2 f^1_{8} - 
          f^2_{14}) V^2_{1} \notag
  \\
  &+ (-b^1_{2} + 2 b^1_{3} - 7 b^1_{4}/4 - 3 b^1_{5} + b^1_{6} - 
          2 f^1_{7} - 4 f^1_{9} - f^2_{25}/2) V^2_{2} \notag
  \\
  &+ (-4 b^1_{2} - 8 b^1_{3} + b^1_{4} + 4 b^1_{5} - 4 b^1_{6} - 
          4 f^1_{6} + 8 f^1_{10} + f^2_{12} - 2 f^2_{23}) V^2_{3} \notag
\end{alignat}
\begin{alignat}{3}
  &+ (-4 b^1_{2} + 8 b^1_{3} - b^1_{4} - 4 b^1_{5} - 4 b^1_{6} - 
          4 f^1_{5} - 8 f^1_{10} - 2 f^1_{19} + f^2_{10} - 2 f^2_{21}) V^2_{4} \notag
  \\
  &+ (-6 b^1_{7} - 2 b^1_{8} + b^1_{9} + 16 b^1_{12} + 4 f^1_{12} + 
          4 f^1_{16} + 2 f^1_{20} + 2 f^2_{23}) V^2_{5} \notag
  \\
  &+ (2 b^1_{7} - 3 b^1_{9} - 16 b^1_{12} - 2 b^1_{13} - 4 f^1_{12} - 
          2 f^1_{16} - 2 f^1_{20} + 2 f^2_{21}) V^2_{6} \notag \\
  &+ (2 b^1_{7} + 2 b^1_{8} - b^1_{9} - 2 b^1_{10} + 2 b^1_{11} + 
          16 b^1_{12} - 8 f^1_{2} + 8 f^1_{11} + 4 f^1_{15} - 2 f^1_{16} - 
          2 f^1_{20} - 4 f^1_{22} \notag \\
          &\quad\;- 2 f^2_{1} - f^2_{5} - f^2_{20} + 
          2 f^2_{22}) V^2_{7} \notag
  \\
  &+ (2 b^1_{7} + 2 b^1_{8} - b^1_{9} + 2 b^1_{10} - 2 b^1_{11} - 
          16 b^1_{12} - 4 b^1_{13} - 8 f^1_{11} - 4 f^1_{15} + 2 f^1_{16}  \notag \\
          &\quad\;+ f^2_{5} - 2 f^2_{22} - 2 f^2_{24}) V^2_{8} \notag \\
  &+ (2 b^1_{7} - b^1_{9} + 2 b^1_{10} - 2 b^1_{11} - 16 b^1_{12} - 
          2 b^1_{13} + 8 f^1_{2} + 2 f^1_{3} - 8 f^1_{11} - 4 f^1_{15} + 
          4 f^1_{22}  \notag \\
          &\quad\;+ 2 f^2_{1} + f^2_{20} - 2 f^2_{22}) V^2_{9} \notag
  \\
  &+ (2 b^1_{10} - 2 b^1_{11} - 32 b^1_{12} + 8 f^1_{2} - 8 f^1_{11} - 
          4 f^1_{12} - 2 f^2_{3} - f^2_{6} + f^2_{19} + f^2_{20} - 2 f^2_{22}) V^2_{10} \notag
  \\
  &+ (2 b^1_{7} + b^1_{9} - 2 b^1_{10} + 2 b^1_{11} + 16 b^1_{12} + 
          2 b^1_{13} + 2 f^1_{3} + 8 f^1_{11} - 2 f^1_{16} - f^2_{5} + 2 f^2_{22} + 
          2 f^2_{24}) V^2_{11} \notag
  \\
  &+ (b^1_{7} + 2 b^1_{8} + b^1_{9}/2 + 2 b^1_{10} - 4 b^1_{11} - 
          b^1_{13} - 4 f^1_{14} - 2 f^1_{18} + f^2_{25}) V^2_{12} \notag
  \\
  &+ (2 b^1_{2} - 4 b^1_{3} - b^1_{7} - b^1_{8} - b^1_{9}/2 - 2 b^1_{10} + 
          4 b^1_{11} + 4 f^1_{7} + 4 f^1_{14} - f^2_{25}) V^2_{13} \notag
  \\
  &+ (-2 b^1_{7} + b^1_{9} + 2 b^1_{10} - 6 b^1_{11} + 2 b^1_{13} - 
          4 f^1_{1} + 4 f^1_{13} + 2 f^1_{17} - f^2_{13}) V^2_{14} \notag
  \\
  &+ (-2 b^1_{7} - 4 b^1_{8} + b^1_{9} - 2 b^1_{10} + 6 b^1_{11} + 
          2 b^1_{13} - 4 f^1_{1} - 4 f^1_{13} - 2 f^1_{17} - 2 f^1_{21} - 
          4 f^1_{23} - f^2_{8}) V^2_{15} \notag
  \\
  &+ (4 b^1_{2} - 8 b^1_{3} + 2 b^1_{7} + 2 b^1_{8} + b^1_{9} - 
          2 b^1_{10} + 6 b^1_{11} + 4 f^1_{5} - 4 f^1_{13} - f^2_{11}) V^2_{16} \notag
  \\
  &+ (4 b^1_{2} + 8 b^1_{3} + 2 b^1_{7} + 2 b^1_{8} + b^1_{9} + 
          2 b^1_{10} - 6 b^1_{11} + 4 f^1_{6} + 4 f^1_{13} + 4 f^1_{23} - f^2_{9} + 
          2 f^2_{15} + 2 f^2_{17}) V^2_{17} \notag
  \\
  &+ (b^1_{4} + 4 b^1_{5} - 4 b^1_{6} + 2 b^1_{7} + 2 b^1_{8} + b^1_{9} - 
          2 b^1_{10} + 6 b^1_{11} - 8 f^1_{2} - 2 f^1_{3} + 8 f^1_{10} - 
          4 f^1_{13} \notag \\
          &\quad\;- 2 f^2_{15} - 2 f^2_{17} + f^2_{19} + f^2_{20} - 
          2 f^2_{22}) V^2_{18} \notag
  \\
  &+ (-b^1_{4} - 4 b^1_{5} - 4 b^1_{6} - 2 b^1_{7} - 2 b^1_{8} - 
          b^1_{9} + 2 b^1_{10} - 6 b^1_{11} + 8 f^1_{2} - 8 f^1_{10} + 4 f^1_{13} - 
          2 f^1_{19} \notag \\
          &\quad\;+ 2 f^2_{15} + 2 f^2_{17} - f^2_{19} - f^2_{20} + 
          2 f^2_{22}) V^2_{19} \notag
  \\
  &+ (-2 b^1_{8} - 2 b^1_{9} - 2 b^1_{13} - 2 f^1_{17} + 
          2 f^2_{22} + 2 f^2_{24}) V^2_{20} \notag
  \\
  &+ (4 b^1_{7} + 6 b^1_{8} - 2 b^1_{13} - 8 f^1_{2} + 2 f^1_{17} + 
          2 f^1_{21} + 4 f^1_{23} + f^2_{20} - 2 f^2_{22}) V^2_{21} \notag
  \\
  &+ (-4 b^1_{7} - 4 b^1_{8} - 2 b^1_{9} + 8 f^1_{2} + 2 f^1_{3} - 
          4 f^1_{23} - f^2_{20} + 2 f^2_{22}) V^2_{22} \notag
  \\
  &+ (-7 b^1_{4}/2 - 6 b^1_{5} + 2 b^1_{6} + b^1_{7} + 
          2 b^1_{8} + b^1_{9}/2 + 2 b^1_{10} - 4 b^1_{11} - b^1_{13} + 4 f^1_{2} + 
          f^1_{3} - 8 f^1_{9} \notag \\
          &\quad\;- 4 f^1_{14} - 2 f^1_{18} + f^2_{2} + f^2_{7}/2 - 
          f^2_{16} - 2 f^2_{18}) V^2_{23} \notag
  \\
  &+ (7 b^1_{4}/2 + 6 b^1_{5} - 2 b^1_{6} - b^1_{7} - b^1_{8} - 
          b^1_{9}/2 - 2 b^1_{10} + 4 b^1_{11} - 4 f^1_{2} + 8 f^1_{9} + 
          4 f^1_{14} \notag \\
          &\quad\;+ f^2_{4} - f^2_{7}/2 + f^2_{16} + 2 f^2_{18}) V^2_{24} \notag
  \\
  &+ (-4 b^1_{3} - f^1_{1} - 2 f^1_{8} - f^2_{16}/2 - f^2_{18}) V^2_{25} \notag
  \\
  &+ (-8 f^1_{24} - 2 f^1_{45} - f^2_{14}/2) V^2_{26} \notag
  \\
  &+ (-4 f^1_{26} - 4 f^1_{55} - 2 f^1_{64} + f^2_{25}) V^2_{27} \notag
  \\
  &+ (4 f^1_{55} - f^2_{25}) V^2_{28} \notag
\end{alignat}
\begin{alignat}{3}
  &+ (-4 f^1_{27} - 2 f^1_{46} - 2 f^1_{57}) V^2_{29} \notag
  \\
  &+ (4 f^1_{31} + 2 f^1_{34} - 4 f^1_{57} - 2 f^1_{65} - 
          2 f^1_{67}) V^2_{30} \notag
  \\
  &+ (-2 f^1_{34} + 2 f^1_{65}) V^2_{31} \notag \\
  &+ (f^1_{26} - 2 f^1_{56} + f^2_{25}/4) V^2_{32} \notag
  \\
  &+ (4 f^1_{44} + 4 f^1_{50} - f^2_{12}/2) V^2_{33} \notag
  \\
  &+ (-4 f^1_{25} - 4 f^1_{50} - 2 f^1_{61} - f^2_{10}/2) V^2_{34} \notag
  \\
  &+ (-f^1_{33} - f^1_{36} + 2 f^1_{66} - 2 f^1_{70} + 
          2 f^1_{72}) V^2_{35} \notag
  \\
  &+ (f^1_{33} - 2 f^1_{66} - f^2_{21} - f^2_{23}) V^2_{36} \notag
  \\
  &+ (2 f^1_{30} + f^1_{33} + f^1_{36} + 2 f^1_{37} - 4 f^1_{51} + 
          4 f^1_{53} - 2 f^1_{66} + 2 f^1_{69} + 2 f^1_{70}) V^2_{37} \notag
  \\
  &+ (-2 f^1_{30} - f^1_{33} - 4 f^1_{53} + 2 f^1_{66} - 
          2 f^1_{69} + f^2_{23}) V^2_{38} \notag
  \\
  &+ (8 f^1_{28} - 4 f^1_{52} + 4 f^1_{54} + 2 f^1_{62} - 
          2 f^1_{63}) V^2_{39} \notag
  \\
  &+ (-16 f^1_{28} - 4 f^1_{35} + 4 f^1_{52} - 8 f^1_{54} + 
          2 f^1_{63} + 2 f^2_{21}) V^2_{40} \notag
  \\
  &+ (8 f^1_{28} + 4 f^1_{54} + 2 f^2_{23}) V^2_{41} \notag
  \\
  &+ (-4 f^1_{38} - 4 f^1_{48} - 2 f^1_{59} - f^2_{8}/2) V^2_{42} \notag
  \\
  &+ (4 f^1_{44} + 4 f^1_{48} - f^2_{9}/2 - f^2_{15}) V^2_{43} \notag
  \\
  &+ (-4 f^1_{25} - 4 f^1_{49} - 2 f^1_{60} - f^2_{11}/2) V^2_{44} \notag
  \\
  &+ (-4 f^1_{38} + 4 f^1_{49} - f^2_{13}/2) V^2_{45} \notag
  \\
  &+ (2 f^1_{42} - 2 f^1_{66} + 2 f^1_{69} + f^2_{5}/2) V^2_{46} \notag
  \\
  &+ (-4 f^1_{32} + 2 f^1_{66} - 2 f^1_{69} - 2 f^1_{70} - 
          f^2_{5}/2 - f^2_{6}/2) V^2_{47} \notag
  \\
  &+ (-8 f^1_{29} - 4 f^1_{32} + 4 f^1_{47} + 2 f^1_{66} - 
          f^2_{5}/2) V^2_{48} \notag
  \\
  &+ (4 f^1_{32} + 4 f^1_{40} - 4 f^1_{47} - 2 f^1_{66} + 2 f^1_{70} - 
          2 f^1_{72} + f^2_{5}/2 + f^2_{6}/2) V^2_{49} \notag
  \\
  &+ (4 f^1_{32} - 2 f^1_{42} + 2 f^1_{58}) V^2_{50} \notag
  \\
  &+ (-8 f^1_{39} - 4 f^1_{47} - 4 f^1_{53} - 2 f^1_{58} - f^2_{3} - 
          f^2_{19}/2 - f^2_{20}/2) V^2_{51} \notag
  \\
  &+ (8 f^1_{39} + 4 f^1_{47} - 4 f^1_{51} + 4 f^1_{53} - f^2_{1} + 
          f^2_{20}/2) V^2_{52} \notag
  \\
  &+ (8 f^1_{28} + 8 f^1_{39} - 4 f^1_{49} - f^2_{24}) V^2_{53} \notag
  \\
  &+ (-8 f^1_{28} - 2 f^1_{35} - 8 f^1_{39} + 4 f^1_{49} + 
          2 f^1_{60} - f^2_{22}) V^2_{54} \notag
  \\
  &+ (2 f^1_{33} + 4 f^1_{43} - 2 f^1_{62} + 2 f^1_{63}) V^2_{55} \notag
  \\
  &+ (-2 f^1_{33} - 2 f^1_{36} - 4 f^1_{43} - 2 f^1_{63} + 
          2 f^2_{17}) V^2_{56} \notag
  \\
  &+ (4 f^1_{43} - 2 f^1_{68} - 2 f^1_{71} - 2 f^1_{73} + 
          2 f^1_{74}) V^2_{57} \notag
  \\
  &+ (-4 f^1_{43} + 2 f^1_{68}) V^2_{58} \notag
  \\
  &+ (2 f^1_{33} + 2 f^1_{68} + 2 f^1_{73} + 2 f^2_{22} + 
          2 f^2_{24}) V^2_{59} \notag
  \\
  &+ (-2 f^1_{33} - 2 f^1_{36} - 2 f^1_{68} + 2 f^1_{71}) V^2_{60} \notag
\end{alignat}
\begin{alignat}{3}
  &+ (4 f^1_{30} - 4 f^1_{43} + 4 f^1_{52} - 4 f^1_{54} - 2 f^1_{68} + 
          2 f^1_{71}) V^2_{61} \notag
  \\
  &+ (-4 f^1_{30} - 4 f^1_{37} + 4 f^1_{43} + 4 f^1_{54} + 
          2 f^1_{68} - 2 f^2_{15} - 2 f^2_{17} - f^2_{19}) V^2_{62} \notag
  \\
  &+ (8 f^1_{41} + 4 f^1_{43} - 4 f^1_{52} + 4 f^1_{54} + 2 f^1_{68} + 
          2 f^1_{73} - 2 f^2_{22}) V^2_{63} \notag \\
  &+ (-8 f^1_{41} - 4 f^1_{43} - 4 f^1_{54} - 2 f^1_{68} - 
          2 f^1_{71} - 2 f^1_{73} + 2 f^1_{74} + 2 f^2_{15} + 2 f^2_{17} + 
          f^2_{19} + f^2_{20}) V^2_{64} \notag
  \\
  &+ (-2 f^1_{30} - 4 f^1_{40} + 4 f^1_{50} + 2 f^1_{61} - 
          f^2_{17}) V^2_{65} \notag
  \\
  &+ (2 f^1_{30} + 2 f^1_{37} + 4 f^1_{40} + 4 f^1_{48} + 
          2 f^1_{59}) V^2_{66} \notag
  \\
  &+ (2 f^1_{30} + f^1_{33} + 4 f^1_{40} + 2 f^1_{42} - 4 f^1_{50} + 
          f^2_{17}) V^2_{67} \notag
  \\
  &+ (-2 f^1_{30} - f^1_{33} - f^1_{36} - 2 f^1_{37} - 4 f^1_{40} - 
          2 f^1_{42} - 4 f^1_{48}) V^2_{68} \notag
  \\
  &+ (8 f^1_{31} - 4 f^1_{41} - 4 f^1_{55} - 2 f^1_{64} - 
          2 f^2_{18}) V^2_{69} \notag
  \\
  &+ (-8 f^1_{31} - 4 f^1_{34} + 4 f^1_{41} + 2 f^1_{43} + 
          4 f^1_{55} + f^2_{16} + 2 f^2_{18}) V^2_{70} \notag
  \\
  &+ (-f^1_{42} - 2 f^1_{65} - 2 f^1_{67} + f^2_{2}/2 + 
          f^2_{4}/2) V^2_{71} \notag
  \\
  &+ (2 f^1_{40} + f^1_{42} - 4 f^1_{46} + 2 f^1_{65} - f^2_{2}/2 - 
          f^2_{16}/2) V^2_{72} \notag
  \\
  &+ (-8 f^1_{27} + 4 f^1_{39} - 4 f^1_{56} - f^2_{7}/4) V^2_{73} \notag
  \\
  &+ (f^1_{26} - f^1_{38} - 2 f^1_{45} + f^2_{18}/2) V^2_{74} \notag
  \\
  &+ (-2 f^1_{78} + 8 f^1_{83} + 2 f^1_{86}) V^2_{75} \notag
  \\
  &+ (2 f^1_{79} - 4 f^1_{82} + 2 f^1_{87} - f^2_{23}/2) V^2_{76} \notag
  \\
  &+ (-2 f^1_{79} - f^1_{80} + 4 f^1_{82} - 2 f^1_{87} + 
          2 f^1_{88} - f^2_{21}/2) V^2_{77} \notag
  \\
  &+ (-4 f^1_{77} - 4 f^1_{82} - 2 f^1_{89} - 2 f^1_{90} - 
          f^2_{15}/2 + f^2_{19}/4) V^2_{78} \notag
  \\
  &+ (-4 f^1_{75} + 4 f^1_{82} + 2 f^1_{89} + f^2_{15}/2 - 
          f^2_{19}/4 - f^2_{20}/4) V^2_{79} \notag
  \\
  &+ (-2 f^1_{84} + f^2_{25}/8) V^2_{80} \notag
  \\
  &+ (f^1_{76} - 2 f^1_{86} - f^2_{16}/4) V^2_{81} \notag
  \\
  &+ (-2 f^1_{76} - f^1_{81} + 2 f^1_{87} - f^2_{17}/2) V^2_{82} \notag
  \\
  &+ (-2 f^1_{76} - 2 f^1_{87} + 2 f^1_{88}) V^2_{83} \notag
  \\
  &+ (-2 f^1_{76} - f^1_{81} - 2 f^1_{89}) V^2_{84} \notag
  \\
  &+ (-2 f^1_{76} + 2 f^1_{89} + 2 f^1_{90}) V^2_{85} \notag
  \\
  &+ (-f^1_{80} - 4 f^1_{85} + f^2_{22}/2 + f^2_{24}/2) V^2_{86} \notag
  \\
  &+ (-4 f^1_{75} + 2 f^1_{79} + f^1_{80} + 4 f^1_{85} - 
          f^2_{22}/2) V^2_{87} \notag
  \\
  &+ (f^1_{75} - f^1_{78} - 2 f^1_{84} + f^2_{18}/4) V^2_{88} \notag
  \\
  &+ (-2 f^1_{4} - 4 f^1_{24} + f^2_{18}/2) V^3_{4} \notag
  \\
  &+ (-2 f^1_{5} - 4 f^1_{25} + f^2_{17} - f^2_{21} + f^2_{22}) V^3_{5} \notag
  \\
  &+ (-2 f^1_{6} - f^2_{15} + f^2_{16} - 2 f^2_{17} - f^2_{23} + 
          f^2_{24} - 8 i_{1}) V^3_{6} \notag
  \\
  &+ (-2 f^1_{7} - 2 f^1_{26} + f^2_{7}/4 - f^2_{18}) V^3_{7} \notag
\end{alignat}
\begin{alignat}{3}
  &+ (-2 f^1_{8} + 2 f^1_{26} - 2 f^2_{14} + f^2_{15}/2 - 
          f^2_{16}/2 - f^2_{18} - 4 i_{1}) V^3_{8} \notag
  \\
  &+ (-2 f^1_{9} - 4 f^1_{27} + f^2_{15}/4 - f^2_{16}/4 - 
          f^2_{18}/2 - 4 i_{2}) V^3_{9} \notag
  \\
  &+ (-2 f^1_{10} - 4 f^1_{28} - f^2_{12}/4 - f^2_{13}/4 + 
          f^2_{15}/2 - f^2_{16}/2 + f^2_{17} + 8 i_{2}) V^3_{10} \notag \\
  &+ (-2 f^1_{11} - 4 f^1_{29} + f^2_{21} - f^2_{22} + f^2_{23}/2 - 
          f^2_{24}/2 + 2 i_{3}) V^3_{11} \notag \\
  &+ (-2 f^1_{12} - 2 f^1_{30} - f^2_{3} - f^2_{4} - f^2_{6}/2 + 
          f^2_{21} - f^2_{22} + f^2_{23} - f^2_{24} + 8 i_{4}) V^3_{12} \notag
  \\
  &+ (-2 f^1_{13} + 2 f^1_{30} - f^2_{10}/2 - f^2_{11}/2 + 
          f^2_{21} - f^2_{22} + 4 i_{3}) V^3_{13} \notag
  \\
  &+ (-2 f^1_{14} - 4 f^1_{31} - f^2_{3}/2 - f^2_{4}/2 + 
          f^2_{7}/4 - 4 i_{4}) V^3_{14} \notag
  \\
  &+ (-2 f^1_{15} - 4 f^1_{32} - f^2_{19}/2 + f^2_{20}/2 + 
          2 i_{5}) V^3_{15} \notag
  \\
  &+ (-2 f^1_{16} - 2 f^1_{33} - f^2_{5} - 2 f^2_{21} + 2 f^2_{22} - 
          2 f^2_{23} + 2 f^2_{24} + 8 i_{6}) V^3_{16} \notag
  \\
  &+ (-2 f^1_{17} + 2 f^1_{33} + f^2_{10} + f^2_{11} + f^2_{12} + 
          f^2_{13} - 2 f^2_{21} + 2 f^2_{22} - 2 f^2_{23} + 2 f^2_{24} + 
          4 i_{5}) V^3_{17} \notag
  \\
  &+ (-2 f^1_{18} - 4 f^1_{34} + f^2_{1} + f^2_{2} + f^2_{3} + f^2_{4} - 
          4 i_{6}) V^3_{18} \notag
  \\
  &+ (-2 f^1_{19} - 4 f^1_{35} + f^2_{10} + f^2_{11} + f^2_{12} + 
          f^2_{13} - 2 f^2_{17}) V^3_{19} \notag
  \\
  &+ (-2 f^1_{20} - 2 f^1_{36} - f^2_{6} + 2 f^2_{21} - 
          2 f^2_{22}) V^3_{20} \notag
  \\
  &+ (-2 f^1_{21} + 2 f^1_{36} - 2 f^2_{8} - 2 f^2_{9} - f^2_{10} - 
          f^2_{11} - f^2_{12} - f^2_{13} + 2 f^2_{21} - 2 f^2_{22} + 2 f^2_{23} - 
          2 f^2_{24}) V^3_{21} \notag
  \\
  &+ (-2 f^1_{22} - 2 f^1_{37} - f^2_{1} - f^2_{2} - f^2_{5}/2 - 
          f^2_{21} + f^2_{22}) V^3_{22} \notag
  \\
  &+ (-2 f^1_{23} + 2 f^1_{37} + f^2_{8}/2 + f^2_{9}/2 + 
          f^2_{10}/2 + f^2_{11}/2 + f^2_{19}/2 - f^2_{20}/2 - f^2_{21} + 
          f^2_{22}) V^3_{23} \notag
  \\
  &+ (-2 f^1_{44} + f^2_{15}/2 - f^2_{16}/2 + 8 i_{7}) V^3_{44} \notag
  \\
  &+ (-2 f^1_{45} - f^2_{14} + f^2_{18}/2 + 4 i_{7}) V^3_{45} \notag
  \\
  &+ (-2 f^1_{46} + f^2_{15}/4 - f^2_{16}/4 - 4 i_{9}) V^3_{46} \notag
  \\
  &+ (-2 f^1_{47} + f^2_{19}/4 - f^2_{20}/4 + 4 i_{8}) V^3_{47} \notag
  \\
  &+ (-2 f^1_{48} - 4 f^1_{77} + f^2_{8}/4 + f^2_{9}/4 + 
          4 i_{8} - 4 i_{12}) V^3_{48} \notag
  \\
  &+ (-2 f^1_{49} + 2 f^1_{79} + f^2_{12}/4 + f^2_{13}/4 + 
          f^2_{23}/2 - f^2_{24}/2 + 4 i_{8}) V^3_{49} \notag
  \\
  &+ (-2 f^1_{50} - 2 f^1_{79} + f^2_{12}/4 + f^2_{13}/4 + 
          f^2_{17} + 8 i_{9}) V^3_{50} \notag
  \\
  &+ (-2 f^1_{51} - 4 f^1_{77} - f^2_{1}/2 - f^2_{2}/2 - 
          f^2_{3}/2 - f^2_{4}/2 - 4 i_{11}) V^3_{51} \notag
  \\
  &+ (-2 f^1_{52} - 4 f^1_{79} - 2 f^1_{80} + f^2_{15} - f^2_{16} + 
          f^2_{17} - f^2_{21} + f^2_{22} - f^2_{23} + f^2_{24} - 8 i_{10}) V^3_{52} \notag
  \\
  &+ (-2 f^1_{53} - f^2_{3}/2 - f^2_{4}/2 - 4 i_{11}) V^3_{53} \notag
  \\
  &+ (-2 f^1_{54} + f^2_{15} - f^2_{16} + 2 f^2_{17} - f^2_{23} + 
          f^2_{24} - 8 i_{10}) V^3_{54} \notag
  \\
  &+ (-2 f^1_{55} + 4 f^1_{78} + f^2_{15}/2 - f^2_{16}/2 - 
          f^2_{18} + 4 i_{10}) V^3_{55} \notag
  \\
  &+ (-2 f^1_{56} + 2 f^1_{78} - f^2_{7}/8 - 2 i_{11}) V^3_{56} \notag
  \\
  &+ (-2 f^1_{57} - 2 f^1_{78} - f^2_{7}/8 - f^2_{18}/2) V^3_{57} \notag
  \\
  &+ (-2 f^1_{58} - f^2_{19}/2 + f^2_{20}/2 - 8 i_{12}) V^3_{58} \notag
  \\
  &+ (-2 f^1_{59} - f^2_{8} - f^2_{9} + 8 i_{12}) V^3_{59} \notag
\end{alignat}
\begin{alignat}{3}
  &+ (-2 f^1_{60} + 2 f^1_{80} - f^2_{10}/2 - f^2_{11}/2 - 
          f^2_{12}/2 - f^2_{13}/2 - f^2_{21} + f^2_{22} - f^2_{23} + f^2_{24} - 
          8 i_{12}) V^3_{60} \notag
  \\
  &+ (-2 f^1_{61} - 2 f^1_{80} - f^2_{10}/2 - f^2_{11}/2 - 
          f^2_{12}/2 - f^2_{13}/2 - f^2_{17}) V^3_{61} \notag
  \\
  &+ (-2 f^1_{62} - 4 f^1_{80} + 2 f^2_{17} - 2 f^2_{21} + 
          2 f^2_{22} - 2 f^2_{23} + 2 f^2_{24} - 8 i_{13}) V^3_{62} \notag \\
  &+ (-2 f^1_{63} + 4 f^2_{17} - 2 f^2_{23} + 2 f^2_{24} - 
          8 i_{13}) V^3_{63} \notag
  \\
  &+ (-2 f^1_{64} - f^2_{15} + f^2_{16} - 4 i_{13}) V^3_{64} \notag
  \\
  &+ (-2 f^1_{65} + f^2_{1}/2 + f^2_{2}/2 - 4 i_{14} - 
          4 i_{15}) V^3_{65} \notag
  \\
  &+ (-2 f^1_{66} + f^2_{5}/2 - f^2_{21} + f^2_{22} - f^2_{23} + 
          f^2_{24} - 8 i_{14}) V^3_{66} \notag
  \\
  &+ (-2 f^1_{67} + f^2_{3}/2 + f^2_{4}/2 + 4 i_{15}) V^3_{67} \notag
  \\
  &+ (-2 f^1_{68} + f^2_{19} - f^2_{20} + 4 i_{16} - 
          8 i_{17}) V^3_{68} \notag
  \\
  &+ (-2 f^1_{69} + 2 f^1_{81} - f^2_{6}/2 - f^2_{21} + 
          f^2_{22}) V^3_{69} \notag
  \\
  &+ (-2 f^1_{70} - 2 f^1_{81} - f^2_{5}/2 - f^2_{6}/2 - f^2_{23} + 
          f^2_{24} + 8 i_{15}) V^3_{70} \notag
  \\
  &+ (-2 f^1_{71} - 4 f^1_{81} + f^2_{19} - f^2_{20} - 8 i_{16} - 
          8 i_{17}) V^3_{71} \notag
  \\
  &+ (-2 f^1_{72} - f^2_{5}/2 + f^2_{21} - f^2_{22}) V^3_{72} \notag
  \\
  &+ (-2 f^1_{73} - f^2_{19} + f^2_{20} + 2 f^2_{23} - 
          2 f^2_{24}) V^3_{73} \notag
  \\
  &+ (-2 f^1_{74} + f^2_{19} - f^2_{20} + 2 f^2_{23} - 2 f^2_{24} - 
          24 i_{17}) V^3_{74} \notag
  \\
  &+ (-2 f^1_{82} - f^2_{15}/4 + f^2_{16}/4 - 8 i_{18}) V^3_{82} \notag
  \\
  &+ (-2 f^1_{83} - 4 f^1_{91} - f^2_{18}/4) V^3_{83} \notag
  \\
  &+ (-2 f^1_{84} - 8 f^1_{91} + f^2_{18}/4 + 4 i_{18}) V^3_{84} \notag
  \\
  &+ (-2 f^1_{85} - 4 f^1_{92} - f^2_{21}/2 + f^2_{22}/2 - 
          f^2_{23}/4 + f^2_{24}/4 - 2 i_{20}) V^3_{85} \notag
  \\
  &+ (-2 f^1_{86} + f^2_{15}/4 - f^2_{16}/4 - 2 i_{19}) V^3_{86} \notag
  \\
  &+ (-2 f^1_{87} + f^2_{17} + f^2_{23}/2 - f^2_{24}/2 + 
          4 i_{19}) V^3_{87} \notag
  \\
  &+ (-2 f^1_{88} - 8 f^1_{92} + f^2_{17}/2 + f^2_{21}/2 - 
          f^2_{22}/2 + f^2_{23}/2 - f^2_{24}/2 + 4 i_{19}) V^3_{88} \notag
  \\
  &+ (-2 f^1_{89} + 4 i_{20}) V^3_{89} \notag
  \\
  &+ (-2 f^1_{90} + f^2_{19}/4 - f^2_{20}/4) V^3_{90}. \notag
\end{alignat}
The local supersymmetry is achieved when all the coefficients of the bases vanish.
Thus we have 264 simultaneous equations for 152 variables, and this requirement
is quite nontrivial\footnote{The coefficient of $V^1_{25}$ is 0, so we have 263 equations in fact.}.

\newpage
\section{Results}\label{sec:res}

The cancellation of the supersymmetric variations (\ref{var}) now gives us 
the 264 simultaneous equations among the 152 variables of $b^1_{1\sim 13}$, $b^2_{1,2}$, 
$f^1_{1\sim 92}$, $f^2_{1\sim 25}$ and $i_{1\sim 20}$.
We solved these equations by using computer programming, and found that solutions are
represented by 15 parameters.

We can choose the following 15 coefficients
\begin{alignat}{3}
  &b^1_3, \quad b^1_5, \quad b^1_8, \quad b^1_{10}, \quad b^1_{11}, \quad b^1_{12}, \quad b^1_{13},
  \quad b^2_{2}, \notag
  \\
  &i_{7}, \quad i_{11}, \quad i_{14}, \quad i_{15}, \quad i_{18}, \quad 
  i_{19}, \quad i_{20}, \label{par}
\end{alignat}
as independent parameters, and the other 137 variables are solved like
\begin{alignat}{3}
  &b^1_{1}= b^2_{2}, \notag \\ 
  &b^1_{2}= 2 b^1_{3} + 32 b^2_{2}, \notag \\ 
  &b^1_{4}= -2 b^1_{5} - 32 b^2_{2}, \notag \\ 
  &b^1_{6}= -16 b^2_{2}, \notag \\ 
  &b^1_{7}= -b^1_{8} - b^1_{10}/2 + b^1_{11} + 2 b^1_{12} + b^1_{13}/2 - 16 b^2_{2}, \notag \\ 
  &b^1_{9}= b^1_{10} - 2 b^1_{11} - 4 b^1_{12} - b^1_{13} - 32 b^2_{2}, \notag 
\\[0.1cm]
  &b^2_{1}=- b^2_{2}/4, \notag
\\[0.1cm]
  &f^1_{1}= b^1_{10}/4 - b^1_{11}/2 - b^1_{12} + 8 b^2_{2} + 8 i_{7} + 4 i_{20}, \notag \\ 
  &f^1_{2}= -b^1_{10}/2 + b^1_{11} + 2 b^1_{12} + 8 b^2_{2} + 2 i_{11} + 
        2 i_{15} + 8 i_{18} - 2 i_{19} - 4 i_{20}, \notag \\ 
  &f^1_{3}= b^1_{10}/2 - b^1_{11} - 2 b^1_{12} + 64 b^2_{2} + 8 i_{14} - 8 i_{15} - 
        16 i_{19} - 8 i_{20}, \notag \\ 
  &f^1_{4}= -b^1_{10}/16 + b^1_{11}/8 + 
        b^1_{12}/4 + 5 b^2_{2} - i_{19} - i_{20}, \label{sol} \\ 
  &f^1_{5}= 
      b^1_{10}/4 - b^1_{11}/2 - b^1_{12} - 16 b^2_{2} + 4 i_{20}, \notag \\ 
  &f^1_{6}= -4 b^1_{3} - b^1_{10}/4 + b^1_{11}/2 + b^1_{12} - 16 b^2_{2} + 
        8 i_{7} - 4 i_{19}, \notag \\ 
  &f^1_{7}= 
      b^1_{10}/4 - b^1_{11}/2 - b^1_{12} - 28 b^2_{2} - 2 i_{11} - 8 i_{18} + 
        4 i_{19} + 4 i_{20}, \notag \\ 
  &f^1_{8}= -2 b^1_{3} - b^1_{10}/8 + 
        b^1_{11}/4 + b^1_{12}/2 - 4 i_{7} - 2 i_{19} - 2 i_{20}, \notag \\ 
  &f^1_{9}= 
      b^1_{5}/8 - b^1_{10}/16 + b^1_{11}/8 + b^1_{12}/4 + 10 b^2_{2} + i_{11} - 
        i_{19} - i_{20}, \notag \\ 
  &f^1_{10}= -b^1_{5}/4 - b^1_{10}/8 + 
        b^1_{11}/4 + b^1_{12}/2 + 12 b^2_{2} + 2 i_{14} + 8 i_{18} - 
        4 i_{19} - 4 i_{20}, \notag \\
  &f^1_{11}= -b^1_{10}/8 + b^1_{11}/2 - 
        b^1_{12}/2 + 2 i_{15}, \notag \\ 
  &f^1_{12}= -b^1_{10}/4 + b^1_{11}/2 - 
        3 b^1_{12} - 16 b^2_{2} + 4 i_{15} - 16 i_{18} + 8 i_{19} + 
        4 i_{20}, \notag \\ 
  &f^1_{13}= 
      b^1_{11}/2 - 2 b^1_{12} - 32 b^2_{2} - 4 i_{11} - 16 i_{18} + 
        4 i_{19} + 8 i_{20}, \notag \\ 
  &f^1_{14}= 
      b^1_{10}/8 - b^1_{11}/4 + 3 b^1_{12}/2 + 16 b^2_{2} + 2 i_{11} + 
        16 i_{18} - 6 i_{19} - 6 i_{20}, \notag 
\end{alignat}
\begin{alignat}{3}
  &f^1_{15}= -b^1_{8}/2 - 
        b^1_{10}/4 + b^1_{11}/2 + b^1_{12} + 16 b^2_{2} + 4 i_{14} - 4 i_{15} - 
        8 i_{19} - 4 i_{20}, \notag \\
  &f^1_{16}= -b^1_{8} + b^1_{13} - 48 b^2_{2} + 
        8 i_{20}, \notag \\ 
  &f^1_{17}= -b^1_{8} - b^1_{10}/2 + b^1_{11} + 2 b^1_{12} + 
        8 i_{19} + 8 i_{20}, \notag \\ 
  &f^1_{18}= 
      b^1_{8}/2 + b^1_{10}/2 - b^1_{11} - 2 b^1_{12} - b^1_{13}/2 - 24 b^2_{2} - 
        4 i_{11} - 16 i_{18} + 8 i_{19} + 8 i_{20}, \notag \\ 
  &f^1_{19}= -b^1_{10}/2 + b^1_{11} + 2 b^1_{12} - 8 i_{14} + 
        8 i_{15} + 16 i_{19} + 8 i_{20}, \notag \\ 
  &f^1_{20}= -b^1_{10} + 
        2 b^1_{11} + 4 b^1_{12} + 64 b^2_{2} - 8 i_{15} - 8 i_{19} - 
        8 i_{20}, \notag \\ 
  &f^1_{21}= 
      b^1_{10} - 2 b^1_{11} - 4 b^1_{12} - 96 b^2_{2} - 8 i_{14} + 8 i_{15} + 
        8 i_{19}, \notag \\ 
  &f^1_{22}= 
      b^1_{10}/4 - b^1_{11}/2 - b^1_{12} - 24 b^2_{2} - 8 i_{14} - 4 i_{15} + 
        4 i_{20}, \notag \\
  &f^1_{23}= -3 b^1_{10}/4 + 3 b^1_{11}/2 + 
        3 b^1_{12} + 72 b^2_{2} + 4 i_{11} + 4 i_{14} + 16 i_{18} - 
        12 i_{19} - 16 i_{20}, \notag \\
  &f^1_{24}= -b^2_{2}/2, \notag \\ 
  &f^1_{25}= 
      4 b^2_{2}, \notag \\ 
  &f^1_{26}= 4 b^2_{2}, \notag \\ 
  &f^1_{27}= -b^2_{2}, \notag \\ 
  &f^1_{28}= 
      0, \notag \\ 
  &f^1_{29}= 
      2 b^2_{2}, \notag \\ 
  &f^1_{30}= -b^1_{10}/4 + b^1_{11}/2 + b^1_{12} - 
        8 b^2_{2}, \notag \\ 
  &f^1_{31}= 2 b^2_{2}, \notag \\
  &f^1_{32}= 0, \notag \\ 
  &f^1_{33}= 
      16 b^2_{2}, \notag \\
  &f^1_{34}= -8 b^2_{2} + 4 i_{19}, \notag \\ 
  &f^1_{
        35}= -8 b^2_{2}, \notag \\ 
  &f^1_{36}= 
      b^1_{10}/2 - b^1_{11} - 2 b^1_{12} - 16 b^2_{2}, \notag \\ 
  &f^1_{37}= 
      16 b^2_{2} - 4 i_{19}, \notag \\ 
  &f^1_{38}= -b^1_{10}/4 + b^1_{11}/2 + 
        b^1_{12} + 12 b^2_{2} + 4 i_{7} - 4 i_{19} - 4 i_{20}, \notag \\ 
  &f^1_{39}= -2 b^2_{2} - i_{11}, \notag \\ 
  &f^1_{40}= 
      8 b^2_{2} - 2 i_{15} - 2 i_{19} - 4 i_{20}, \notag \\ 
  &f^1_{41}= 
      b^1_{10}/4 - b^1_{11}/2 - b^1_{12} - 12 b^2_{2} - 8 i_{18} + 4 i_{19} + 
        4 i_{20}, \notag \\ 
  &f^1_{42}= 
      b^1_{10}/4 - b^1_{11}/2 - b^1_{12} - 32 b^2_{2} - 4 i_{14} + 4 i_{15} + 
        8 i_{19} + 12 i_{20}, \notag \\ 
  &f^1_{43}= -b^1_{10}/4 + b^1_{11}/2 + 
        b^1_{12} - 4 i_{20}, \notag \\ 
  &f^1_{44}= -b^1_{10}/4 + b^1_{11}/2 + 
        b^1_{12} + 12 b^2_{2} + 4 i_{7} - 2 i_{19} - 4 i_{20}, \notag 
\end{alignat}
\begin{alignat}{3}
  &f^1_{45}= 
      b^1_{10}/16 - b^1_{11}/8 - b^1_{12}/4 - 2 i_{7} + i_{19} + i_{20}, \notag \\ 
  &f^1_{46}= -2 b^2_{2} - i_{11} - 4 i_{18}, \notag \\ 
  &f^1_{47}= -b^1_{10}/8 + b^1_{11}/4 + b^1_{12}/2 + 8 b^2_{2} + 
        2 i_{14} - 2 i_{19} - 4 i_{20}, \notag \\
  &f^1_{48}= -b^1_{10}/8 + 
        b^1_{11}/4 + b^1_{12}/2 + 4 b^2_{2} + 2 i_{14} - 2 i_{20}, \notag \\ 
  &f^1_{49}= -8 b^2_{2} - 2 i_{11} - 8 i_{18}, \notag \\ 
  &f^1_{50}= 
      b^1_{10}/8 - b^1_{11}/4 - b^1_{12}/2 - 12 b^2_{2} - 2 i_{14} + 2 i_{19} + 
        4 i_{20}, \notag \\ 
  &f^1_{51}= 
      8 b^2_{2} - 2 i_{14} + 8 i_{18} - 2 i_{19}, \notag \\
  &f^1_{52}= -8 b^2_{2} - 
        16 i_{18} + 4 i_{19} + 4 i_{20}, \notag \\
  &f^1_{53}= 
      4 b^2_{2} + 2 i_{15} + 8 i_{18}, \notag \\
  &f^1_{54}= 
      b^1_{10}/4 - b^1_{11}/2 - b^1_{12} - 16 b^2_{2} - 16 i_{18} + 4 i_{19} + 
        8 i_{20}, \notag \\ 
  &f^1_{55}= -b^1_{10}/8 + b^1_{11}/4 + b^1_{12}/2 + 
        12 b^2_{2} + 8 i_{18} - 2 i_{19} - 2 i_{20}, \notag \\ 
  &f^1_{56}= -b^1_{10}/16 + b^1_{11}/8 + b^1_{12}/4 + 8 b^2_{2} + 
        4 i_{18} - i_{19} - i_{20}, \notag \\ 
  &f^1_{57}= 
      4 b^2_{2} + i_{11} + 4 i_{18}, \notag \\ 
  &f^1_{58}= 
      b^1_{10}/4 - b^1_{11}/2 - b^1_{12} - 32 b^2_{2} - 4 i_{14} + 4 i_{15} + 
        8 i_{19} + 12 i_{20}, \notag \\ 
  &f^1_{59}= 
      b^1_{10}/2 - b^1_{11} - 2 b^1_{12} - 32 b^2_{2} - 4 i_{14} + 4 i_{15} + 
        8 i_{19} + 12 i_{20}, \notag \\ 
  &f^1_{60}= 
      b^1_{10}/4 - b^1_{11}/2 - b^1_{12} - 24 b^2_{2} + 4 i_{19} + 
        4 i_{20}, \notag \\ 
  &f^1_{61}= -b^1_{10}/4 + b^1_{11}/2 + b^1_{12} + 
        24 b^2_{2} + 4 i_{14} - 4 i_{15} - 8 i_{19} - 12 i_{20}, \notag \\ 
  &f^1_{62}= 
      8 i_{20}, \notag \\
  &f^1_{63}= 
      b^1_{10}/2 - b^1_{11} - 2 b^1_{12} - 16 b^2_{2} + 16 i_{20}, \notag \\ 
  &f^1_{
        64}= -8 b^2_{2}, \notag \\ 
  &f^1_{65}= -8 b^2_{2} + 4 i_{19}, \notag \\ 
  &f^1_{66}= 
      16 b^2_{2} + 4 i_{20}, \notag \\ 
  &f^1_{67}= -4 b^2_{2} - 2 i_{11} - 
        8 i_{18}, \notag \\ 
  &f^1_{68}= -b^1_{10}/2 + b^1_{11} + 2 b^1_{12} - 
        8 i_{20}, \notag \\ 
  &f^1_{69}= 
      24 b^2_{2} - 4 i_{15} - 4 i_{19} - 4 i_{20}, \notag \\ 
  &f^1_{70}= -8 b^2_{2} - 4 i_{14} + 4 i_{19} + 8 i_{20}, \notag \\ 
  &f^1_{71}= -8 i_{20}, \notag \\ 
  &f^1_{72}= 
      b^1_{10}/4 - b^1_{11}/2 - b^1_{12} - 24 b^2_{2} - 4 i_{14} + 4 i_{19} + 
        4 i_{20}, \notag \\ 
  &f^1_{73}= 16 b^2_{2} - 8 i_{19}, \notag \\ 
  &f^1_{74}= 
      16 b^2_{2} - 8 i_{19} - 8 i_{20}, \notag  
\end{alignat}
\begin{alignat}{3}
  &f^1_{75}= 
      2 b^2_{2} + 4 i_{18}, \notag \\ 
  &f^1_{76}= -b^1_{10}/8 + b^1_{11}/4 + 
        b^1_{12}/2 - 2 i_{20}, \notag \\ 
  &f^1_{77}= 2 b^2_{2} - i_{19}, \notag \\ 
  &f^1_{78}= 
      0, \notag \\ 
  &f^1_{79}= 0, \notag \\
  &f^1_{80}= 0, \notag \\
  &f^1_{81}= 
      b^1_{10}/4 - b^1_{11}/2 - b^1_{12}, \notag \\ 
  &f^1_{82}= 
      b^1_{10}/8 - b^1_{11}/4 - b^1_{12}/2 - 6 b^2_{2} - 4 i_{18} + i_{19} + 
        2 i_{20}, \notag \\
  &f^1_{83}= 
      b^1_{10}/32 - b^1_{11}/16 - b^1_{12}/8 - 3 b^2_{2}/2 + i_{19}/2 + 
        i_{20}/2, \notag \\ 
  &f^1_{84}= -b^1_{10}/32 + b^1_{11}/16 + b^1_{12}/8 + 
        3 b^2_{2} + 2 i_{18} - i_{19}/2 - i_{20}/2, \notag \\ 
  &f^1_{85}= 
      b^1_{10}/16 - b^1_{11}/8 - b^1_{12}/4 - 4 b^2_{2} + i_{19} + i_{20}, \notag \\ 
  &f^1_{86}= -b^1_{10}/8 + b^1_{11}/4 + b^1_{12}/2 + 6 b^2_{2} - 
        2 i_{19} - 2 i_{20}, \notag \\ 
  &f^1_{87}= 
      b^1_{10}/8 - b^1_{11}/4 - b^1_{12}/2 - 4 b^2_{2}, \notag \\ 
  &f^1_{88}= -4 b^2_{2} - 
        2 i_{20}, \notag \\ 
  &f^1_{89}= 
      2 i_{20}, \notag \\ 
  &f^1_{90}= -b^1_{10}/8 + b^1_{11}/4 + b^1_{12}/2 - 
        4 i_{20}, \notag \\ 
  &f^1_{91}= -b^2_{2}/4, \notag \\ 
  &f^1_{92}= b^2_{2}, \notag
\\[0.1cm]
  &f^2_{1}= -32 b^2_{2} - 8 i_{11} + 16 i_{14} + 8 i_{15} - 32 i_{18} + 8 i_{19}, \notag \\ 
  &f^2_{2}= 8 i_{11} - 8 i_{14} + 32 i_{18} + 8 i_{19}, \notag \\ 
  &f^2_{3}= -b^1_{10}/2 + b^1_{11} + 2 b^1_{12} + 96 b^2_{2} + 8 i_{11} - 
        16 i_{15} + 32 i_{18} - 24 i_{19} - 24 i_{20}, \notag \\ 
  &f^2_{4}= b^1_{10}/2 - b^1_{11} - 2 b^1_{12} - 112 b^2_{2} - 16 i_{11} + 8 i_{15} - 
        64 i_{18} + 24 i_{19} + 24 i_{20}, \notag \\ 
  &f^2_{5}= -b^1_{10} + 2 b^1_{11} + 4 b^1_{12} + 96 b^2_{2} + 16 i_{14} - 16 i_{19} - 
        16 i_{20}, \notag \\ 
  &f^2_{6}= 
      b^1_{10} - 2 b^1_{11} - 4 b^1_{12} - 96 b^2_{2} + 16 i_{15} + 16 i_{19} + 
        16 i_{20}, \notag \\ 
  &f^2_{7}= 
      b^1_{10} - 2 b^1_{11} - 4 b^1_{12} - 128 b^2_{2} - 16 i_{11} - 64 i_{18} + 
        16 i_{19} + 16 i_{20}, \notag \\ 
  &f^2_{8}= 
      b^1_{10} - 2 b^1_{11} - 4 b^1_{12} - 32 i_{7} - 16 i_{15}, \notag \\ 
  &f^2_{9}= -2 b^1_{10} + 4 b^1_{11} + 8 b^1_{12} + 128 b^2_{2} + 32 i_{7} + 
        16 i_{14} - 32 i_{19} - 32 i_{20}, \notag \\ 
  &f^2_{10}= -32 b^2_{2} + 
        16 i_{15} + 16 i_{19} + 16 i_{20}, \notag \\ 
  &f^2_{11}= -b^1_{10} + 
        2 b^1_{11} + 4 b^1_{12} + 128 b^2_{2} + 16 i_{11} + 64 i_{18} - 
        16 i_{19} - 16 i_{20}, \notag \\ 
  &f^2_{12}= -b^1_{10} + 2 b^1_{11} + 
        4 b^1_{12} + 32 i_{7} - 16 i_{14}, \notag 
\end{alignat}
\begin{alignat}{3}
  &f^2_{13}= 
      2 b^1_{10} - 4 b^1_{11} - 8 b^1_{12} - 160 b^2_{2} - 32 i_{7} - 
        16 i_{11} - 64 i_{18} + 32 i_{19} + 32 i_{20}, \notag \\ 
  &f^2_{14}= -b^1_{10}/4 + b^1_{11}/2 + b^1_{12} + 8 b^2_{2} + 8 i_{7} - 
        4 i_{19} - 4 i_{20}, \notag \\
  &f^2_{15}= -b^1_{10}/2 + b^1_{11} + 
        2 b^1_{12} + 8 i_{19} - 8 i_{20}, \notag \\ 
  &f^2_{16}= b^1_{10}/2 - b^1_{11} - 2 b^1_{12} - 48 b^2_{2} + 16 i_{19} + 
        8 i_{20}, \notag \\ 
  &f^2_{17}= b^1_{10}/2 - b^1_{11} - 2 b^1_{12} - 16 b^2_{2} + 8 i_{20}, \notag \\ 
  &f^2_{18}= -b^1_{10}/4 + b^1_{11}/2 + b^1_{12} + 16 b^2_{2} - 
        4 i_{19} - 4 i_{20}, \notag \\
  &f^2_{19}= -64 b^2_{2} - 64 i_{18} + 
        16 i_{19}, \notag \\ 
  &f^2_{20}= b^1_{10} - 2 b^1_{11} - 4 b^1_{12} - 64 b^2_{2} - 64 i_{18} + 16 i_{19} + 
        32 i_{20}, \notag \\
  &f^2_{21}= b^1_{10}/2 - b^1_{11} - 2 b^1_{12} - 48 b^2_{2} - 32 i_{18} + 8 i_{19} + 
        8 i_{20}, \notag \\ 
  &f^2_{22}= b^1_{10}/2 - b^1_{11} - 2 b^1_{12} - 48 b^2_{2} - 32 i_{18} + 8 i_{19} + 
        8 i_{20}, \notag \\ 
  &f^2_{23}= -b^1_{10}/2 + b^1_{11} + 2 b^1_{12} + 
        32 b^2_{2} + 32 i_{18} - 8 i_{19} - 16 i_{20}, \notag \\ 
  &f^2_{24}= 16 b^2_{2} + 32 i_{18}, \notag \\ 
  &f^2_{25}= -b^1_{10}/2 + b^1_{11} + 
        2 b^1_{12} + 48 b^2_{2} + 32 i_{18} - 8 i_{19} - 8 i_{20}, \notag 
\\[0.1cm]
  &i_{1}= b^1_{3} + b^1_{10}/8 - b^1_{11}/4 - b^1_{12}/2 - 2 i_{7} + 3 i_{19} + 
        2 i_{20}, \notag \\ 
  &i_{2}= -b^1_{5}/16 - 3 b^2_{2} - i_{11}/2 + i_{19}/2, \notag \\ 
  &i_{3}= b^1_{11}/4 - b^1_{12} + 2 i_{15} + 2 i_{19} + 4 i_{20}, \notag \\ 
  &i_{4}= -b^1_{12} - 16 b^2_{2} - i_{11} + i_{15} - 8 i_{18} + 
        4 i_{19} + 4 i_{20}, \notag \\ 
  &i_{5}= -b^1_{8}/2 - b^1_{10}/2 + 
        b^1_{11} + 2 b^1_{12} + 16 b^2_{2} + 4 i_{14} - 4 i_{15} - 8 i_{19} - 
        12 i_{20}, \notag \\ 
  &i_{6}= -b^1_{8}/4 - b^1_{10}/4 + b^1_{11}/2 + 
        b^1_{12} + b^1_{13}/4 + 8 b^2_{2} + 2 i_{14} - 4 i_{19} - 4 i_{20}, \notag \\
  &i_{8}= 4 b^2_{2} + i_{14} - i_{19}, \notag \\ 
  &i_{9}= -b^1_{10}/16 + b^1_{11}/8 + 
        b^1_{12}/4 + 4 b^2_{2} + i_{11}/2 + 2 i_{18} - i_{19}/2 - 
        i_{20}, \notag \\ 
  &i_{10}= 4 b^2_{2} + 4 i_{18} - i_{19}, \notag \\ 
  &i_{12}= 
      8 b^2_{2} + i_{14} - i_{15} - 2 i_{19} - i_{20}, \notag \\ 
  &i_{13}= 
      b^1_{10}/4 - b^1_{11}/2 - b^1_{12} - 8 b^2_{2} + 2 i_{19} + 4 i_{20}, \notag \\ 
  &i_{16}= -b^1_{10}/6 + b^1_{11}/3 + 2 b^1_{12}/3, \notag \\ 
  &i_{17}= -b^1_{10}/12 + b^1_{11}/6 + b^1_{12}/3 - 2 i_{20}. \notag
\end{alignat}
These solutions are the main result of our paper.
We started from the higher derivative effective action which contains 132 parameters.
From the requirement of the local supersymmetry, the number of these parameters are reduced to 15.
Therefore the higher derivative effective action has 15 parameters at this stage.

The result (\ref{sol}) is obtained by employing the computer programming, so it is 
important to justify it by comparing with the results established so far.
Let us focus on the bosonic part of the higher derivative effective action in more detail.
By inserting the result (\ref{sol}) into the effective action, the purely bosonic
part of it can be written as
\begin{alignat}{3}
  &\mathcal{L}[eR^4]_{\text{pure}} + \mathcal{L}[e\ep_{11}AR^4] \notag 
  \\
  =&+ e R_{abcd} R_{abcd} R_{efgh} R_{efgh} \times (a) \notag
  \\&
  + e R_{abcd} R_{abce} R_{dfgh} R_{efgh} \times (- 16 a) \notag
  \\&
  + e R_{abcd} R_{abef} R_{cdgh} R_{efgh} \times (2 a) \notag
  \\&
  + e R_{abcd} R_{aecg} R_{bfdh} R_{efgh} \times (16 a) 
  \\&
  + e R_{abce} R_{abdg} R_{cfdh} R_{efgh} \times (-32 a) \notag
  \\&
  + e R_{abce} R_{abdf} R_{cdgh} R_{efgh} \times
  (-\tfrac{1}{4}b) \notag
  \\&
  + e R_{abce} R_{adcg} R_{bfdh} R_{efgh} \times 
  (32 a + b) \notag
  \\&
  + e \epsilon_{11}^{\mu_1\cdots\mu_{11}} A_{\mu_1\mu_2\mu_3}
  R_{ab\mu_4\mu_5}R_{ab\mu_6\mu_7}R_{cd\mu_8\mu_9}R_{cd\mu_{10}\mu_{11}} \times
  (\tfrac{1}{24} a) \notag
  \\&
  + e \epsilon_{11}^{\mu_1\cdots\mu_{11}} A_{\mu_1\mu_2\mu_3}
  R_{ab\mu_4\mu_5}R_{bc\mu_6\mu_7}R_{cd\mu_8\mu_9}R_{da\mu_{10}\mu_{11}} \times
  (-\tfrac{1}{6} a), \notag
\end{alignat}
where we defined
\begin{alignat}{3}
  a &= b^2_2, \qquad
  b &= - b^1_{10} + 2 b^1_{11} + 4 b^1_{12}.
\end{alignat}
It is very surprising that the bosonic part of the higher derivative corrections
are controlled by only two parameters. This means that there are at most two 
superinvariants for the higher derivative corrections.
The remaining 13 parameters are redundant and will be related to some linear combinations
of $a$ and $b$ when the cancellation which include the 4-form field strength is examined.

The bosonic part of the higher derivative effective action with the parameter $a$,
which is noted $\mc{L}_a$, is deformed as
\begin{alignat}{3}
  \mathcal{L}_{a} &=
  a \big( + e R_{abcd} R_{abcd} R_{efgh} R_{efgh}
  - 16 e R_{abcd} R_{abce} R_{dfgh} R_{efgh} \notag
  \\&\qquad\;
  + 2 e R_{abcd} R_{abef} R_{cdgh} R_{efgh}
  + 16 e R_{abcd} R_{aecg} R_{bfdh} R_{efgh} \notag
  \\&\qquad\;
  - 32 e R_{abce} R_{abdg} R_{cfdh} R_{efgh}
  + 32 e R_{abce} R_{adcg} R_{bfdh} R_{efgh} \notag
  \\&\qquad\;
  + \tfrac{1}{24} \epsilon_{11}^{\mu_1\cdots\mu_{11}} A_{\mu_1\mu_2\mu_3}
  R_{ab\mu_4\mu_5}R_{ab\mu_6\mu_7}R_{cd\mu_8\mu_9}R_{cd\mu_{10}\mu_{11}} \notag
  \\&\qquad\;
  - \tfrac{1}{6} \epsilon_{11}^{\mu_1\cdots\mu_{11}} A_{\mu_1\mu_2\mu_3}
  R_{ab\mu_4\mu_5}R_{bc\mu_6\mu_7}R_{cd\mu_8\mu_9}R_{da\mu_{10}\mu_{11}} \big) \notag
  \\[0.1cm]
  &= \tfrac{1}{12}a \big( t_8 t_8 e R^4 - \tfrac{1}{12} \epsilon_{11} t_8 A R^4 \big),
\end{alignat}
where $t_8$ is a tensor with 8 indices and defined in the appendix \ref{CC}.
As discussed in the introduction, this form precisely matches with the result obtained 
by evaluating the one-loop scattering amplitude of massless closed strings.
Though we do not explain explicitly, it is also checked that the result in ref. \cite{PVW}, 
which includes the bilinear terms of the Majorana gravitino, can be reproduced by appropriately
choosing the remaining 13 parameters.

The bosonic part of the higher derivative effective action with the parameter $b$,
which is noted $\mc{L}_b$, is transformed into
\begin{alignat}{3}
\label{final}
  \mathcal{L}_{b} &=
  b \big(- \tfrac{1}{4} e R_{abce} R_{abdf} R_{cdgh} R_{efgh}
  + e R_{abce} R_{adcg} R_{bfdh} R_{efgh} \big) \notag
  \\
  &= \tfrac{1}{24 \times 32}b \big( t_8 t_8 e R^4 
  + \tfrac{1}{4!} \epsilon_{11} \epsilon_{11} e R^4 \big).
\end{alignat}
Again as discussed in the introduction, this form precisely matches with the result obtained 
by evaluating the tree level scattering amplitude of massless closed strings.

Therefore we could derive the bosonic terms of the two superinvariants completely by imposing
the $\mathcal{N}=1$ local supersymmetry in eleven dimensions.
Note that the Noether method is very sensitive to miscalculations. One error is fatal
to the result. So the expected conclusion here implies the correctness of our procedure.

Let us remind the discussions in the section \ref{sec:ov}.
So far we have investigated the cancellation of $V=0$ in the eq.~(\ref{cond1}),
where the terms which are proportional to the field equations are neglected.
Our final task is to go into a question for the modifications of 
the supersymmetric transformation rules in the eq.~(\ref{cond}).
Here we only consider the modification of the transformation rule for the Majorana gravitino.

The variations which contain the field equation of the Majorana gravitino 
only come from those of $\mc{L}[eR^3\bar{\psi}\psi_{(2)}]$ and are written as
\begin{alignat}{3}
&\bar{X}_x E(\psi)^x \subset \notag \\
&-2 e(f^1_1 R_{afbg}R_{acde}R_{bcde} + f^1_2 R_{afbc}R_{agde}R_{bcde}
+ f^1_3 R_{bfad}R_{cgae}R_{bcde}) \bar{\epsilon} \gamma_f D_h \psi_{gh} \notag
\\[0.1cm]
& -2 e(f^1_{24} R_{efhi}R_{abcd}R_{abcd} + f^1_{25} R_{efah}R_{bicd}R_{abcd}
  + f^1_{26} R_{hiae}R_{bfcd}R_{abcd} \notag \\
&\quad\;\, + f^1_{27} R_{efab}R_{hicd}R_{abcd} 
  + f^1_{28} R_{ehab}R_{ficd}R_{abcd} + f^1_{29} R_{efab}R_{ahcd}R_{bicd}
 \notag\\
&\quad\;\, + f^1_{30} R_{ehab}R_{afcd}R_{bicd} + f^1_{31} R_{hiab}R_{aecd}R_{bfcd} 
 + f^1_{32} R_{efac}R_{bhad}R_{bicd} \notag\\
&\quad\;\,+ f^1_{33} R_{ehac}R_{bfad}R_{bicd}
  + f^1_{34} R_{hiac}R_{bead}R_{bfcd} + f^1_{35} R_{aech}R_{bfdi}R_{abcd} \notag \\
&\quad\;\, + f^1_{36} R_{aech}R_{bfad}R_{bicd} + f^1_{37} R_{aebh}R_{facd}R_{ibcd}
  ) \bar{\epsilon} \gamma_{ef}\gamma_{g} D_g \psi_{hi} 
\\[0.1cm]
& -2 e (f^1_{38} R_{efah}R_{bgcd}R_{abcd} + f^1_{39} R_{efab}R_{ghcd}R_{abcd}
  + f^1_{40} R_{efab}R_{agcd}R_{bhcd}\notag\\ 
&\quad\;\,+ f^1_{41} R_{ehab}R_{afcd}R_{bgcd} 
+ f^1_{42} R_{efac}R_{bgad}R_{bhcd} + f^1_{43} R_{ehac}R_{bfad}R_{bgcd}
  ) \bar{\epsilon} \gamma_{efg} D_i \psi_{hi} \notag 
\\[0.1cm]
& -2 e (f^1_{75} R_{deai}R_{fgbc}R_{ahbc} + f^1_{76} R_{deab}R_{fgac}R_{bhci}
  ) \bar{\epsilon} \gamma_{defgh} D_j \psi_{ij} \notag 
\\[0.1cm]
&-2 e(f^1_{77} R_{deai}R_{fgbc}R_{ajbc} + f^1_{78} R_{ijad}R_{efbc}R_{agbc}
  + f^1_{79} R_{deai}R_{fjbc}R_{agbc}  \notag \\
&\quad\;\,+ f^1_{80} R_{debi}R_{afcj}R_{agbc} 
 + f^1_{81} R_{deab}R_{fiac}R_{bgcj} ) \bar{\epsilon} 
\gamma_{defg}\gamma_{h} D_h \psi_{ij} \notag 
\\[0.1cm]
& -2 e(f^1_{91} R_{cdjk}R_{efab}R_{ghab} + f^1_{92} R_{cdaj}R_{efbk}R_{ghab}
  ) \bar{\epsilon} \gamma_{cdefgh}\gamma_{i} D_i \psi_{jk} . \notag
\end{alignat}
Then by using the eq.~(\ref{feq2}), we can read off the modification of 
supersymmetric transformation rule for the Majorana gravitino, $\delta_1 \psi_x = - X_x$,
as follows.
\begin{alignat}{3}
  &\delta_1 \psi_x = \notag 
\\[0.1cm]
  &+ \big(- \eta_{xg} \gamma_h + \tfrac{1}{9} \gamma_x \gamma_{gh} \big) \gamma_f  \notag\\
  &\quad D_h \{  (f^1_1 R_{afbg}R_{acde}R_{bcde} + f^1_2 R_{afbc}R_{agde}R_{bcde}  
  + f^1_3 R_{bfad}R_{cgae}R_{bcde}) \epsilon \} \notag
\\[0.1cm]
  & +  \big( -2 \eta_{k[h} \eta_{i]x} - 
  \tfrac{2}{9} \gamma_x \gamma_{[h} \eta_{i]k} \big) \gamma_{ef} \notag \\
  &\quad D_k \{  ( f^1_{24} R_{efhi}R_{abcd}R_{abcd} + f^1_{25} R_{efah}R_{bicd}R_{abcd}
    + f^1_{26} R_{hiae}R_{bfcd}R_{abcd} \notag \\
  &\qquad + f^1_{27} R_{efab}R_{hicd}R_{abcd} 
    + f^1_{28} R_{ehab}R_{ficd}R_{abcd} + f^1_{29} R_{efab}R_{ahcd}R_{bicd}
  \notag \\
  &\qquad + f^1_{30} R_{ehab}R_{afcd}R_{bicd} + f^1_{31} R_{hiab}R_{aecd}R_{bfcd} 
   + f^1_{32} R_{efac}R_{bhad}R_{bicd} \notag \\
  &\qquad + f^1_{33} R_{ehac}R_{bfad}R_{bicd}
    + f^1_{34} R_{hiac}R_{bead}R_{bfcd} + f^1_{35} R_{aech}R_{bfdi}R_{abcd} 
  \notag \\
  &\qquad + f^1_{36} R_{aech}R_{bfad}R_{bicd} + f^1_{37} R_{aebh}R_{facd}R_{ibcd}
  ) \epsilon \}  \notag
\\[0.1cm]
  &- \big(  -\eta_{xh} \gamma_i + \tfrac{1}{9} \gamma_x \gamma_{hi} \big) \gamma_{efg} \notag\\  
  &\quad  D_i \{ (f^1_{38} R_{efah}R_{bgcd}R_{abcd} + f^1_{39} R_{efab}R_{ghcd}R_{abcd}
  + f^1_{40} R_{efab}R_{agcd}R_{bhcd} \notag\\  
  &\qquad + f^1_{41} R_{ehab}R_{afcd}R_{bgcd} 
  + f^1_{42} R_{efac}R_{bgad}R_{bhcd} + f^1_{43} R_{ehac}R_{bfad}R_{bgcd}
  ) \epsilon   \}  \label{mtr}
\\[0.1cm]
  & + ( - \eta_{xi} \gamma_j + \tfrac{1}{9} \gamma_x \gamma_{ij} ) \gamma_{defgh} \notag\\ 
  &\quad  D_j \{ (f^1_{75} R_{deai}R_{fgbc}R_{ahbc} + f^1_{76} R_{deab}R_{fgac}R_{bhci} )
  \epsilon \} \notag
\\[0.1cm]
  &- \big( -2 \eta_{k[i} \eta_{j]x} - \tfrac{2}{9} \gamma_x \gamma_{[i} \eta_{j]k} \big) 
  \gamma_{defg} \notag \\
  &\quad  D_k \{  (f^1_{77} R_{deai}R_{fgbc}R_{ajbc} + f^1_{78} R_{ijad}R_{efbc}R_{agbc}
  + f^1_{79} R_{deai}R_{fjbc}R_{agbc} \notag \\
  &\qquad + f^1_{80} R_{debi}R_{afcj}R_{agbc} 
  + f^1_{81} R_{deab}R_{fiac}R_{bgcj} ) \epsilon \} \notag
\\[0.1cm]
  &+   \big( -2 \eta_{m[j} \eta_{k]x} - \tfrac{2}{9} \gamma_x \gamma_{[j} \eta_{k]m} \big) 
  \gamma_{cdefgh} \notag \\
  &\quad  D_m \{  ( f^1_{91} R_{cdjk}R_{efab}R_{ghab} + f^1_{92} R_{cdaj}R_{efbk}R_{ghab}
  ) \epsilon  \} \notag.
\end{alignat} 
In this expression we neglect the torsion terms.  
The coefficients are chosen as the eq.~(\ref{sol}).

\section{Conclusions and Discussions}\label{sec:con}

In this paper we constructed the part of the higher derivative effective action of the M-theory 
by applying the Noether method with respect to the $\mathcal{N}=1$ local supersymmetry. 
The Noether method also makes it possible to derive the modifications of the supersymmetric
transformation rules.

The ansatz for the higher derivative effective action is given by the sum of
$\mc{L}[eR^4]$, $\mc{L}[e\ep_{11}AR^4]$, $\mc{L}[eR^3\bar{\psi}\psi_{(2)}]$ and
$\mc{L}[eR^2\bar{\psi}_{(2)}D\psi_{(2)}]$, which are given by
the eqs.~(\ref{b1}), (\ref{b2}), (\ref{f1}) and (\ref{f2}) respectively.
The ansatz contains totally 132 terms.
The variations of the ansatz are expanded by the 264 bases of 
$V[eR^4\bar{\ep}\psi]$, $V[eR^2DR\bar{\ep}\psi_{(2)}]$ and $V[eR^3\bar{\ep}D\psi_{(2)}]$,
which are given by the eqs.~(\ref{v1}), (\ref{v2}) and (\ref{v3}) respectively.
The terms of $V[eR^4\bar{\ep}\psi]$ and $V[eR^3\bar{\ep}D\psi_{(2)}]$ 
are related by the 20 identities (\ref{id}).
The results of the variations of the ansatz are expanded by the 264 bases as 
the eqs.~(\ref{vb1}), (\ref{vb2}), (\ref{vf1}) and (\ref{vf2}). 
Notice that these variations do not contain terms which are proportional to the 
field equations. 
The $\mc{N}=1$ local supersymmetry requires the cancellation of the variations,
and it gives us 264 simultaneous equations among the coefficients of the terms
in the ansatz. We found that the 15 parameters (\ref{par}) are not determined
and the other coefficients are solved like (\ref{sol}).

Among 15 parameters, only two are related to the coefficients of the bosonic part.
As a result, we obtained two candidates of the superinvariants which completely
match with the results obtained by type IIA string perturbative calculations.
We also derived the higher derivative modifications to
the supersymmetric transformation rules (\ref{mtr}).
Thus it seems that the local supersymmetry is powerful enough to determine the structure
of the higher derivative effective action.

As a next future work, we will try to construct
the effective action of the M-theory which
includes the terms which depend on the 4-form field strength\cite{DS3}.
Though there are difficulties to treat the vast number of 4-form field strength terms as well,
the techniques here, namely the calculations by hand and computer programming, 
will lead us to the complete effective action of the M-theory. 

Following works in refs.\cite{CWKM,OSV},
it has recently been shown that if we take into account higher
derivative corrections to the effective action of heterotic string theory,
the entropy of the black hole computed from the degeneracy of elementary
string states agrees with the entropy computed from the classical
calculation\cite{Dabholkar}.
So if we can obtain the higher derivative corrections including
the R-R potential terms in the type IIA superstring theory, it is interesting
to apply them to the special cases of black hole whose entropy vanishes
in the classical supergravity, for example, the supertube solutions with two charges.
The corrected supersymmetric transformations will also be useful
to investigate them.  
Applications to black hole physics, brane solutions or cosmology are also important
directions\cite{My,BCN,MO}.

\section*{Acknowledgements}

We would like to thank Gary Horowitz and Pierre Vanhove for useful discussions.
YH would like to thank Keisuke Ohashi, Kasper Peeters and Yuuichirou Shibusa for valuable communications.
This work was supported in part by the Japan Society for the Promotion of Science.
This research was also supported in part by the National Science Foundation under Grant No. PHY99-07949.

\newpage
\appendix
\addcontentsline{toc}{part}{Appendix}
\part*{Appendix}

\section{Notations}\label{notations}

Field content of the 11 dimensional supergravity consists of the vielbein $e^\mu{}_a$,
the Majorana gravitino $\psi_\mu$ and the three-form potential $A_{\mu\nu\rho}$.
The spin connection $\omega_\mu{}^{ab}$ is the gauge field for the local Lorentz group
and expressed by other fields as
\begin{alignat}{3}
  \omega_{\mu}{}^{ab} &= - e^{a\nu} e^{b\rho} e_{\mu c} \partial_{[\nu} e^c{}_{\rho]} +
  e^{b\rho} \partial_{[\rho} e^a{}_{\mu]} - e^{a\rho} \partial_{[\rho} e^b{}_{\mu]} 
  \notag\\&\quad\,
  + e^{a\nu} e^{b\rho} (\Gamma_{\mu[\nu\rho]} - \Gamma_{\nu[\rho\mu]} + \Gamma_{\rho[\nu\mu]}).
\end{alignat}
To derive this equation we used the vielbein postulate 
$D_\mu e^\nu{}_a + \Gamma^\nu{}_{\mu\rho} e^\rho{}_a = 0$.
The $\Gamma^\rho{}_{\mu\nu}$ is the connection and the 
$T^\rho{}_{\mu\nu} = 2 \Gamma^\rho{}_{[\mu\nu]}$ is the torsion.
The indices in the brackets are antisymmetrized completely.
By solving the equation of motion for the spin connection, 
the torsion is written by a bilinear of the Majorana gravitino.
\begin{alignat}{3}
  \Gamma^\rho{}_{[\mu\nu]} &= \tfrac{1}{4} \bar{\psi}_{[\mu} \gamma^\rho \psi_{\nu]}
  - \tfrac{1}{8} \bar{\psi}_\al \gamma_{\mu\nu}{}^{\rho\al\bt} \psi_\bt.
\end{alignat}
The field strengths are defined as follows.
\begin{alignat}{3}
  R^{ab}{}_{\mu\nu} &= 2 \partial_{[\mu} \omega_{\nu]}{}^{ab} 
  + 2 \omega_{[\mu}{}^a{}_c \omega_{\nu]}{}^{cb}, \qquad&
  R_{abcd} &= e^\mu{}_c e^\nu{}_d R_{ab\mu\nu},
  \notag\\
  \psi_{\mu\nu} &= 2 D_{[\mu} \psi_{\nu]}, \qquad&
  \psi_{ab} &= 2 D_{[a} \psi_{b]},
  \\
  F_{\mu\nu\rho\sigma} &= 4 \partial_{[\mu} A_{\nu\rho\sigma]}, \qquad&
  F_{abcd} &= e^\mu{}_a e^\nu{}_b e^\rho{}_c e^\sigma{}_d F_{\mu\nu\rho\sigma}.
  \notag
\end{alignat}
For the calculation, we use following notations.
\bea
&&\bar{\psi}_{\mu} = -\psi_{\mu}^T C^{-1},\quad C^{T}=-C,\quad
\bar{X}= C X^T C^{-1},\quad \bar{\gamma^{n} }=(-1)^{\frac{n(n+1)}{2}} \gamma
^{n}.\nn
&& \bar{\psi}_1 X \psi_2 =\bar{\psi}_2 \bar{X} \psi_1 ,\quad
\bar{D\ep}X\psi \simeq -\bar{\ep}D(X\psi),\quad
\bar{\psi}_1 X D \psi_2 \simeq - \bar{\psi}_2 D (\bar{X} \psi_1 )
\eea
The last approximations are the equations up to the total derivative.

The supersymmetry transformations of the building blocks are given by
\begin{alignat}{3}
  \delta e &= \bar{\ep} \gamma_a \psi_a,
  \notag\\
  \delta R_{abcd} &= - R_{abce} \bar{\ep} \gamma_e \psi_d + R_{abde} \bar{\ep} \gamma_e \psi_c
  + 2 D_{[c} \delta \omega_{d]ab} + T_{ecd} \delta \omega_{eab}.
  \notag\\
  \delta \psi_a &= d_1 D_a \ep 
  + d_2 F_{ajkl} \gamma_{jkl} \epsilon
  + d_3 F_{ijkl} \gm_{ijkla} \epsilon
  + \mathcal{O}(\psi^2), 
  \notag\\
  \delta \psi_{ab} &= \tfrac{1}{4} d_1 R_{xyab} \gamma^{xy} \epsilon 
  + 2 d_2 D_{[a} (F_{b]jkl} \gm^{jkl} \epsilon)
  + 2 d_3 D_{[a} (F^{ijkl} \gm^{ijkl}{}_{b]} \epsilon)
  + \mathcal{O}(\psi^2),
  \\
  \delta F_{abcd} &= - 4 F_{e[bcd} \bar{\epsilon} \gamma^e \psi_{a]} 
  + 4 d_4 D_{[a}(\bar{\epsilon} \gamma_{bc} \psi_{d]})
  + 4 d_5 D_{[a}(\bar{\epsilon} \gamma_{bcde} \psi_{e]}) + \mathcal{O}(\psi^3),
  \notag
\end{alignat}
where $d_1=2$, $d_2=-\tfrac{1}{18}$, $d_3=\tfrac{1}{144}$, $d_4=-3$ and $d_5=0$.

\newpage
\section{The structure of $t_8 t_8 R^4$ and $\epsilon_{11}\epsilon_{11} R^4$ \label{CC}}

In this section, we show the structure of $t_8 t_8 R^4$ and $\epsilon_{11}\epsilon_{11} R^4$
explicitly.

The definition of $t_8^{abcdefgh} R_{1 ab} R_{2 cd} R_{3 ef} R_{4 gh}$
is given by
\begin{alignat}{3}
  &t_8^{abcdefgh} R_{1 ab} R_{2 cd} R_{3 ef} R_{4 gh} \notag
  \\
  &= - 2 ( \text{tr} R_1 R_2 \text{tr} R_3 R_4
  + \text{tr} R_1 R_3 \text{tr} R_2 R_4
  + \text{tr} R_1 R_4 \text{tr} R_2 R_3 ) \notag
  \\
  &\quad\, + 8 ( \text{tr} R_1 R_2 R_3 R_4 + \text{tr} R_1 R_3 R_2 R_4 +
  \text{tr} R_1 R_3 R_4 R_2 )
  \\
  &= - 2 ( R_{1 ab} R_{2 ba} R_{3 cd} R_{4 dc}
  + R_{1 ab} R_{3 ba} R_{2 cd} R_{4 dc} + R_{1 ab} R_{4 ba} R_{2 cd} R_{3 dc} ) \notag
  \\
  &\quad\, + 8 ( R_{1 ab} R_{2 bc} R_{3 cd} R_{4 da}
  + R_{1 ab} R_{3 bc} R_{2 cd} R_{4 da} + R_{1 ab} R_{3 bc} R_{4 cd} R_{2 da} ). \notag
\end{alignat}
Then the 4-point amplitude term $t_8 t_8 R^4$ are
\begin{alignat}{3}
  &t_8^{ijklmnpq} t_8^{abcdefgh} R_{ij ab} R_{kl cd} R_{mn ef} R_{pq gh} \notag
  \\
  &= - 2 t_8^{ijklmnpq} ( R_{ij ab} R_{kl ba} R_{mn cd} R_{pq dc}
  + R_{ij ab} R_{mn ba} R_{kl cd} R_{pq dc} + R_{ij ab} R_{pq ba} R_{kl cd} R_{mn dc} )
  \notag
  \\
  &\quad\, + 8 t_8^{ijklmnpq} ( R_{ij ab} R_{kl bc} R_{mn cd} R_{pq da}
  + R_{ij ab} R_{mn bc} R_{kl cd} R_{pq da} + R_{ij ab} R_{mn bc} R_{pq cd} R_{kl da} )
  \notag
  \\
  &= + 12 (R_{abcd} R^{abcd})^2 + 24 R_{ij ab} R_{ij cd} R_{mn ab} R_{mn cd} \notag
  \\
  &\quad\, - 96 R_{ij ab} R_{ij cd} R_{mn ad} R_{mn cb}
  - 192 R_{ij ab} R_{ij bc} R_{mn cd} R_{mn da}
  \\
  &\quad\, + 192 R_{ij ab} R_{jk bc} R_{kl cd} R_{li da}
  + 384 R_{ij ab} R_{kl bc} R_{jk cd} R_{li da} \notag
  \\
  &= 12 (A_1 - 16 A_2 + 2 A_3 - 32 A_5 + 16 A_6 + 32 A_7) \; .
  \notag
\end{alignat}
Here we give definitions again. $\mu,\nu$ are general coordinates indices
and $a,b$ are Lorentz indices. $A_1, \cdots A_7$ are given by eq.(\ref{eq:pboson}).

The topological term, $\epsilon_{11} \epsilon_{11} R^4 =
\epsilon_{11}^{\alpha\beta\gamma ijklmnpq}
  {\epsilon_{11 \, \alpha\beta\gamma}}^{abcdefgh}
  R_{ij ab} R_{kl cd} R_{mn ef} R_{pq gh}$, is expressed as
\begin{alignat}{3}
 &- \tfrac{1}{3! \cdot 8!} \epsilon_{11}^{\alpha\beta\gamma ijklmnpq}
  {\epsilon_{11 \, \alpha\beta\gamma}}^{abcdefgh}
  R_{ij ab} R_{kl cd} R_{mn ef} R_{pq gh} \notag
  \\
  &= \delta^{ijklmnpq}_{[abcdefgh]} R_{ij ab} R_{kl cd} R_{mn ef} R_{pq gh}
  \notag \\
  &= \tfrac{1}{7 \cdot 5!} ( A_1 - 16 A_2 + 2 A_3 + 16 A_4 - 32 A_5
  + 16 A_6 - 32 A_7 ).
\end{alignat}
Finally we obtain
\begin{alignat}{3}
  t_8 t_8 R^4 + \tfrac{1}{4 \cdot 3!} \epsilon_{11} \epsilon_{11} R^4
  &= 12 (A_1 - 16 A_2 + 2 A_3 - 32 A_5 + 16 A_6 + 32 A_7) \notag
  \\
  &\, - 12 ( A_1 - 16 A_2 + 2 A_3 - 32 A_5 + 16 A_6 - 32 A_7 + 16 A_4 ) \notag
  \\
  &= 3 \cdot 2^8 ( A_7 - \tfrac{1}{4} A_4 ) \; ,
\end{alignat}
which corresponds to eq.(\ref{final}).

\newpage
\section{Classifications of $[R^2]$, $[R^3]$, $[R^4]$ and $[R^2DR]$} \label{BB}

\begin{itemize}
\item Classification of $[R^2]$.
\end{itemize}
The types of $[R^2]$ are classified by the positions of the contracted indices.
As an example, let us consider a quadratic term $R_{\s bcd}R_{\s bcd}$
where $b$, $c$ and $d$ are the contracted indices and blanks are arbitrary. 
This term is classified by the positions of the contracted indices as $\{3,3\}\{3\}$.
The $\{3,3\}$ shows that the number of the contracted indices in each Riemann tensor.
That is, the first and the second Riemann tensor contains three contracted indices, respectively.
The contracted index $b$ is contained in the first and the second Riemann tensor, so
the numbers $(1,2)$ are assigned for this index. Similarly for the indices $c$ and $d$, the numbers
$(1,2)$ are assigned, and totally this example has the numbers of $(1,2)^3$.
The $\{3\}$ represents the number of the power of $(1,2)^3$.
The numbers are aligned in order of rising.
Thus the example $R_{\s bcd}R_{\s bcd}$ is classified by the numbers 
of $\{3,3\}\{3\}$ which are not affected by the properties of the Riemann tensor.
The types of $[R^2]$ are classified in this way and the complete list is given as follows.
\begin{alignat}{3}
  &\{1,1\}\{1\} \qquad&& R \s\s\s _d R \s\s\s _d, \notag \\
  &\{2,2\}\{2\} \qquad&& R \s _c \s _d R \s _c \s _d, \notag \\
  &\{3,3\}\{3\} \qquad&& R \s _{bcd} R \s _{bcd}, \\
  &\{4,4\}\{4\} \qquad&& R_{abcd} R_{abcd}. \notag 
\end{alignat}
The above result is checked both by hand and by the computer programming independently.
\\
\begin{itemize}
\item Classification of $[R^3]$. 
\end{itemize}
The types of $[R^3]$ are classified by the positions of the contracted indices.
As an example, let us consider a cubic term $R_{\s\s\s c}R_{\s b \s d}R_{\s bcd}$
where $b$, $c$ and $d$ are the contracted indices and blanks are arbitrary. 
This term is classified by the positions of the contracted indices as $\{1,2,3\}\{1,2\}$.
The $\{1,2,3\}$ shows that the number of the contracted indices in each Riemann tensor.
That is, the first Riemann tensor contains one contracted index, the second does two
and the third does three.
The contracted index $c$ is contained in the first and the third Riemann tensor, so
the numbers $(1,3)$ are assigned for this index. Similarly for the indices $b$ and $d$, the numbers
$(2,3)$ are assigned, and totally this example has the numbers of $(1,3)^1(2,3)^2$.
The $\{1,2\}$ represents the numbers of the powers of $(1,3)^1$ and $(2,3)^2$.
The numbers are aligned in order of rising.
Thus the example $R_{\s\s\s c}R_{\s b \s d}R_{\s bcd}$ is classified by the numbers 
of $\{1,2,3\}\{1,2\}$ which are not affected by the properties of the Riemann tensor.
The types of $[R^3]$ are classified in this way and the complete list is given as follows. 
\begin{alignat}{3}
  &\{0,1,1\}\{1\} \qquad&& R \s\s\s\s R \s\s\s _d R \s\s\s _d, \notag \\[0.2cm]
  &\{0,2,2\}\{2\} \qquad&& R \s\s\s\s  R \s _c \s _d R \s _c \s _d, \notag \\
  &\{1,1,2\}\{1,1\} \qquad&& R \s\s\s _c R \s\s\s _d   R \s _c \s _d, \notag \\[0.2cm]
  &\{0,3,3\}\{3\} \qquad&& R \s\s\s\s R \s _{bcd} R \s _{bcd}, \notag \\
  &\{1,2,3\}\{1,2\} \qquad&& R \s\s\s _c R \s _b \s _d R \s _{bcd}, \notag \\
  &\{2,2,2\}\{1,1,1\} \qquad&& R \s _b \s _c R \s _b \s _{d}  R \s _{c} \s _{d}, \notag \\[0.2cm]
  &\{0,4,4\}\{4\} \qquad&& R \s\s\s\s R_{abcd} R_{abcd}, \\
  &\{1,3,4\}\{1,3\} \qquad&& R \s\s\s _a R \s _{bcd} R_{abcd}, \notag \\
  &\{2,2,4\}\{2,2\} \qquad&& R \s _a \s _c R \s _b \s _d R_{abcd}, \notag \\
  &\{2,3,3\}\{1,1,2\} \qquad&& R \s _a \s _b R \s _{acd} R \s _{bcd}, \quad
  R \s _a \s _c R \s _{bad} R \s _{bcd}, \notag \\[0.2cm]
  &\{2,4,4\}\{1,1,3\} \qquad&& R \s _e \s _a R_{ebcd} R_{abcd}, \notag \\
  &\{3,3,4\}\{1,2,2\} \qquad&& R \s _{eab} R \s _{ecd} R_{abcd}, \quad
  R \s _{aec} R \s _{bed} R_{abcd}, \notag \\[0.2cm]
  &\{4,4,4\}\{2,2,2\} \qquad&& R_{efab} R_{efcd} R_{abcd}, \quad
  R_{eafc} R_{ebfd} R_{abcd}. \notag
\end{alignat}
The above result is checked both by hand and by the computer programming independently.
\\
\begin{itemize}
\item Classification of $[R^4]$. 
\end{itemize}
The types of $[R^4]$ are classified by the positions of the contracted indices.
As an example, let us consider a quartic term $R \s _e \s _f R \s _{eaf} R \s _{bcd} R_{abcd}$
where $a$, $b$, $c$, $d$, $e$ and $f$ are contracted by the flat metric and blanks are arbitrary. 
This term is classified by the positions of the contracted indices as $\{2,3,3,4\}\{1,2,3\}$.
The $\{2,3,3,4\}$ shows that the number of the contracted indices in each Riemann tensor.
That is, the first Riemann tensor contains two contracted indices,
the second does three, the third does three and the fourth does four.
The contracted index $a$ is contained in the second and the fourth Riemann tensor, so
the numbers $(2,4)$ are assigned for this index. In a similar way the numbers
$(3,4)$, $(3,4)$, $(3,4)$, $(1,2)$ and $(1,2)$ are assigned for the 
indices $b$, $c$, $d$, $e$ and $f$, respectively, 
and totally this example has the numbers of $(2,4)^1(1,2)^2(3,4)^3$.
The $\{1,2,3\}$ represents the numbers of the powers of $(2,4)^1$, $(1,2)^2$ and $(3,4)^3$.
The numbers are aligned in order of rising.
Thus the example $R \s _e \s _f R \s _{eaf} R \s _{bcd} R_{abcd}$ is classified by the numbers 
of $\{2,3,3,4\}\{1,2,3\}$ which are not affected by the properties of the Riemann tensor.
The types of $[R^4]$ are classified in this way and the complete list is given as follows.
\begin{alignat}{3}
  &\{0,0,1,1\}\{1\} \qquad&& R \s\s\s\s R \s\s\s\s R \s\s\s _d R \s\s\s _d, \notag 
  \\[0.2cm]
  &\{0,0,2,2\}\{2\} \qquad&& R \s\s\s\s R \s\s\s\s  R \s _c \s _d R \s _c \s _d, \notag \\
  &\{0,1,1,2\}\{1,1\} \qquad&& R \s\s\s\s R \s\s\s _c R \s\s\s _d   R \s _c \s _d, \notag \\
  &\{1,1,1,1\}\{1,1\} \qquad&& R \s\s\s _c R \s\s\s _c R \s\s\s _d   R \s\s\s _d, \notag 
  \\[0.2cm]
  &\{0,0,3,3\}\{3\} \qquad&& R \s\s\s\s R \s\s\s\s R \s _{bcd} R \s _{bcd}, \notag \\
  &\{0,1,2,3\}\{1,2\} \qquad&& R \s\s\s\s R \s\s\s _c R \s _b \s _d R \s _{bcd}, \notag \\
  &\{1,1,1,3\}\{1,1,1\} \qquad&& R \s\s\s _b R \s\s\s _c R \s\s\s _d R \s _{bcd}, \notag \\
  &\{0,2,2,2\}\{1,1,1\} \qquad&& R \s\s\s\s R \s _b \s _c R \s _b \s _d  R \s _c \s _d, \notag \\
  &\{1,1,2,2\}\{1,1,1\} \qquad&& R \s\s\s _b R \s\s\s _c R \s _b \s _d  R \s _c \s _d, \notag \\
  &\{1,1,2,2\}\{1,2\} \qquad&& R \s\s\s _b R \s\s\s _b R \s _c \s _d R \s _c \s _d, \notag \\[0.2cm]
  &\{0,0,4,4\}\{4\} \qquad&& R \s\s\s\s R \s\s\s\s R_{abcd} R_{abcd}, \notag \\
  &\{0,1,3,4\}\{1,3\} \qquad&& R \s\s\s\s R \s\s\s _a R \s _{bcd} R_{abcd}, \notag \\
  &\{0,2,2,4\}\{2,2\} \qquad&& R \s\s\s\s R \s _a \s _c R \s _b \s _d R_{abcd}, \notag \\
  &\{1,1,2,4\}\{1,1,2\} \qquad&& R \s\s\s _a R \s\s\s _c R \s _b \s _d R_{abcd}, \notag \\
  &\{0,2,3,3\}\{1,1,2\} \qquad&& R \s\s\s\s R \s _a \s _b R \s _{acd} R \s _{bcd}, \quad
  R \s\s\s\s R \s _a \s _c R \s _{bad} R \s _{bcd}, \notag \\
  &\{1,1,3,3\}\{1,3\} \qquad&& R \s\s\s _a R \s\s\s _a R \s _{bcd} R \s _{bcd}, \notag \\
  &\{1,1,3,3\}\{1,1,2\} \qquad&& R \s\s\s _a R \s\s\s _b R \s _{acd} R \s _{bcd}, \quad 
  R \s\s\s _a R \s\s\s _b R \s _{cad} R \s _{cbd}, \\
  &\{1,2,2,3\}\{1,1,2\} \qquad&& R \s\s\s _a R \s _a \s _c R \s _b \s _d R \s _{bcd}, \notag \\
  &\{1,2,2,3\}\{1,1,1,1\} \qquad&& R \s\s\s _c R \s _a \s _b R \s _a \s _d R \s _{bcd}, \notag \\
  &\{2,2,2,2\}\{1,1,1,1\} \qquad&& R \s _a \s _b R \s _a \s _c R \s _b \s _d R \s _c \s _d, \notag \\
  &\{2,2,2,2\}\{2,2\} \qquad&& R \s _a \s _b R \s _a \s _b R \s _c \s _d R \s _c \s _d, \notag
  \\[0.2cm]
  &\{0,2,4,4\}\{1,1,3\} \qquad&& R \s\s\s\s R \s _a \s _b R_{aecd} R_{becd}, \notag \\
  &\{1,1,4,4\}\{1,4\} \qquad&& R \s\s\s _e R \s\s\s _e R_{abcd} R_{abcd}, \notag \\ 
  &\{1,1,4,4\}\{1,1,3\} \qquad&& R \s\s\s _a R \s\s\s _b R_{aecd} R_{becd}, \notag \\
  &\{0,3,3,4\}\{1,2,2\} \qquad&& R \s\s\s\s R \s _{eab} R \s _{ecd} R_{abcd}, \quad  
  R \s\s\s\s R \s _{aeb} R \s _{ced} R_{acbd}, \notag \\
  &\{1,2,3,4\}\{1,1,3\} \qquad&& R \s\s\s _e R \s _e \s _a R \s _{bcd} R_{abcd}, \notag \\ 
  &\{1,2,3,4\}\{1,2,2\} \qquad&& R \s\s\s _e R \s _a \s _b R \s _{ced} R_{acbd}, \notag \\ 
  &\{1,2,3,4\}\{1,1,1,2\} \qquad&& R \s\s\s _a R \s _e \s _b R \s _{ecd} R_{abcd}, \quad 
  R \s\s\s _a R \s _e \s _b R \s _{ced} R_{acbd}, \notag \\ 
  &\{2,2,2,4\}\{1,1,1,2\} \qquad&& R \s _e \s _a R \s _e \s _b R \s _c \s _d R_{acbd}, \notag
\end{alignat}
\begin{alignat}{3}
  &\{1,3,3,3\}\{1,1,1,2\} \qquad&& R \s\s\s _e R \s _{aeb} R \s _{acd} R \s _{bcd}, \quad 
  R \s\s\s _e R \s _{aeb} R \s _{cad} R \s _{cbd}, \notag \\
  &\{2,2,3,3\}\{2,3\} \qquad&& R \s _e \s _a R \s _e \s _a R \s _{bcd} R \s _{bcd}, \notag \\
  &\{2,2,3,3\}\{1,1,1,2\} \qquad&& R \s _e \s _a R \s _e \s _b R \s _{acd} R \s _{bcd}, \quad 
  R \s _e \s _a R \s _e \s _b R \s _{cad} R \s _{cbd}, \notag \\  
  &\{2,2,3,3\}\{1,2,2\} \qquad&& R \s _a \s _b R \s _c \s _d R \s _{aeb} R \s _{ced}, \notag \\
  &\{2,2,3,3\}\{1,1,1,1,1\} \qquad&& R \s _a \s _c R \s _b \s _d R \s _{aeb} R \s _{ced}, \quad 
  R \s _a \s _d R \s _b \s _c R \s _{aeb} R \s _{ced}, \notag \\
  &\{0,4,4,4\}\{2,2,2\} \qquad&& R \s\s\s\s R_{efab} R_{efcd} R_{abcd}, \quad
  R \s\s\s\s R_{eafb} R_{ecfd} R_{acbd}, \notag \\
  &\{1,3,4,4\}\{1,1,1,3\} \qquad&& R \s\s\s _e R \s _{aeb} R_{afcd} R_{bfcd}, \notag \\   
  &\{1,3,4,4\}\{1,1,2,2\} \qquad&& R \s\s\s _e R \s _{fab} R_{efcd} R_{abcd}, \quad
  R \s\s\s _e R \s _{afb} R  _{ecfd} R_{acbd}, \notag \\
  &\{2,2,4,4\}\{2,4\} \qquad&& R \s _e \s _f R \s _e \s _f R_{abcd} R_{abcd}, \notag \\ 
  &\{2,2,4,4\}\{1,1,1,3\} \qquad&& R \s _e \s _a R \s _e \s _b R_{afcd} R_{bfcd}, \notag \\    
  &\{2,2,4,4\}\{2,2,2\} \qquad&& R \s _e \s _f  R \s _a \s _b  R_{ecfd} R_{acbd}, \notag \\
  &\{2,2,4,4\}\{1,1,1,1,2\} \qquad&& R \s _e \s _a  R \s _f \s _b  R_{efcd} R_{abcd}, \quad    
  R \s _e \s _a  R \s _f \s _b  R_{ecfd} R_{acbd}, \notag \\
  &\{2,3,3,4\}\{1,2,3\} \qquad&& R \s _e \s _f R \s _{eaf} R \s _{bcd} R_{abcd}, \notag \\
  &\{2,3,3,4\}\{1,1,2,2\} \qquad&& R \s _e \s _f R \s _{eab} R \s _{fcd} R_{abcd}, \quad
  R \s _e \s _f R \s _{aeb} R \s _{cfd} R_{acbd}, \notag \\
  &&& R \s _a \s _b R \s _{cef} R \s _{def} R_{acbd}, \quad
  R \s _a \s _b R \s _{ecf} R \s _{edf} R_{acbd}, \notag \\
  &\{2,3,3,4\}\{1,1,1,1,2\} \qquad&& R \s _e \s _a R \s _{ebf} R \s _{fcd} R_{abcd}, \quad 
  R \s _e \s _a R \s _{fbe} R \s _{fcd} R_{abcd}, \notag \\
  &&& R \s _e \s _a R \s _{ebf} R \s _{cfd} R_{acbd}, \quad
  R \s _e \s _a R \s _{fbe} R \s _{cfd} R_{acbd}, \notag \\ 
  &\{3,3,3,3\}\{3,3\} \qquad&& R \s _{eab} R \s _{eab} R \s _{fcd} R \s _{fcd}, \notag \\  
  &\{3,3,3,3\}\{1,1,2,2\} \qquad&& R \s _{eab} R \s _{fab} R \s _{ecd} R \s _{fcd}, \quad
  R \s _{aeb} R \s _{afb} R \s _{ecd} R \s _{fcd}, \notag \\
  &&& R \s _{aeb} R \s _{afb} R \s _{ced} R \s _{cfd}, \notag \\
  &\{3,3,3,3\}\{1,1,1,1,1,1\} \qquad&& R \s _{aec} R \s _{bed} R \s _{afb} R \s _{cfd}, \quad
  R \s _{eac} R \s _{bed} R \s _{afb} R \s _{cfd}, \notag \\
  &&& R \s _{aec} R \s _{bed} R \s _{afb} R \s _{dfc}, \quad
  R \s _{eac} R \s _{bed} R \s _{afb} R \s _{dfc}, \notag \\[0.2cm]
  &\{2,4,4,4\}\{1,1,2,3\} \qquad&& R \s _e \s _f R_{eafb} R_{agcd} R_{bgcd}, \notag \\   
  &\{2,4,4,4\}\{1,1,1,2,2\} \qquad&& R \s _e \s _f R_{egab} R_{fgcd} R_{abcd}, \quad  
  R \s _e \s _f R_{eagb} R_{fcgd} R_{acbd}, \notag \\
  &\{3,3,4,4\}\{3,4\} \qquad&& R \s _{efg} R \s _{efg} R_{abcd} R_{abcd}, \notag \\ 
  &\{3,3,4,4\}\{1,1,2,3\} \qquad&& R \s _{aef} R \s _{bef} R_{agcd} R_{bgcd}, \quad
  R \s _{eaf} R \s _{ebf} R_{agcd} R_{bgcd}, \notag \\ 
  &\{3,3,4,4\}\{1,2,2,2\} \qquad&& R \s _{efg} R \s _{eab} R_{fgcd} R_{abcd}, \quad     
  R \s _{feg} R \s _{aeb} R_{fcgd} R_{acbd}, \notag \\ 
  &\{3,3,4,4\}\{1,1,1,1,1,2\} \qquad&& R \s _{aef} R \s _{beg} R_{fgcd} R_{abcd}, \quad    
  R \s _{aef} R \s _{geb} R_{fgcd} R_{abcd}, \notag \\  
  &&& R \s _{aef} R \s _{beg} R_{fcgd} R_{acbd}, \quad  
  R \s _{aef} R \s _{geb} R_{fcgd} R_{acbd}, \notag \\
  &\{3,3,4,4\}\{1,3,3\} \qquad&& R \s _{afg} R \s _{bcd} R_{eafg} R_{ebcd}, \notag 
\end{alignat}
\begin{alignat}{3}
  &\{3,3,4,4\}\{1,1,1,2,2\} \qquad&& R \s _{agb} R \s _{dfc} R_{eafb} R_{ecgd}, \quad    
  R \s _{agb} R \s _{cfd} R_{eafb} R_{ecgd}, \notag \\
  &&& R \s _{bga} R \s _{dfc} R_{eafb} R_{ecgd}, \notag 
  \\[0.2cm]
 &\{4,4,4,4\}\{4,4\} \qquad&& R_{abcd} R_{abcd} R_{efgh} R_{efgh}, \notag \\   
 &\{4,4,4,4\}\{1,1,3,3\} \qquad&& R_{abcd} R_{abce} R_{dfgh} R_{efgh}, \notag \\
 &\{4,4,4,4\}\{2,2,2,2\} \qquad&& R_{abcd} R_{abef} R_{cdgh} R_{efgh}, \quad
 R_{abcd} R_{aecg} R_{bfdh} R_{efgh}, \notag \\ 
 &\{4,4,4,4\}\{1,1,1,1,2,2\} \qquad&& R_{abce} R_{abdg} R_{cfdh} R_{efgh}, \quad
 R_{abce} R_{abdf} R_{cdgh} R_{efgh}, \notag \\
 &&& R_{abce} R_{adcg} R_{bfdh} R_{efgh}. \notag   
\end{alignat}
The result there is checked both by hand and by the computer programming independently.
\\
\begin{itemize}
\item Classification of $[R^2DR]$ with the index of the covariant derivative unfilled. 
\end{itemize}
The types of $[R^2DR]$ are classified by the positions of the contracted indices.
As an example, let us consider a quartic term $R \s _c \s _d R \s _{cbd} D_b R \s\s\s\s$,
where $b$, $c$ and $d$ are contracted by the flat metric and blanks are arbitrary. 
This term is classified by the positions of the contracted indices as $\{2,3,1\}\{1,2\}$.
The numbers of 2 and 3 in $\{2,3,1\}$ represent the numbers of the contracted indices in 
the first and the second Riemann tensor, respectively.
The last number 1 in $\{2,3,1\}$ represents the number of the contracted indices in 
the covariant derivative and the third Riemann tensor,
because the indices of the covariant derivative and the third Riemann tensor
can be exchanged by using the Bianchi identity of the Riemann tensor.
Thus the index of the covariant derivative is grouped into the position of the third Riemann tensor.
The contracted index $b$ is contained in the second Riemann tensor and the third position, so
the numbers $(2,3)$ are assigned for this index. In a similar way the numbers
$(1,2)$ are assigned both for the indices $c$ and $d$, 
and totally this example has the numbers of $(2,3)^1(1,2)^2$.
The $\{1,2\}$ represents the numbers of the powers of $(2,3)^1$ and $(1,2)^2$,
where the numbers are aligned in order of rising.
Thus the example, $R \s _c \s _d R \s _{cbd} D_b R \s\s\s\s$, is classified by the numbers 
of $\{2,3,1\}\{1,2\}$ which are not affected by the properties of the Riemann tensor.
The types of $[R^2DR]$ with the index of the covariant derivative unfilled 
are classified in this way and the complete list is given as follows.
\begin{alignat}{3}
  &\{0,1,1\}\{1\} \qquad&& R \s\s\s\s R \s\s\s _d D \s R \s\s\s _d, \notag \\
  &\{1,1,0\}\{1\} \qquad&& R \s\s\s _d R \s\s\s _d D \s R \s\s\s\s, \notag \\[0.2cm]
  &\{0,2,2\}\{2\} \qquad&& R \s\s\s\s  R \s _c \s _d D \s R \s _c \s _d, \notag \\
  &\{1,1,2\}\{1,1\} \qquad&& R \s\s\s _c R \s\s\s _d D \s R \s _c \s _d, \notag \\
  &\{1,2,1\}\{1,1\} \qquad&& R \s\s\s _c R \s _c \s _d D \s R \s\s\s _d, \notag \\
  &\{2,2,0\}\{2\} \qquad&& R \s _c \s _d R \s _c \s _d D \s R \s\s\s\s, \notag \\[0.2cm]
  &\{0,3,3\}\{3\} \qquad&& R \s\s\s\s R \s _{bcd} D \s R \s _{bcd}, \notag \\
  &\{1,2,3\}\{1,2\} \qquad&& R \s\s\s _b R \s _c \s _d D \s R \s _{cbd}, \notag \\
  &\{1,3,2\}\{1,2\} \qquad&& R \s\s\s _b R \s _{cbd} D \s R \s _c \s _d, \notag \\
  &\{2,2,2\}\{1,1,1\} \qquad&& R \s _b \s _c R \s _b \s _d D \s R \s _c \s _d, \notag \\
  &\{2,3,1\}\{1,2\} \qquad&& R \s _c \s _d R \s _{cbd} D \s R \s\s\s _b, \notag \\
  &\{3,3,0\}\{3\} \qquad&& R \s _{bcd} R \s _{bcd} D \s R \s\s\s\s, \notag \\[0.2cm]
  &\{0,4,4\}\{4\} \qquad&& R \s\s\s\s R _{abcd} D \s R _{abcd}, \\
  &\{1,3,4\}\{1,3\} \qquad&& R \s\s\s _a R \s _{bcd} D \s R _{abcd}, \notag \\
  &\{2,2,4\}\{2,2\} \qquad&& R \s _a \s _c R \s _b \s _d D \s R _{abcd}, \notag \\
  &\{1,4,3\}\{1,3\} \qquad&& R \s\s\s _a R _{abcd} D \s R \s _{bcd}, \notag \\
  &\{2,3,3\}\{1,1,2\} \qquad&& R \s _a \s _b R \s _{cad} D \s R \s _{cbd}, \quad
  R \s _a \s _b R \s _{dac} D \s R \s _{cbd}, \notag \\
  &\{2,4,2\}\{2,2\} \qquad&& R \s _a \s _c R _{abcd} D \s R \s _b \s _d, \notag \\
  &\{3,3,2\}\{1,1,2\} \qquad&& R \s _{cad} R \s _{cbd} D \s R \s _a \s _b, \quad
  R \s _{dac} R \s _{cbd} D \s R \s _a \s _b, \notag \\
  &\{3,4,1\}\{1,3\} \qquad&& R \s _{bcd} R _{abcd} D \s R \s\s\s _a, \notag \\
  &\{4,4,0\}\{4\} \qquad&& R _{abcd} R _{abcd} D \s R \s\s\s\s, \notag \\[0.2cm]
  &\{2,4,4\}\{1,1,3\} \qquad&& R \s _a \s _b R _{aecd} D \s R _{becd}, \notag \\
  &\{3,3,4\}\{1,2,2\} \qquad&& R \s _{aec} R \s _{bed} D \s R _{abcd}, \quad
  R \s _{cea} R \s _{bed} D \s R _{abcd}, \notag \\
  &\{3,4,3\}\{1,2,2\} \qquad&& R \s _{aec} R _{abcd} D \s R \s _{bed}, \quad
  R \s _{cea} R _{abcd} D \s R \s _{bed}, \notag \\
  &\{4,4,2\}\{1,1,3\} \qquad&& R _{aecd} R _{becd} D \s R \s _a \s _b, \notag 
\end{alignat}
The result above is checked both by hand and by the computer programming independently.
\\
\begin{itemize}
\item Classification of $[R^2 DR]$ with the index of the covariant derivative filled. 
\end{itemize}
The types of $[R^2DR]$ with the index of the covariant derivative filled 
are classified as like the above and the complete list is given as follows.
\begin{alignat}{3}
  &\{1,1,2\}\{1,1\} \qquad&& R \s\s\s _c R \s\s\s _d D_c R \s\s\s _d, \notag \\
  &\{0,2,2\}\{2\} \qquad&& R \s\s\s\s R \s _c \s _d D_c R \s\s\s _d, \notag \\
  &\{1,2,1\}\{1,1\} \qquad&& R \s\s\s _d R \s _c \s _d D_c R \s\s\s\s, \notag \\[0.2cm]
  &\{1,2,3\}\{1,2\} \qquad&& R \s\s\s _b  R \s _c \s _d D_b R \s _c \s _d, \quad
  R \s\s\s _c R \s _b \s _d D_b R \s _c \s _d, \notag \\
  &\{0,3,3\}\{3\} \qquad&& R \s\s\s\s  R \s _{cbd} D_b R \s _c \s _d, \notag \\
  &\{2,2,2\}\{1,1,1\} \qquad&& R \s _b \s _c R \s _c \s _d D_b R \s\s\s _d, \notag \\
  &\{1,3,2\}\{1,2\} \qquad&& R \s\s\s _c R \s _{cbd} D_b R \s\s\s _d, \quad
  R \s\s\s _c R \s _{dbc} D_b R \s\s\s _d, \notag \\
  &\{2,3,1\}\{1,2\} \qquad&& R \s _c \s _d R \s _{cbd} D_b R \s\s\s\s, \notag \\[0.2cm]
  &\{1,3,4\}\{1,3\} \qquad&& R \s\s\s _b R \s _{cad} D_a R \s _{cbd}, \quad
  R \s\s\s _b R \s _{dac} D_a R \s _{cbd}, \notag \\
  &\{0,4,4\}\{4\} \qquad&& R \s\s\s\s R_{abcd} D_a R \s _{bcd}, \notag \\
  &\{2,2,4\}\{2,2\} \qquad&& R \s _a \s _b R \s _c \s _d D_a R \s _{cbd}, \notag \\
  &\{2,3,3\}\{1,1,2\} \qquad&& R \s _a \s _b R \s _{cbd} D_a R \s _c \s _d, \quad
  R \s _b \s _c R \s _{bad} D_a R \s _c \s _d, \notag \\
  &&& R \s _b \s _c R \s _{dab} D_a R \s _c \s _d, \\
  &\{1,4,3\}\{1,3\} \qquad&& R \s\s\s _b R_{acbd} D_a R \s _c \s _d, \notag \\
  &\{3,3,2\}\{1,1,2\} \qquad&& R \s _{cad} R \s _{cbd} D_a R \s\s\s _b, \quad
  R \s _{dac} R \s _{cbd} D_a R \s\s\s _b, \notag \\
  &\{2,4,2\}\{2,2\} \qquad&& R \s _c \s _d R_{acbd} D_a R \s\s\s _b, \notag \\
  &\{3,4,1\}\{1,3\} \qquad&& R \s _{bcd} R_{abcd} D_a R \s\s\s\s, \notag \\[0.2cm]
  &\{1,4,5\}\{1,4\} \qquad&& R \s\s\s _e R _{abcd} D_e R _{abcd}, \notag \\
  &\{2,3,5\}\{2,3\} \qquad&& R \s _e \s _a R \s _{bcd} D_e R _{abcd}, \notag \\
  &\{2,4,4\}\{1,1,3\} \qquad&& R \s _a \s _b R_{ecad} D_e R \s _{cbd}, \quad
  R \s _a \s _b R_{edac} D_e R \s _{cbd}, \notag \\
  &\{3,3,4\}\{1,2,2\} \qquad&& R \s _{aeb} R \s _{ced} D_a R \s _{cbd}, \quad
  R \s _{aeb} R \s _{dec} D_a R \s _{cbd}, \notag \\
  &&& R \s _{bea} R \s _{ced} D_a R \s _{cbd}, \quad
  R \s _{bea} R \s _{dec} D_a R \s _{cbd}, \notag \\
  &\{3,4,3\}\{1,2,2\} \qquad&& R \s _{cad} R_{ecbd} D_e R \s _a \s _b, \quad
  R \s _{dac} R_{ecbd} D_e R \s _a \s _b, \notag \\
  &&& R \s _{aec} R _{abcd} D_e R \s _b \s _d, \notag \\
  &\{4,4,2\}\{1,1,3\} \qquad&& R_{ebcd} R_{abcd} D_e R \s\s\s _a, \notag \\[0.2cm]
  &\{3,4,5\}\{1,2,3\} \qquad&& R \s _{afb} R_{aecd} D_f R_{becd}, \quad
  R \s _{bfa} R _{aecd} D_f R _{becd}, \notag \\
  &\{4,4,4\}\{2,2,2\} \qquad&& R_{faec} R_{abcd} D_f R \s _{bed}, \quad
  R_{fcea} R_{abcd} D_f R \s _{bed}. \notag
\end{alignat}
The result above is checked both by hand and by the computer programming independently.

\end{document}